\documentclass[a4paper, 10pt]{article}
\usepackage{geometry}
\usepackage{amsmath}
\usepackage{amssymb}
\usepackage{amsthm}
\usepackage{amsfonts}
\usepackage{tikz}
\usepackage{graphicx}
\usepackage{xcolor}
\usepackage{enumerate}
\usepackage{subcaption}
\usepackage{microtype}
\usepackage[font=small,labelfont=bf, margin=1cm]{caption}
\usepackage{url}
\usepackage{hyperref}
\hypersetup{
    colorlinks, 
    citecolor=black,
    filecolor=black,
    linkcolor=black,
    urlcolor=black
}
\usepackage[most]{tcolorbox} 

\usepackage{fancyhdr}
\newcommand{\xmin}{x_\text{min}}
\newcommand{\xmax}{x_\text{max}}

\newtcolorbox{helpquotebox}[1][]{%
    colback=green!50!black!5,
    colframe=green!50!black!5,
    notitle,
    sharp corners,
    borderline west={2pt}{0pt}{green!80!black},
    enhanced,
    breakable,
}

\newtcolorbox{warningquotebox}[1][]{%
    colback=red!50!black!5,
    colframe=red!50!black!5,
    notitle,
    sharp corners,
    borderline west={2pt}{0pt}{red!80!black},
    enhanced,
    breakable, 
}

\newtcolorbox{msgquotebox}[1][]{%
    colback=cyan!50!black!5,
    colframe=cyan!50!black!5,
    notitle,
    sharp corners,
    borderline west={2pt}{0pt}{cyan!80!black},
    enhanced,
    breakable,
}

\newenvironment{helpquote}[1]
    {
        \vspace{3ex}
        \begin{helpquotebox}
            \textbf{#1}

            \vspace{1ex}
    }{
        \end{helpquotebox}
    }

\newenvironment{warningquote}[1]
    {
        \vspace{3ex}
        \begin{warningquotebox}
            \textbf{#1}

            \vspace{1ex}
    }{
        \end{warningquotebox}
    }

\newenvironment{msgquote}[1]
    {
        \vspace{3ex}
        \begin{msgquotebox}
            \textbf{#1}

            \vspace{1ex}
    }{
        \end{msgquotebox}
    }

\title{Venture Capital Portfolio Construction \\
\begin{small}
\textit{And the Main Factors Impacting the Optimal Strategy}
\end{small}
}
\author{Francesco Farina, Mike Arpaia, Harpal Khing, Jonas Vetterle \\ \normalsize{Moonfire Ventures} \\ \small{\texttt{\{francesco, mike, harpal, jonas\}@moonfire.com}}}
\date{}

\begin{document}
    \maketitle
    \setcounter{tocdepth}{2}
    \tableofcontents
    \newpage

    \section{Introduction}
    \label{sec:intro}

    The optimal portfolio size for a venture capital (VC) fund is a topic often debated, but there is no consensus on the best strategy.

    There are many successful VCs implementing small portfolios (like Cherry Ventures, LocalGlobe, Backed), while many others succeed by implementing large portfolios (Possible Ventures, YCombinator, Kima Ventures, 500 Startups).
    
    This is because the optimal portfolio size is a function of many factors. It is not easy to find a general formula that can be applied to all situations, and it largely depends on the goal of the fund.

    In this report, we will go through the different factors step by step, studying how they affect fund returns and the optimal portfolio size, starting with some basic assumptions and then increasing the complexity of the model.

    To explore our models in a more interactive way and to see how the different parameters we discuss in this paper jointly affect returns, one can use the portfolio simulator application~\cite{2023_portfolio_simulator} that we've created.
    
    \subsection{Probability distribution of early-stage investment returns}
    \label{sec:prob}
    Early-stage VC investments returns follow a power law distribution. This has been shown by various studies over the years. Correlation Ventures conducted two over the years by looking at over 21,000 investments in the period 2004-2013~\cite{2014_correlation_ventures} and over 27,000 in the period 2009-2018~\cite{2018_correlation_ventures}. AngelList performed a similar analysis~\cite{2019othman,2020_angellist} looking at a universe of 1,808 investments made on the platform. They also looked at the distribution for investments returning 0-1x~\cite{2019othman,2022_angellist}, showing an average -81\% performance. Horsley Bridge, an LP in several US VC funds, looked at 7,000 of its investments over the course of 1975-2014~\cite{2017_alex_graham}.

    Data in these studies, corroborated by internal proprietary data, suggest that a power law distribution with parameter $\alpha=2.05$ and minimum return $\xmin=0.35$ would be a good fit to macroscopically simulate real world returns~\cite{2018_crossan}. 

    However, there are two main issues with this assumption when looking at the microscopic behaviour:
    \begin{enumerate}
        \item the minimum return of $0.35$ is usually unrealistic as most of investments with a negative return will return 0. In fact, distribution for investments returning 0-1x is skewed towards the 0x~\cite{2019othman,2022_angellist}.
        \item the power law distribution allows for unbounded returns, which in practice is not realistic. In fact, in the last two decades we've seen maximum returns ranging from 50x to 1000x in the venture capital industry.
    \end{enumerate}

    \section{Mathematics of the power law} 
    \label{sec:math}
    \subsection{Power law distribution}
    A power law distribution is defined via the probability density function
    \begin{equation}
        \label{eq:powerlaw}
        f(x) = \frac{\alpha-1}{\xmin} \left(\frac{x}{\xmin}\right)^{-\alpha}, \qquad x\geq \xmin
    \end{equation}
    with $\xmin>0$ the minimum value (or return in this context) and $\alpha>1$. 
    The cumulative density function can be easily derived as
    \begin{equation}
        \label{eq:powerlawcdf}
        F(x) = P[X<x]= 1 - \left(\frac{x}{\xmin}\right)^{1-\alpha}, \qquad x\geq \xmin
    \end{equation}

    \subsection{Squashing the power law to $0$}
    To make the minimum return equal to $0$, one can't just set $\xmin=0$ in~\eqref{eq:powerlaw}, as this would lead to an ill-defined distribution. 
    Instead, one can scale~\eqref{eq:powerlaw} in the range $x\in[\xmin, 1)$ such that the minimum return is equal to $0$.
    Mathematically, we make the transformation $x\mapsto \frac{x-\xmin}{1-\xmin}$, for $x\in[\xmin, 1]$, which, by applying the change of variables, results in
    \begin{equation}
        \label{eq:powerlaw_rescaled}
        f_0(x) = \begin{cases}
            \frac{(1-\xmin)(\alpha-1)}{\xmin} \left(\frac{(1-\xmin)x + \xmin}{\xmin}\right)^{-\alpha} & \text{if } x\in[0,1),\\
            \frac{\alpha-1}{\xmin} \left(\frac{x}{\xmin}\right)^{-\alpha} & \text{if } x\geq1.
        \end{cases}
    \end{equation}
    Unfortunately this rescaling creates a discontinuity in the distribution at $x=1$. However, we found it to be empirically acceptable for our use case as the macrobehaviour is not affected.

    \subsection{Bounding the maximum value}
    In order to limit the maximum value $\xmax$ that samples drawn from the distribution can attain, it is sufficient to truncate the probability distribution such that all the mass for $r>M$ is concentrated in $r=M$. This can be done by setting $f(x)$ for $x=\xmax$ at
    \begin{align*}
        \int_{\xmax}^\infty \frac{\alpha-1}{\xmin} \left(\frac{x}{\xmin}\right)^{-\alpha} \partial x =
        \left[\frac{\alpha-1}{\xmin} \frac{x\left(\frac{x}{\xmin}\right)^{-\alpha}}{1-\alpha}\right]_{x=\xmax}^\infty=\frac{\xmax}{\xmin}\left(\frac{\xmax}{\xmin}\right)^{-\alpha}
    \end{align*}
    which results in the overall distribution
    \begin{equation}
        \label{eq:powerlaw_bounded}
        f_0^M(x) = \begin{cases}
            \frac{(1-\xmin)(\alpha-1)}{\xmin} \left(\frac{(1-\xmin)x + \xmin}{\xmin}\right)^{-\alpha} & \text{if } x\in[0,1),\\
            \frac{\alpha-1}{\xmin} \left(\frac{x}{\xmin}\right)^{-\alpha} & \text{if }x\in[1,\xmax),\\
            \frac{\xmax}{\xmin}\left(\frac{\xmax}{\xmin}\right)^{-\alpha} & \text{if } x= \xmax,\\
            0 & \text{if } x>\xmax.
        \end{cases}
    \end{equation}

    \section{Simulating the average VC fund}
    According to Crunchbase data, in the last few years we've seen about 5,000 early-stage investments per quarter. If we assume to deploy the entire fund in 3 years, there's a pool of 60,000 investments VCs have access to.

    To simulate an average VC fund, we can draw 60,000 samples from the power law distribution~\eqref{eq:powerlaw_rescaled}, with $\alpha=2.05$, $x_{min}=0.35$. 

    We consider different portfolio sizes $N$ ranging from $1$ to $300$. In the case of 1 investment, the entire fund's capital will be invested in a single deal, while in the case of $300$ investments, each one will receive $1/300$ of the total capital.
    For each portfolio size, we simulate 100,000 different funds with investments randomly drawn from the 60,000 samples. To take into account the effect of the particular set of 60,000 samples and average it out, we repeat the experiment for 20 different sets of samples and compute aggregate statistics.

    \subsection{Risk profile}
    \label{sec:risk}
\begin{figure}[h!]
        \centering
        \begin{subfigure}[b]{0.49\linewidth}
            \includegraphics[width=\textwidth]{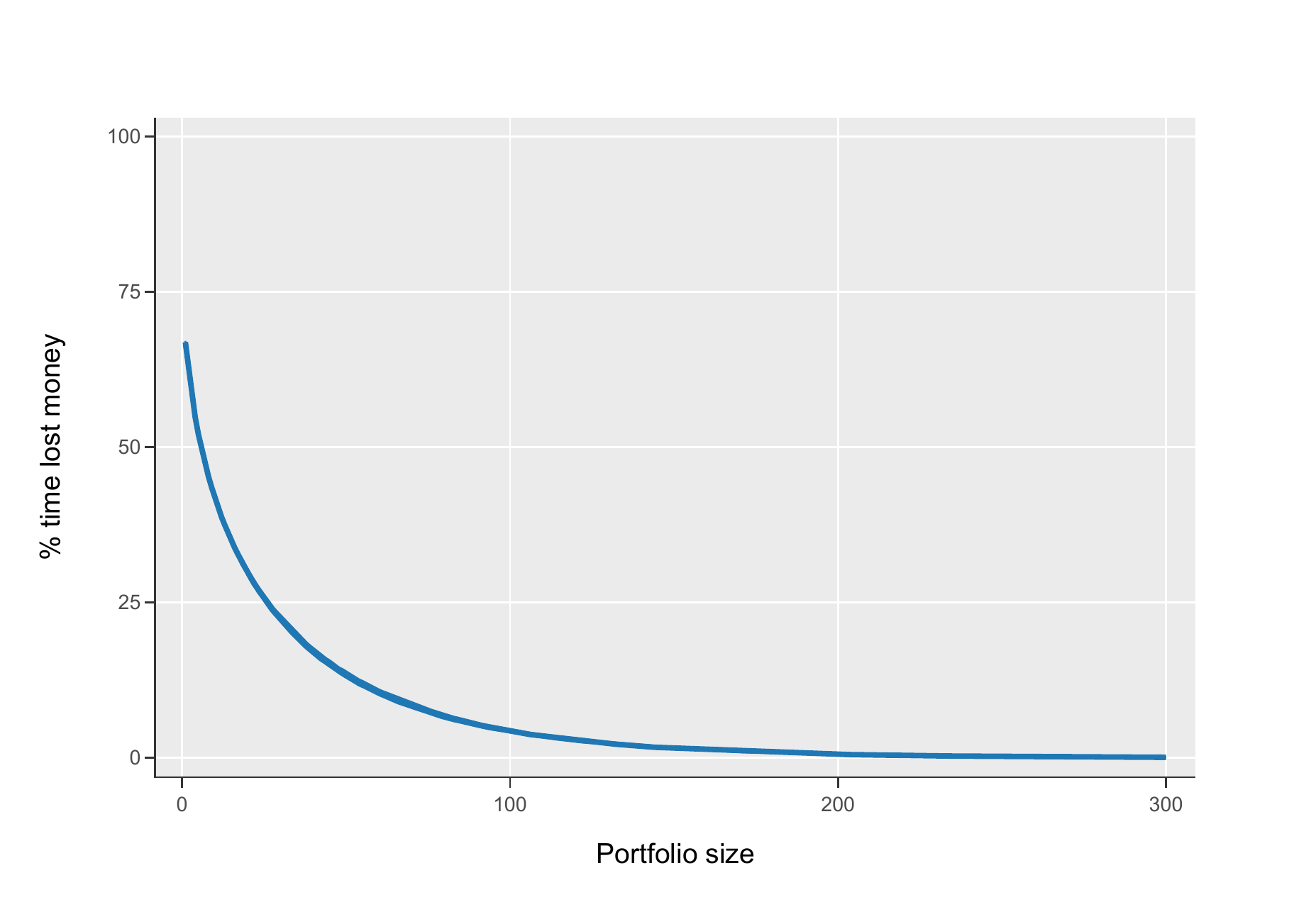}
            \caption{Percentage of portfolios losing money}
            \label{fig:pc_time_lost_money_avg}
        \end{subfigure}
        \begin{subfigure}[b]{0.49\linewidth}
            \includegraphics[width=\textwidth]{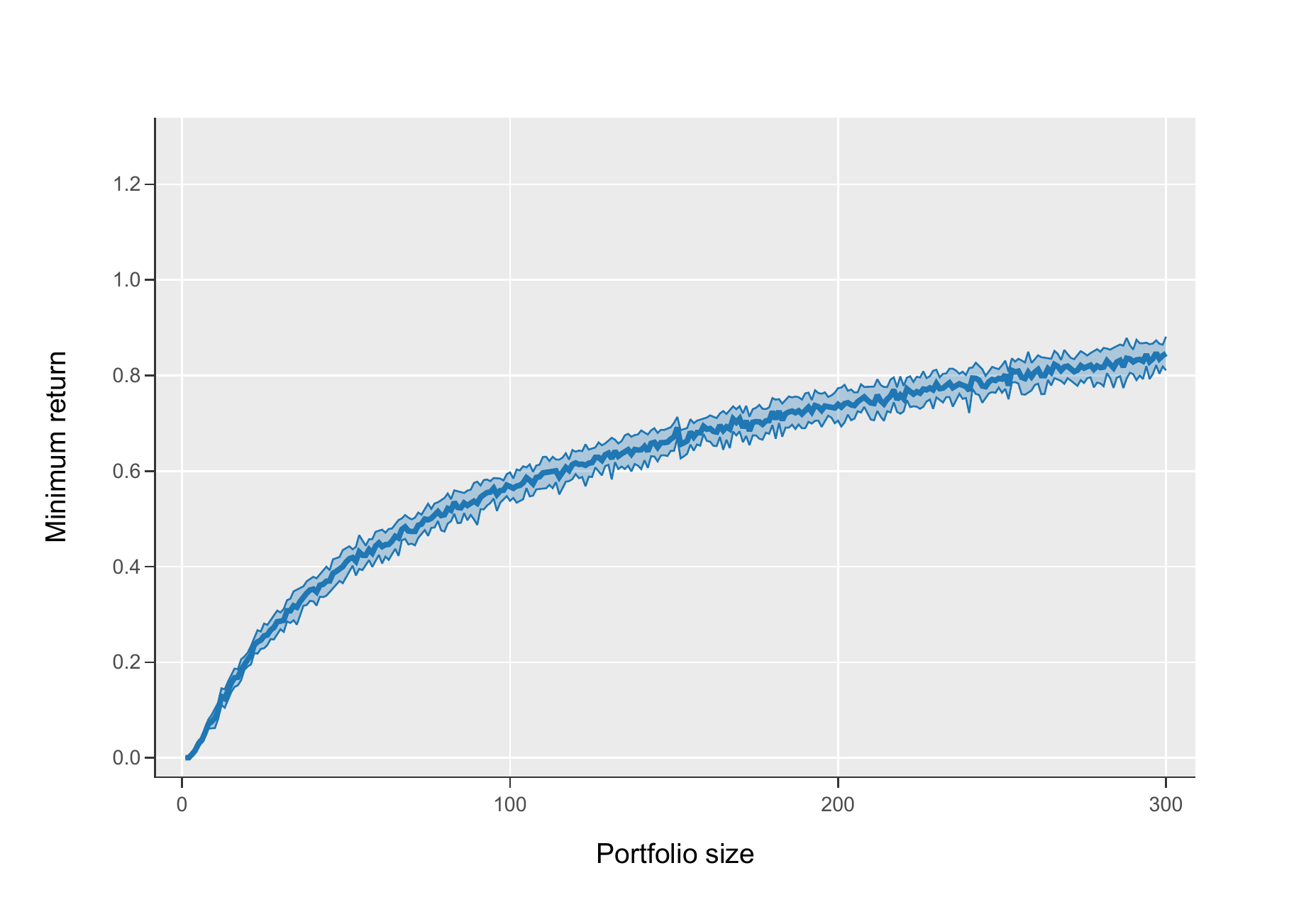}
            \caption{Minimum return}
            \label{fig:min_return_avg}
        \end{subfigure}
        \caption{Risk profile for different portfolio sizes (mean and standard deviation).}
    \end{figure}

    For each portfolio size, we estimate the associated risk profile via two main factors:
    \begin{enumerate}[(i)]
        \item the chances of losing money by returning less than the initial invested capital (Figure~\ref{fig:pc_time_lost_money_avg})
        \item the minimum return achieved across all the simulations (Figure~\ref{fig:min_return_avg})
    \end{enumerate}

    On the one hand, the probability of returning less than 1x decreases as the portfolio size increases, becoming close to 0 for a portfolio size $N>200$.
    On the other hand, the minimum return achieved over the set of simulations increases with the portfolio size. For $N=100$, over 100,000 simulated portfolios, none of them returned less than 0.55x. For $N=300$ none of them returned less than 0.8x.

    \begin{msgquote}{Take-home} 
        Risk and portfolio size are inversely correlated. The larger the portfolio, the lower the risk of losing money.
    \end{msgquote}

    \begin{helpquote}{Why does this happen?}
        A single home run (high-return investment) can offset many strikeouts. Increasing the number of investments increases the probability of hitting more home runs, each of which may offset more strikeouts than those in the entire portfolio. Also, having a larger number of investments returning more than 1x offsets more and more strikeouts.
    \end{helpquote}

    \begin{warningquote}{Reduced risk is correlated with reduced maximum attainable returns}
        A larger portfolio size also reduces the maximum return one can attain on a fund. In fact, a 100x return on an investment will result in a 100x return on the fund for $N=1$, while in a large portfolio it's impact will be diluted.
        However, while stellar returns may happen for small portfolios, they're very unlikely. In fact as from the results above, the probability of losing money is very high for small portfolios.
        \begin{center}
            \captionsetup{type=figure}
            \centering
            \includegraphics[width=0.5\textwidth]{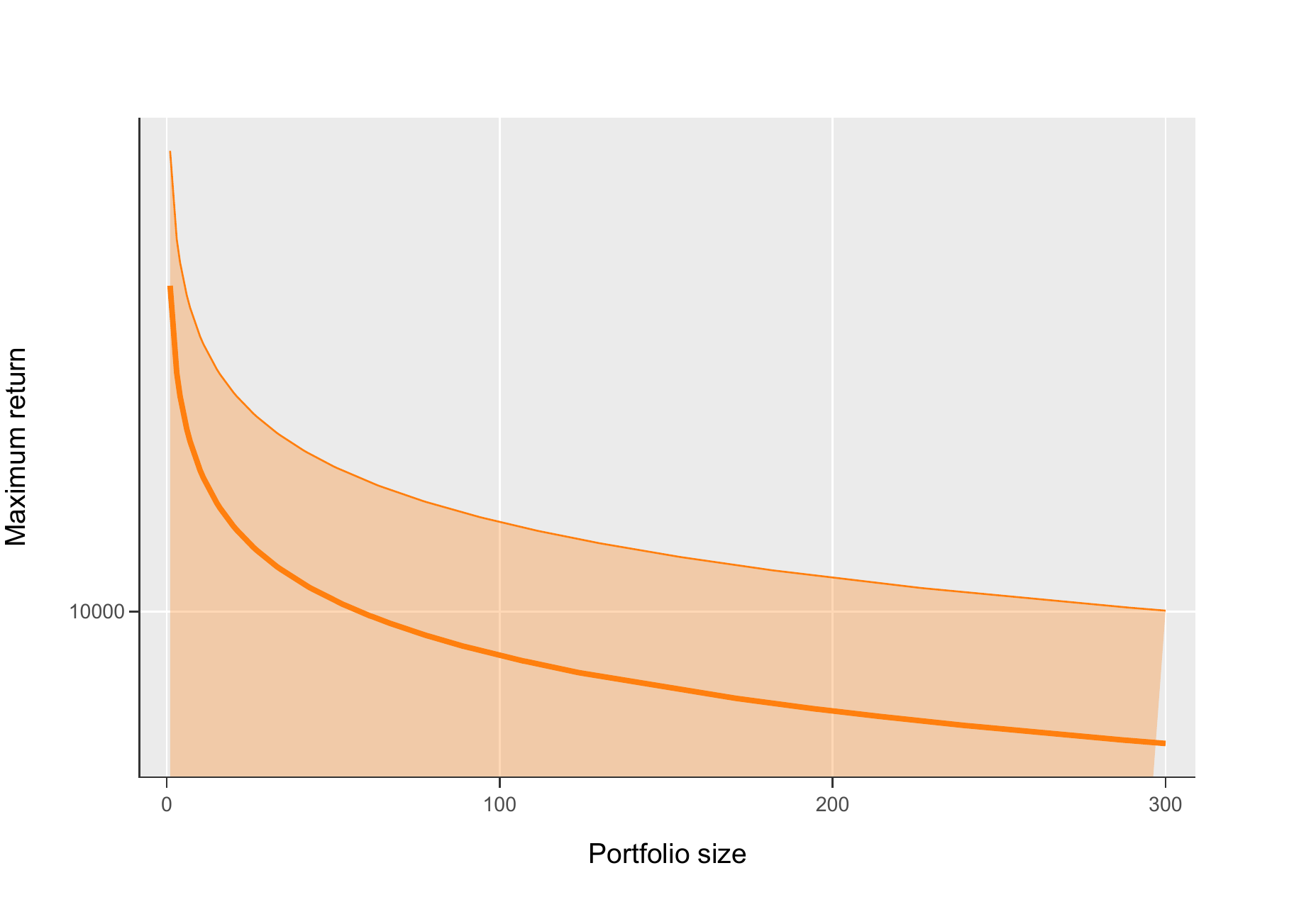}
            \caption{Maximum returns for different portfolio sizes (mean and standard deviation).}
        \end{center}
    \end{warningquote}

    \subsection{Return profile}
    Let's now look at the frequency of returning 2-10x the invested capital over the simulated portfolios.

    \begin{figure}[h!]
        \centering
        \includegraphics[width=0.6\textwidth]{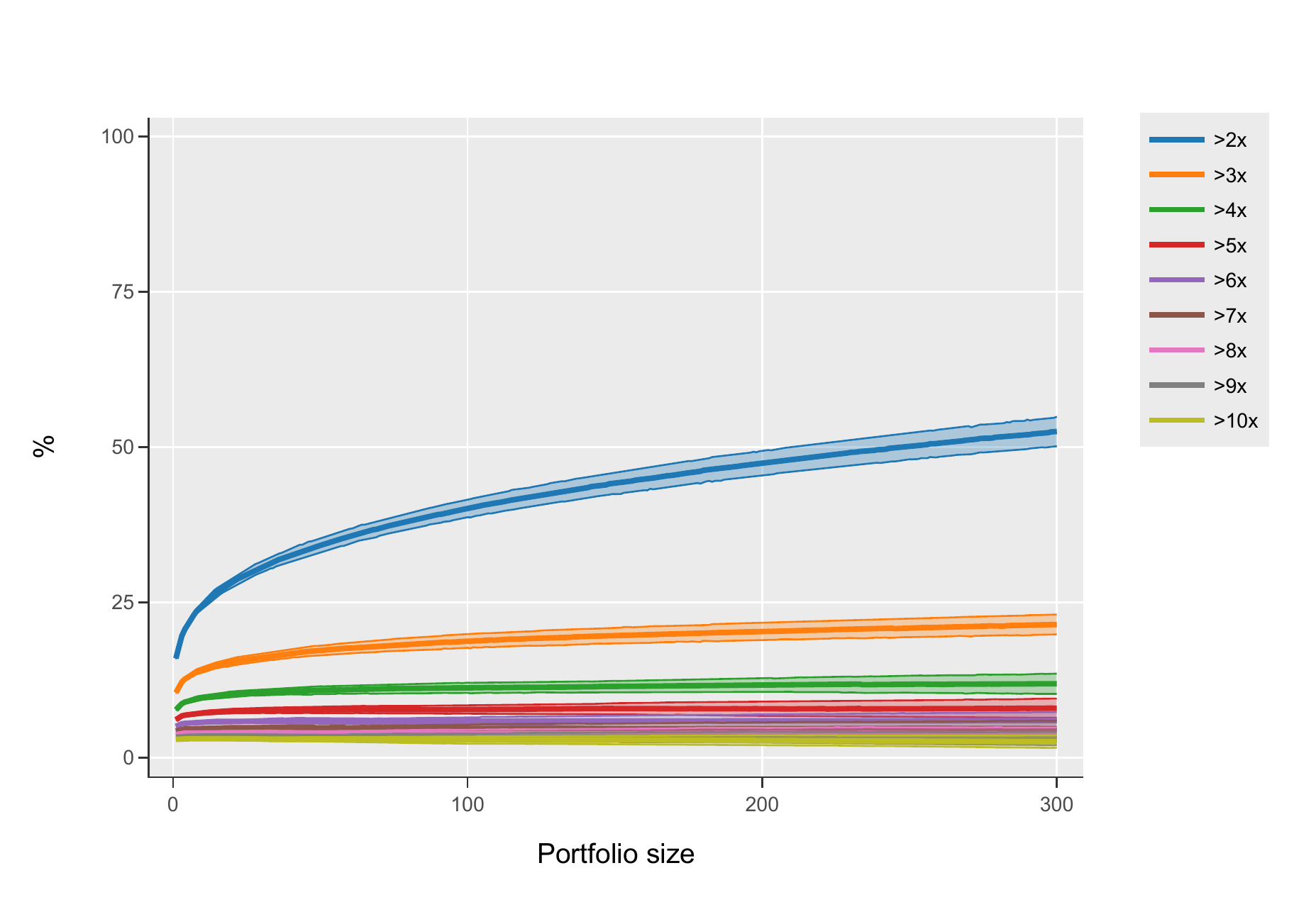}
        \caption{Frequency of returns in the range 2-10x for different portfolio sizes (mean and standard deviation).}
        \label{fig:x_returns_avg}
    \end{figure}

    In Figure~\ref{fig:x_returns_avg}, we can see that the probability of doubling the initial investment increases quite dramatically as the portfolio size increases. A similar but less dramatic increase happens to the probability of tripling the fund. If we look at the frequencies of 5x-ing or 10x-ing the fund, their average value stays pretty constant over different portfolio sizes across all the simulations.

    \begin{msgquote}{Take-home}
        Larger portfolio sizes increase the probability of returning 2-5x the invested capital.
    \end{msgquote}

    \section{Impact of bounded return on investment (ROI)}

    By definition, the power law distribution allows for arbitrarily large returns to happen with a non-zero probability. However, in practice, returns are not unbounded. One of the largest returns in recent history is believed to be the one of the first angel investment in Google which is estimated to have returned $\sim$20,000x. A large recent return is the $\sim$400x achieved by Index ventures on their investment in Figma (shares bought at $\sim$0.09 and sold at $\sim$40). In general, over the past years, we've seen maximum returns ranging from 50x to more than 1,000x~\cite{venture_returns}.

    While limiting the ROI doesn't affect the risk profile over the set of simulations, it has important effects on the probability of positive returns on the fund.

\begin{figure}[h!]
        \centering
        \begin{subfigure}{0.32\linewidth}
            \includegraphics[width=\textwidth]{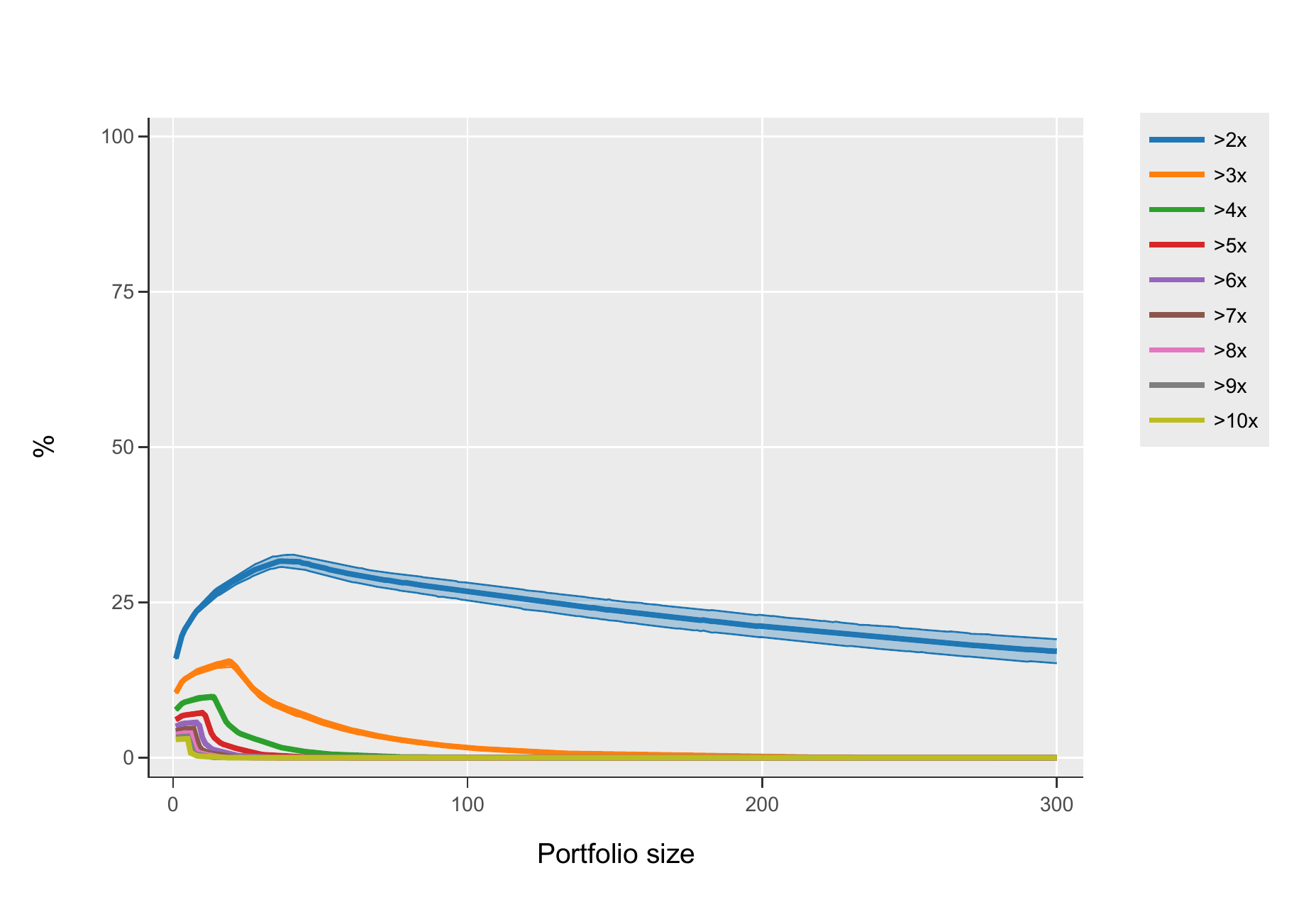}
            \caption{50x bound}
            \label{fig:x_returns_avg_50}
        \end{subfigure}
        \begin{subfigure}{0.32\linewidth}
            \includegraphics[width=\textwidth]{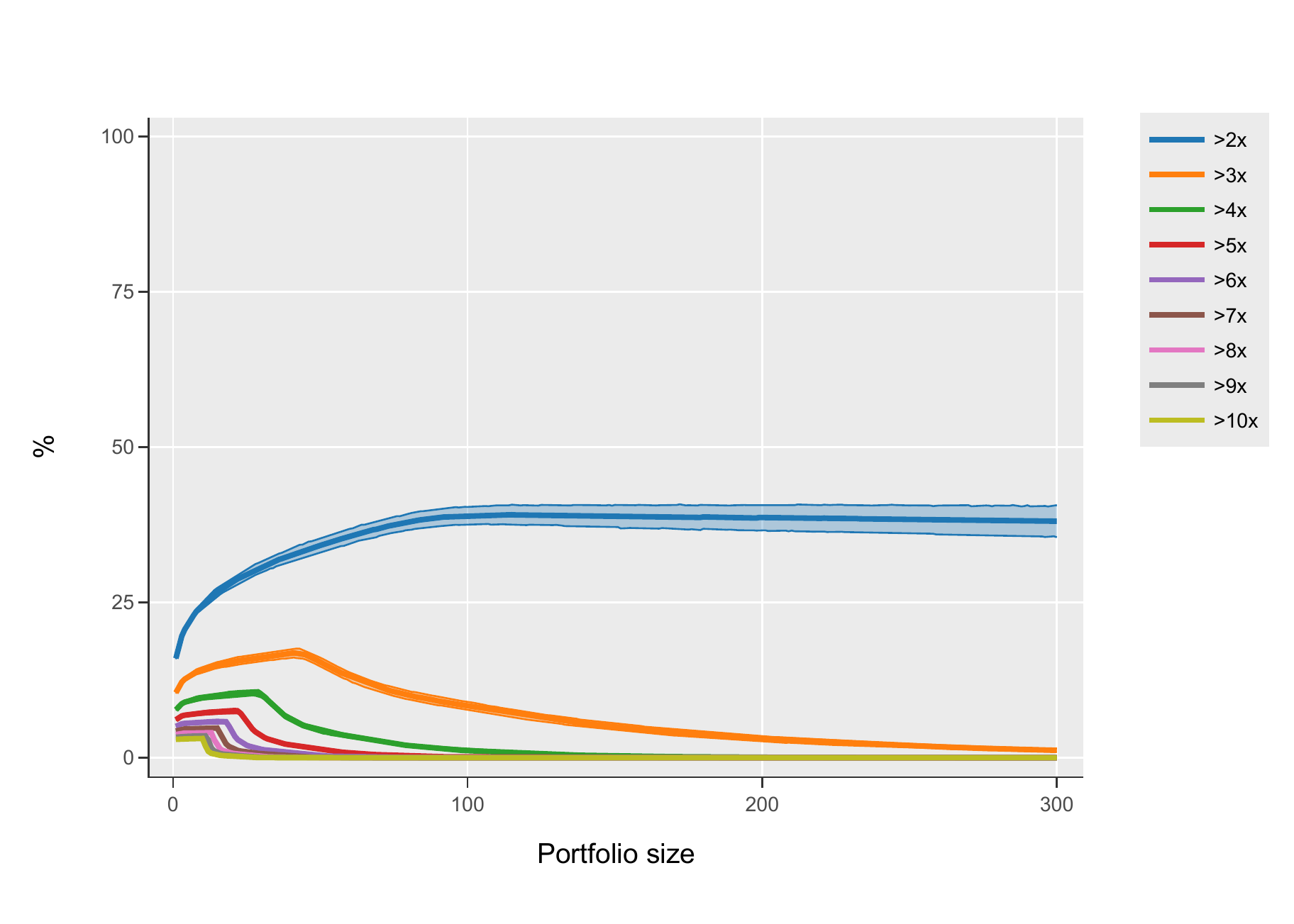}
            \caption{100x bound}
            \label{fig:x_returns_avg_100}
        \end{subfigure}
        \begin{subfigure}{0.32\linewidth}
            \includegraphics[width=\textwidth]{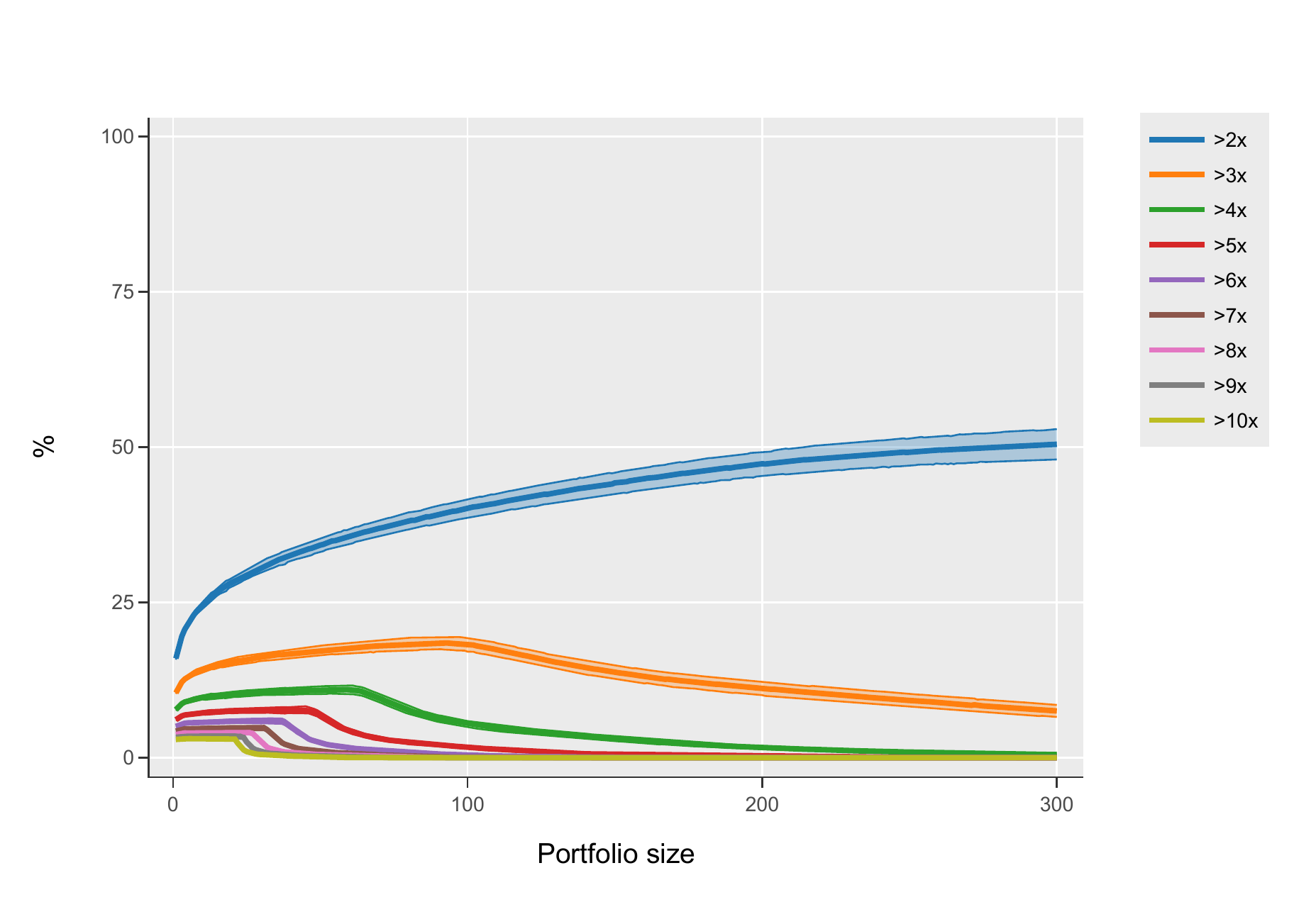}
            \caption{200x bound}
            \label{fig:x_returns_avg_200}
        \end{subfigure}
        \begin{subfigure}{0.32\linewidth}
            \includegraphics[width=\textwidth]{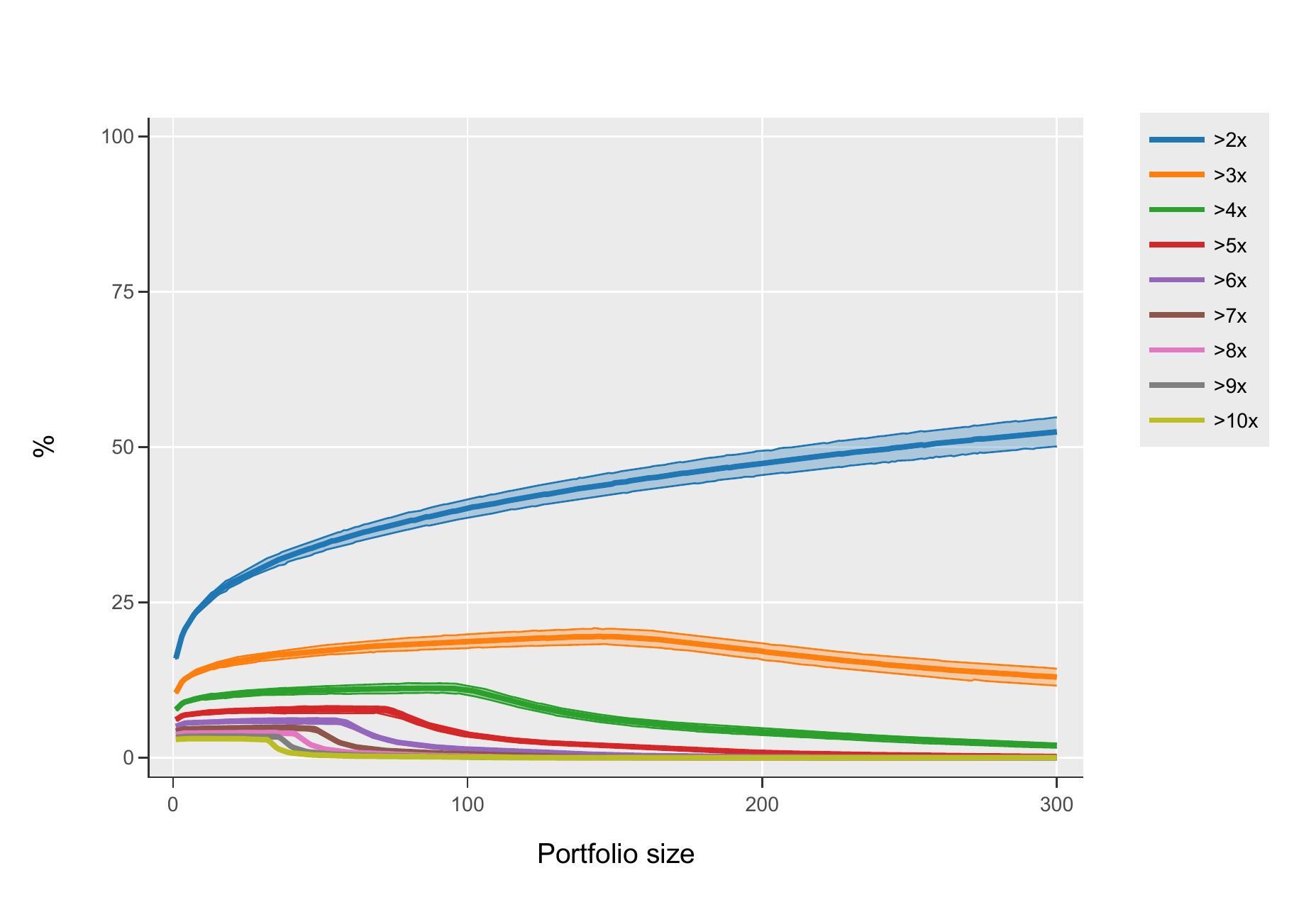}
            \caption{300x bound}
            \label{fig:x_returns_avg_300}
        \end{subfigure}
        \begin{subfigure}{0.32\linewidth}
            \includegraphics[width=\textwidth]{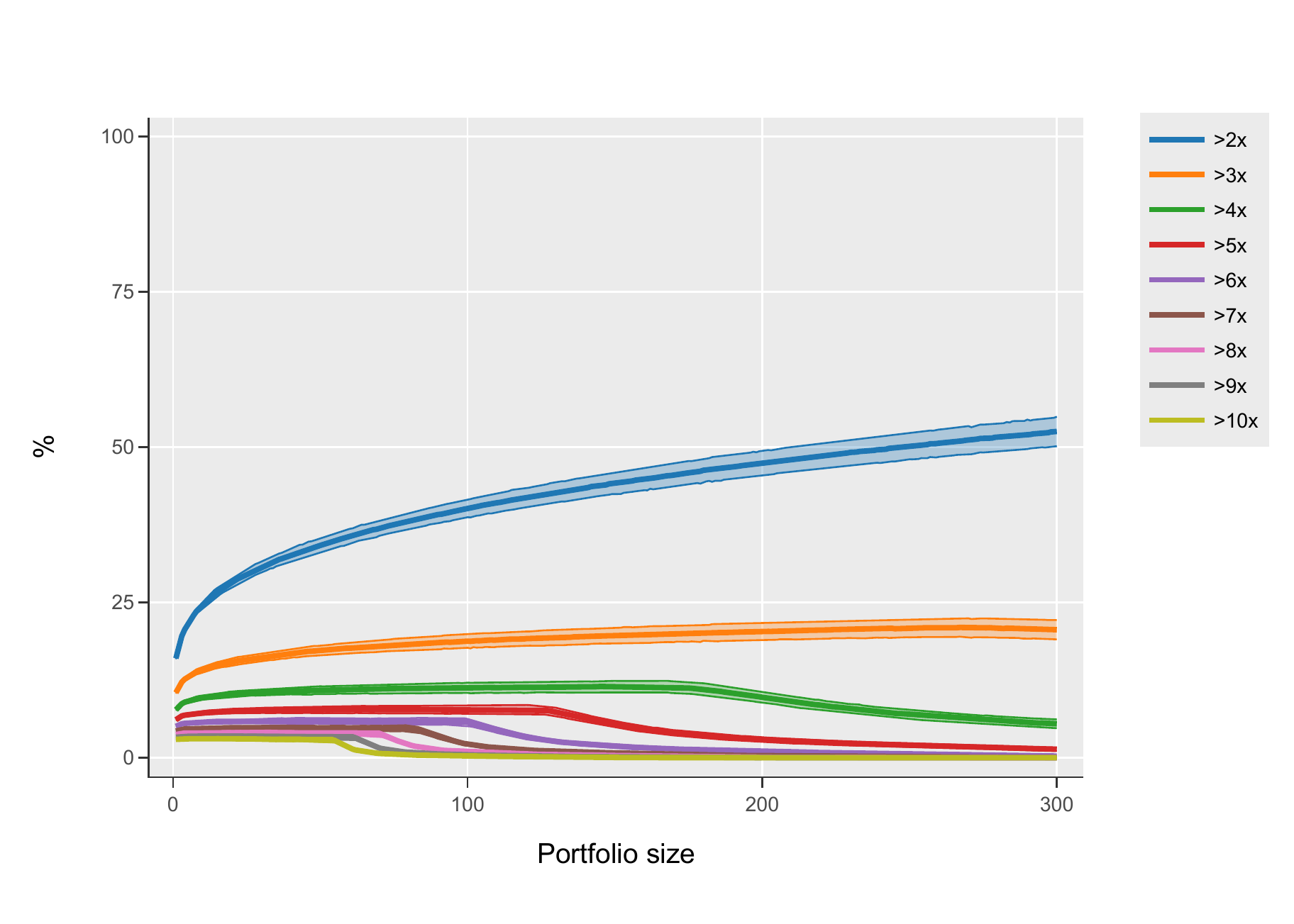}
            \caption{500x bound}
            \label{fig:x_returns_avg_500}
        \end{subfigure}
        \begin{subfigure}{0.32\linewidth}
            \includegraphics[width=\textwidth]{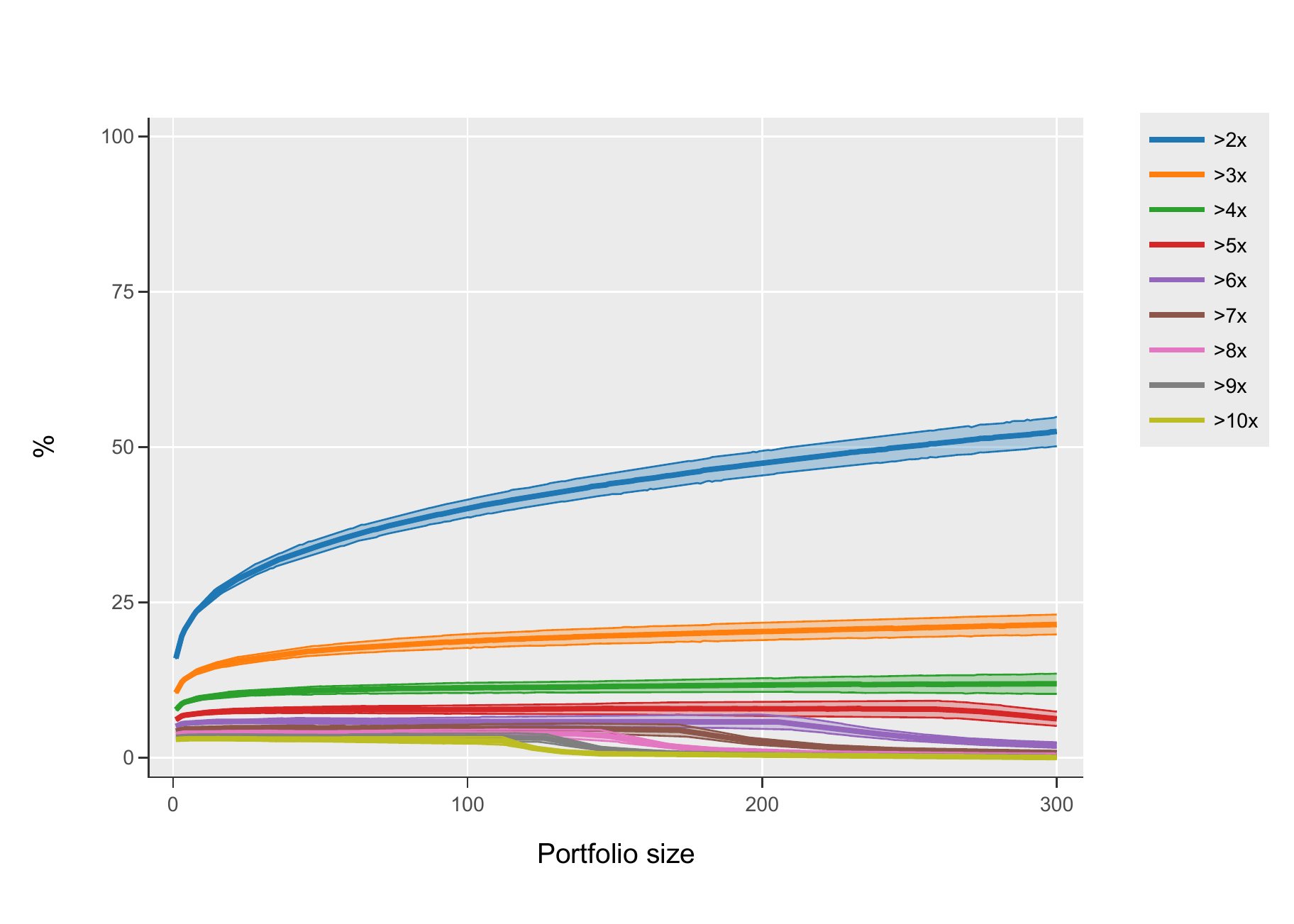}
            \caption{1000x bound}
            \label{fig:x_returns_avg_1000}
        \end{subfigure}
        \caption{Frequency of returns in the range 2-10x for different portfolio sizes and different bounds on the maximum return per investment (mean and standard deviation).}
        \label{fig:x_returns_avg_bounds}
    \end{figure}

    In order to simulate portfolios where the ROI is bounded, we can use~\eqref{eq:powerlaw_bounded} and consider bounds at 50, 100, 200, 300, 500 and 1,000. Figure~\ref{fig:x_returns_avg_bounds} shows frequency of returning 2-10x the fund for different portfolio sizes across the set of simulations.

    As we can see, stricter bounds affect the probability of high returns on the fund. If the maximum possible return on a single investment is 50x, then the probability of returning 2x the fund increases until a portfolio size of 40, and then it starts to decrease. If the bound is 100, the probability of a 2x return on the fund increases until $N=100$ and then it stays constant, while the probability of doing more than 3x starts to decrease after $N=40$. As we increase the bound on the maximum possible return per investment, the benefits of larger portfolios restart to appear along with their probability of achieving higher returns on the fund.

    \begin{helpquote}{Why does this happen?}
        By limiting the maximum return, one is intrinsically reducing the number of strikeout that a home run can offset.     
    \end{helpquote}

    \section{Impact of decision quality}
    In a real scenario, not all VCs are the same. Some are able to select better investments more often and some are not. An overperforming VC will pick good deals with a higher frequency – this can be modelled by decreasing the value of $\alpha$ for the considered power law distribution: we'll pick $\alpha=1.85$. On the contrary, an underperforming VC picks bad deals with a higher frequency, resulting in a higher value of $\alpha$, say $\alpha=2.5$.

    \subsection{Overperfomers}
\begin{figure}[h!]
        \centering
        \includegraphics[width=0.45\textwidth]{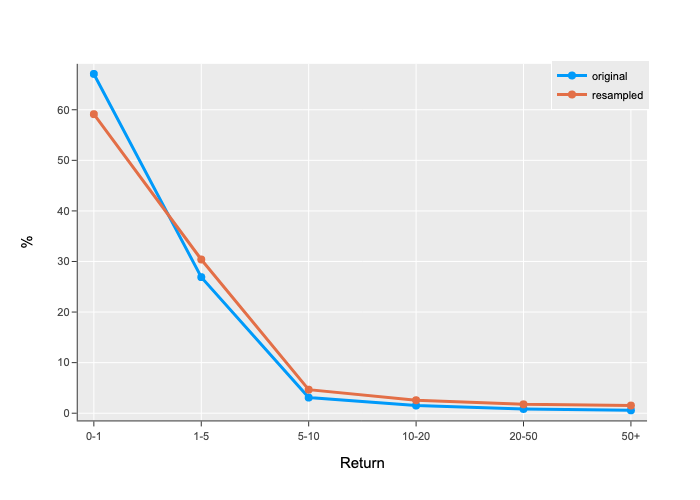}
        \caption{Overperformer vs world distribution.}
        \label{fig:overperformer_distribution}
    \end{figure}

    While investment returns are distributed according to the same probability distribution we used so far (blue in Figure~\ref{fig:overperformer_distribution}), an over-performing VC can screen out some bad investments while picking good ones more often (drawing the according to the red distribution).

    \subsubsection{Risk profile}
\begin{figure}[htbp!]
        \centering
        \begin{subfigure}[b]{0.45\linewidth}
            \includegraphics[width=\textwidth]{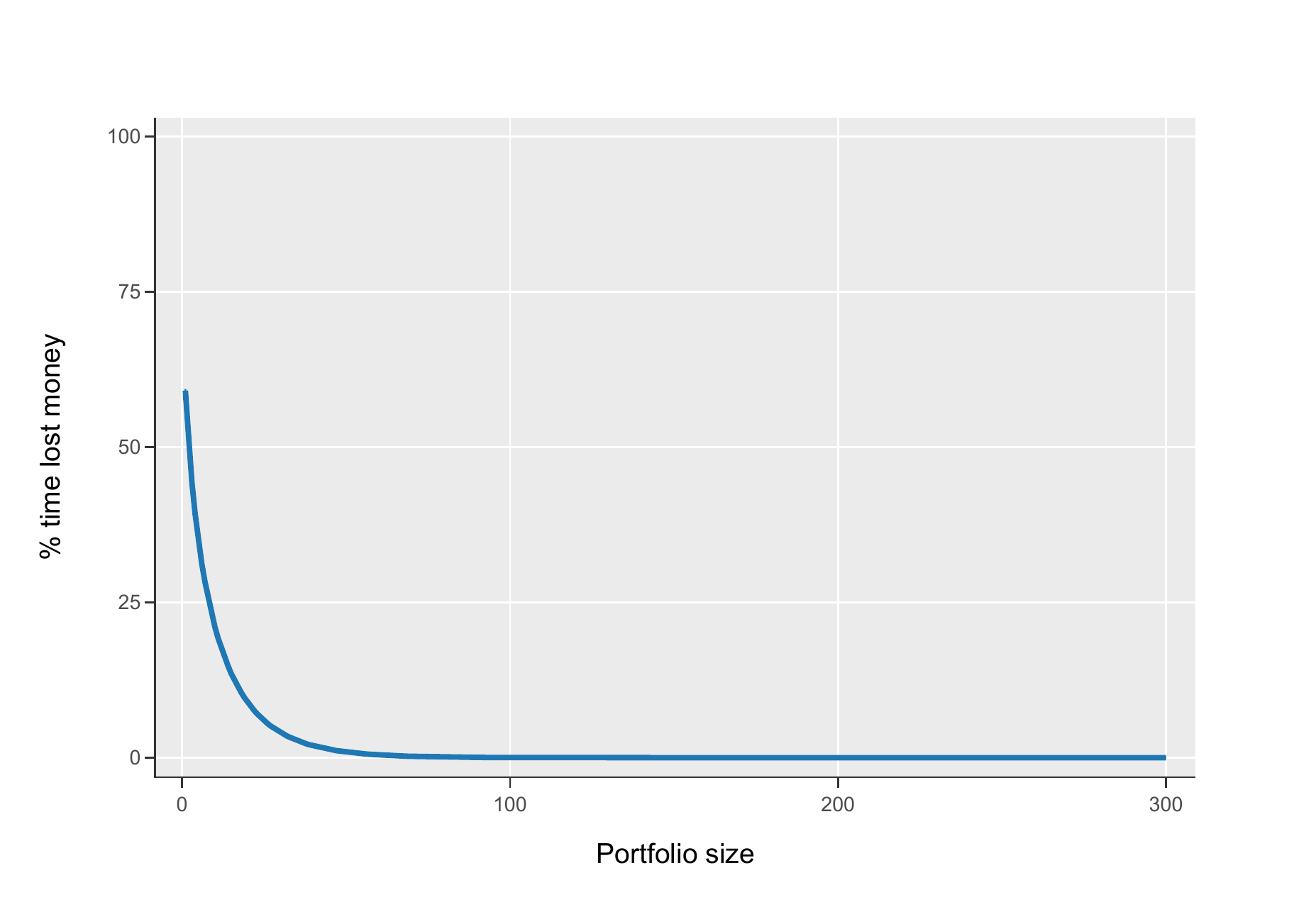}
            \caption{Percentage of portfolios losing money}
            \label{fig:pc_time_lost_money_overperformer}
        \end{subfigure}
        \begin{subfigure}[b]{0.45\linewidth}
            \includegraphics[width=\textwidth]{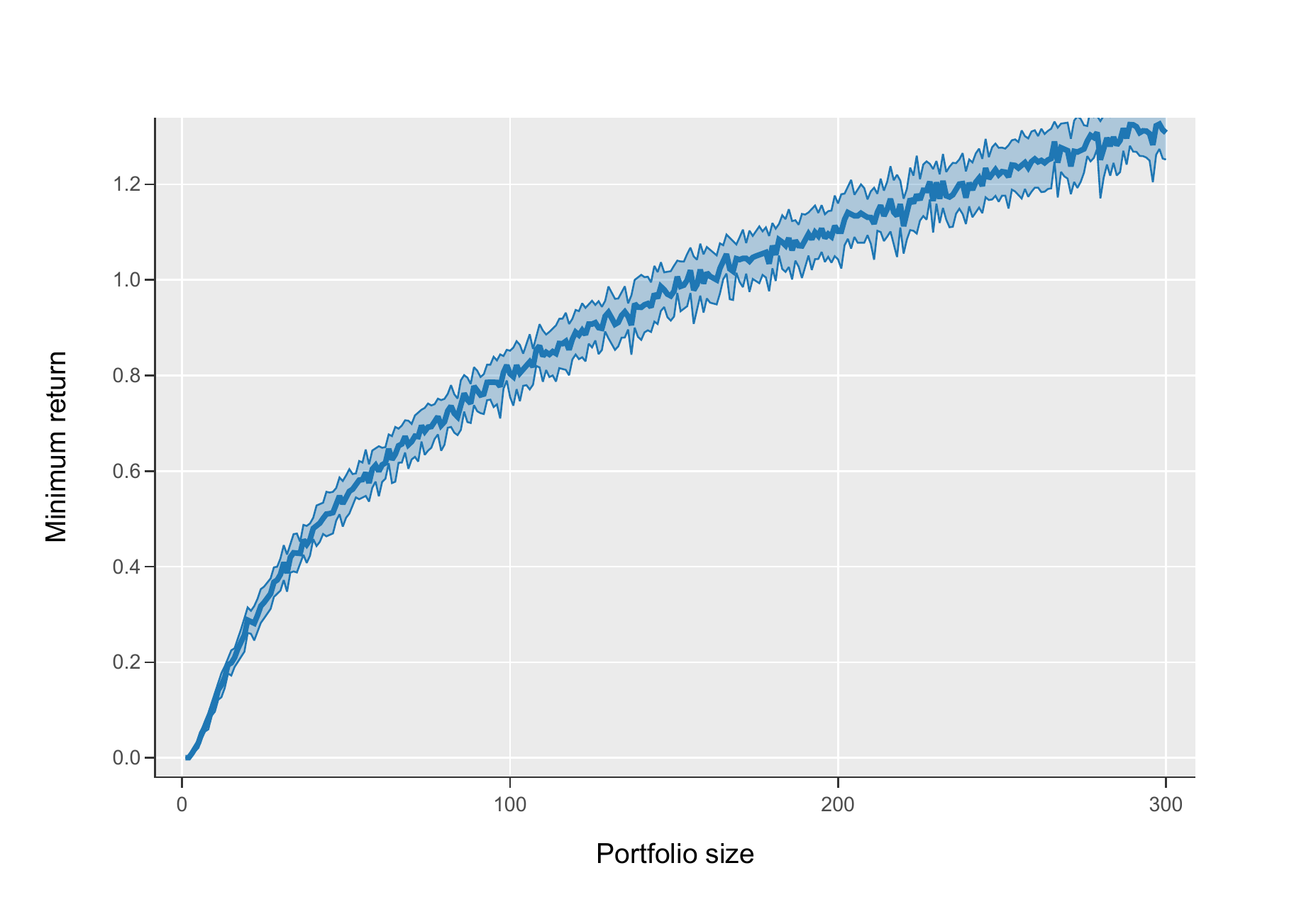}
            \caption{Minimum return}
            \label{fig:min_return_overperformer}
        \end{subfigure}
        \caption{Risk profile for an overperformer (mean and standard deviation).}
        \label{fig:overperformer_risk_profile}
    \end{figure}

    Being an overperformer allows you to achieve a minimum 1x expected return much earlier, at about $N=100$. The probability of losing money also decreases more rapidly as the portfolio size increases. See Figure~\ref{fig:overperformer_risk_profile}. 

    \subsubsection{Expected returns}
\begin{figure}[h!]
        \centering
        \begin{subfigure}{0.32\linewidth}
            \includegraphics[width=\textwidth]{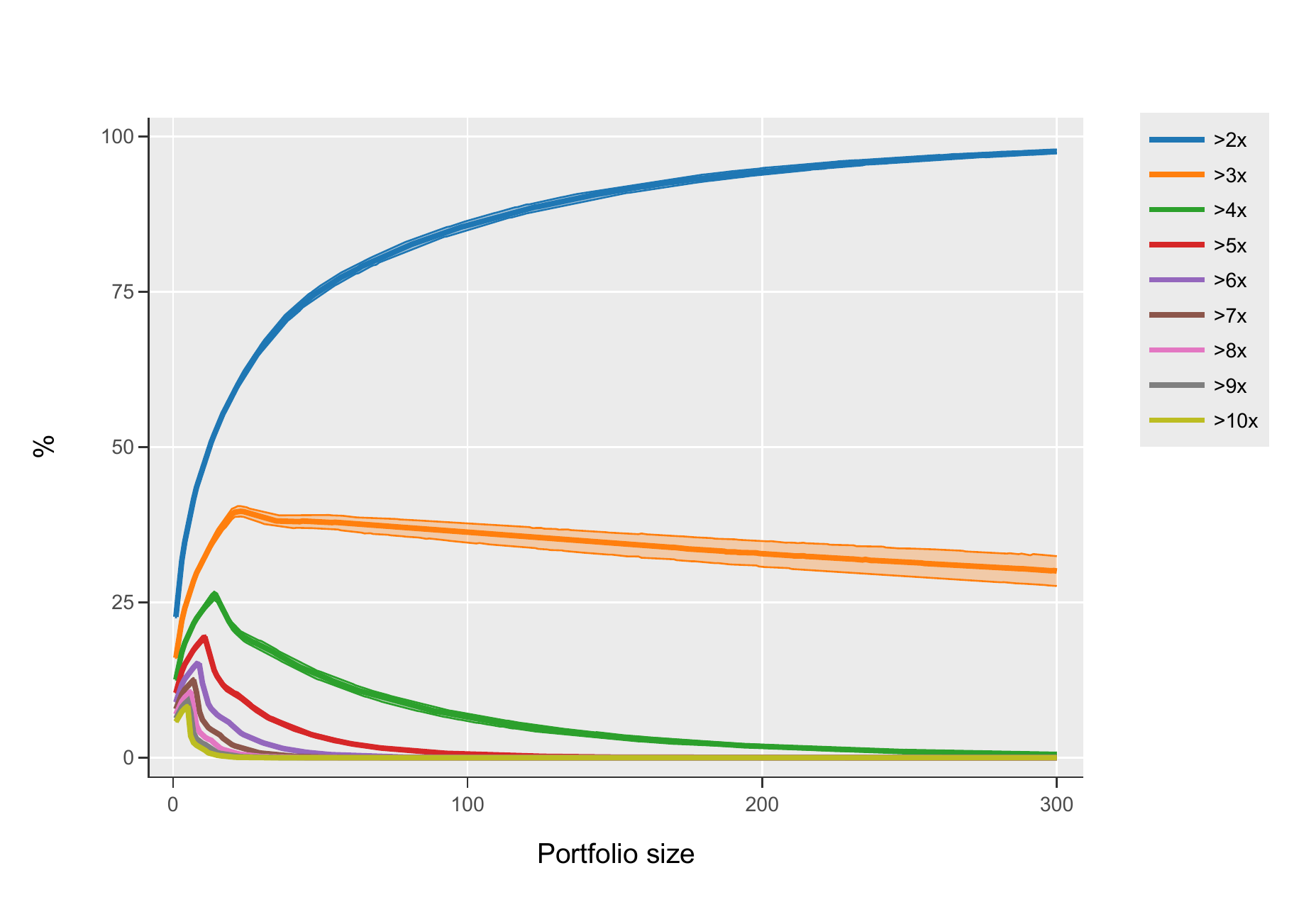}
            \caption{50x bound}
            \label{fig:x_returns_overperformer_50}
        \end{subfigure}
        \begin{subfigure}{0.32\linewidth}
            \includegraphics[width=\textwidth]{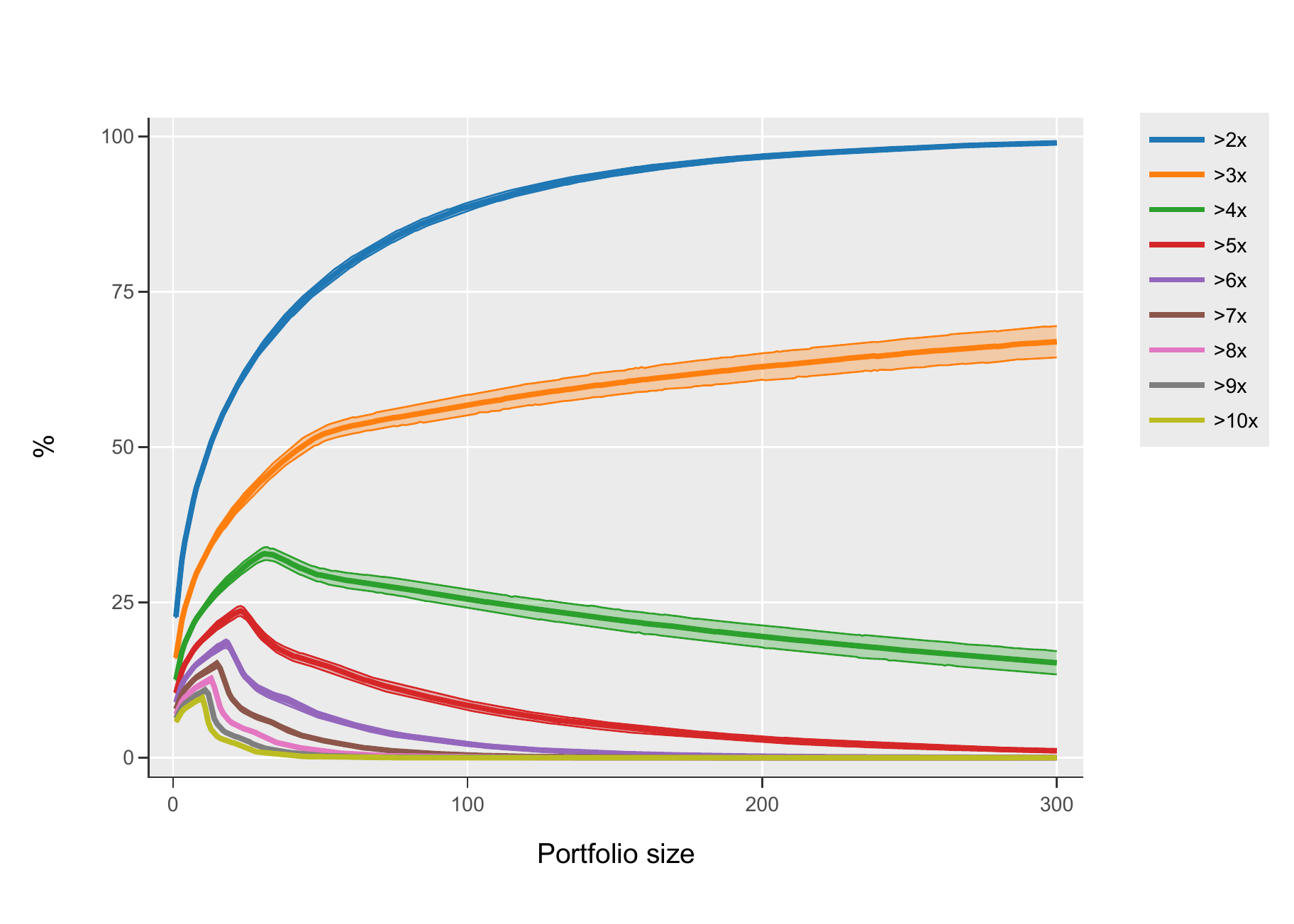}
            \caption{100x bound}
            \label{fig:x_returns_overperformer_100}
        \end{subfigure}
        \begin{subfigure}{0.32\linewidth}
            \includegraphics[width=\textwidth]{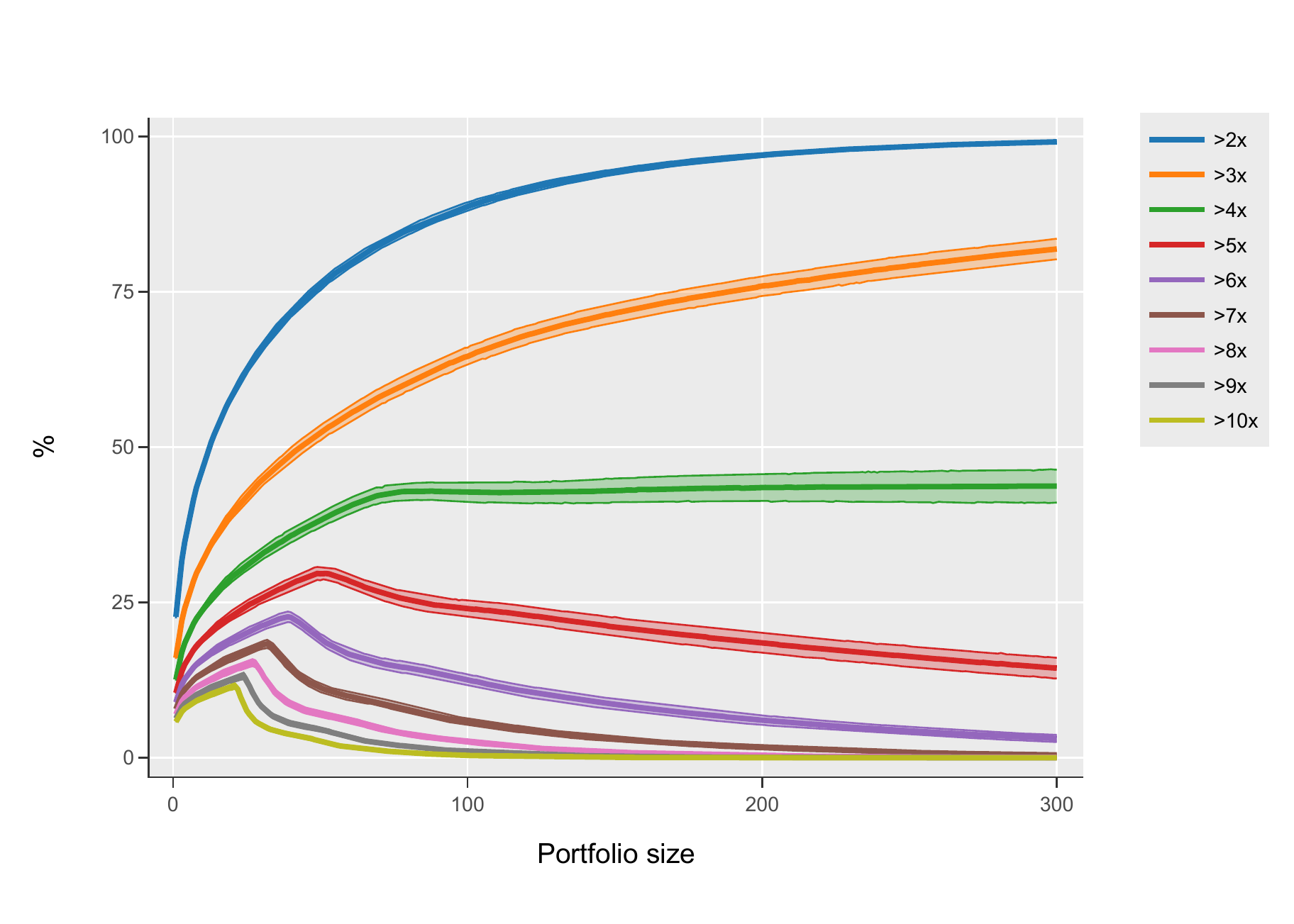}
            \caption{200x bound}
            \label{fig:x_returns_overperformer_200}
        \end{subfigure}
        \begin{subfigure}{0.32\linewidth}
            \includegraphics[width=\textwidth]{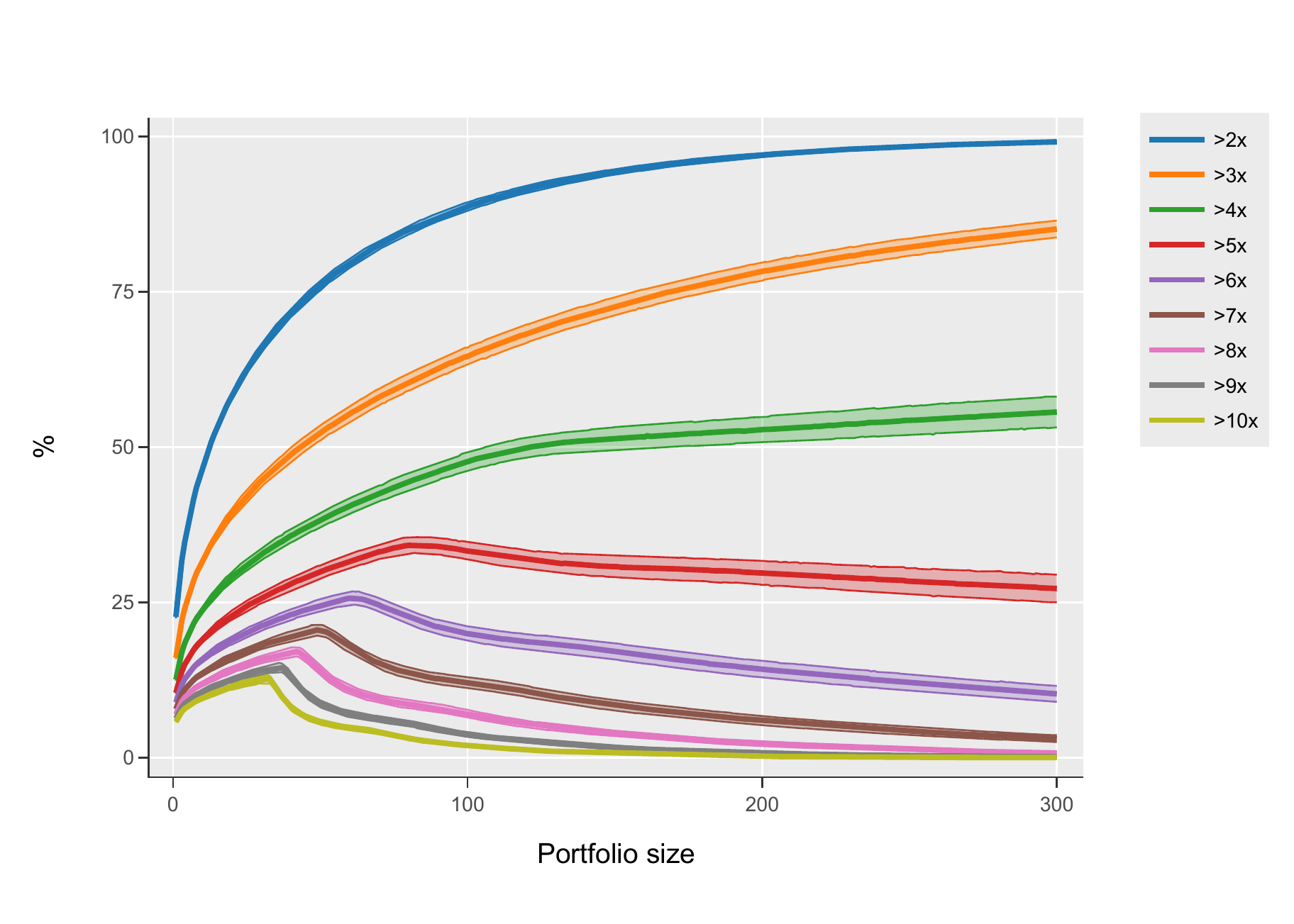}
            \caption{300x bound}
            \label{fig:x_returns_overperformer_300}
        \end{subfigure}
        \begin{subfigure}{0.32\linewidth}
            \includegraphics[width=\textwidth]{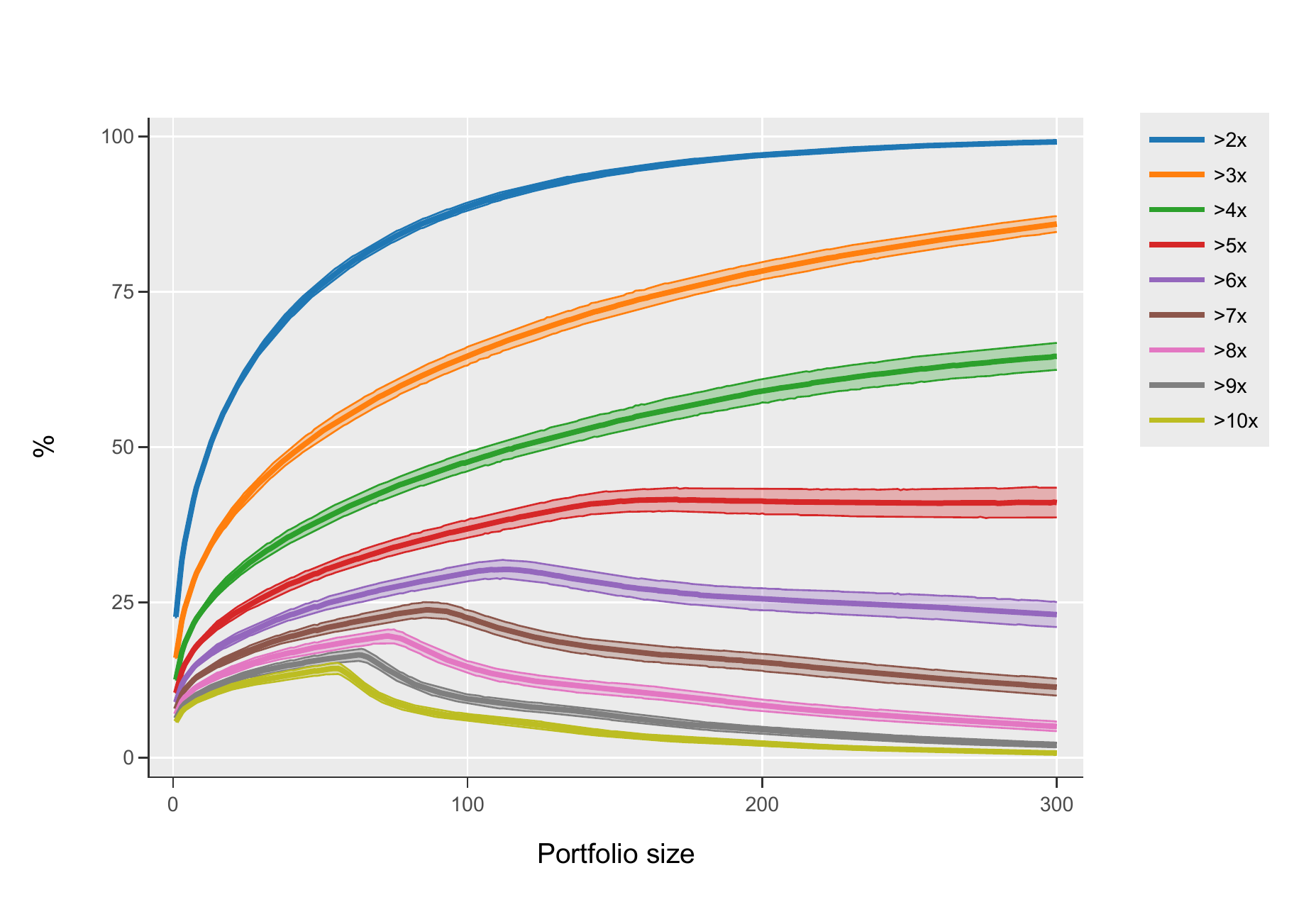}
            \caption{500x bound}
            \label{fig:x_returns_overperformer_500}
        \end{subfigure}
        \begin{subfigure}{0.32\linewidth}
            \includegraphics[width=\textwidth]{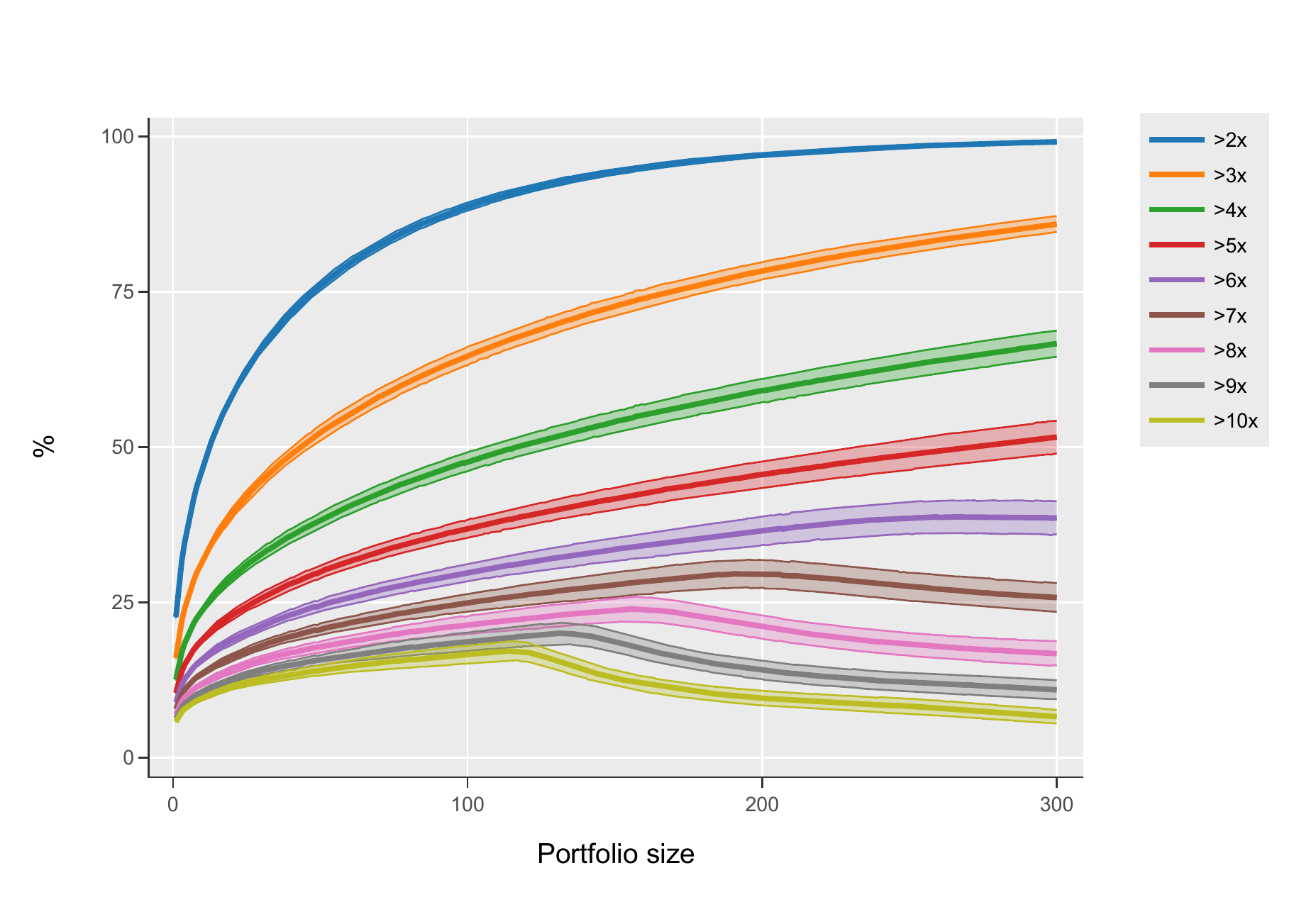}
            \caption{1,000x bound}
            \label{fig:x_returns_overperformer_1000}
        \end{subfigure}
        \caption{Frequency of returns in the range 2-10x for different portfolio sizes and different bounds on the maximum return per investment for an overperformer (mean and standard deviation).}
        \label{fig:x_returns_overperformer_bounds}
    \end{figure}

    As one would expect, the probability of returning a multiple of the fund is much higher for an overperformer. Doubling the fund is almost a certainty for $N>200$, regardless of the bound on the ROI. Being an overperformer also tends to mitigate the effect of bounded returns (compare Figure~\ref{fig:x_returns_overperformer_bounds} with Figure~\ref{fig:x_returns_avg_bounds}).

    \subsection{Underperformers}
\begin{figure}[h!]
        \centering
        \includegraphics[width=0.6\textwidth]{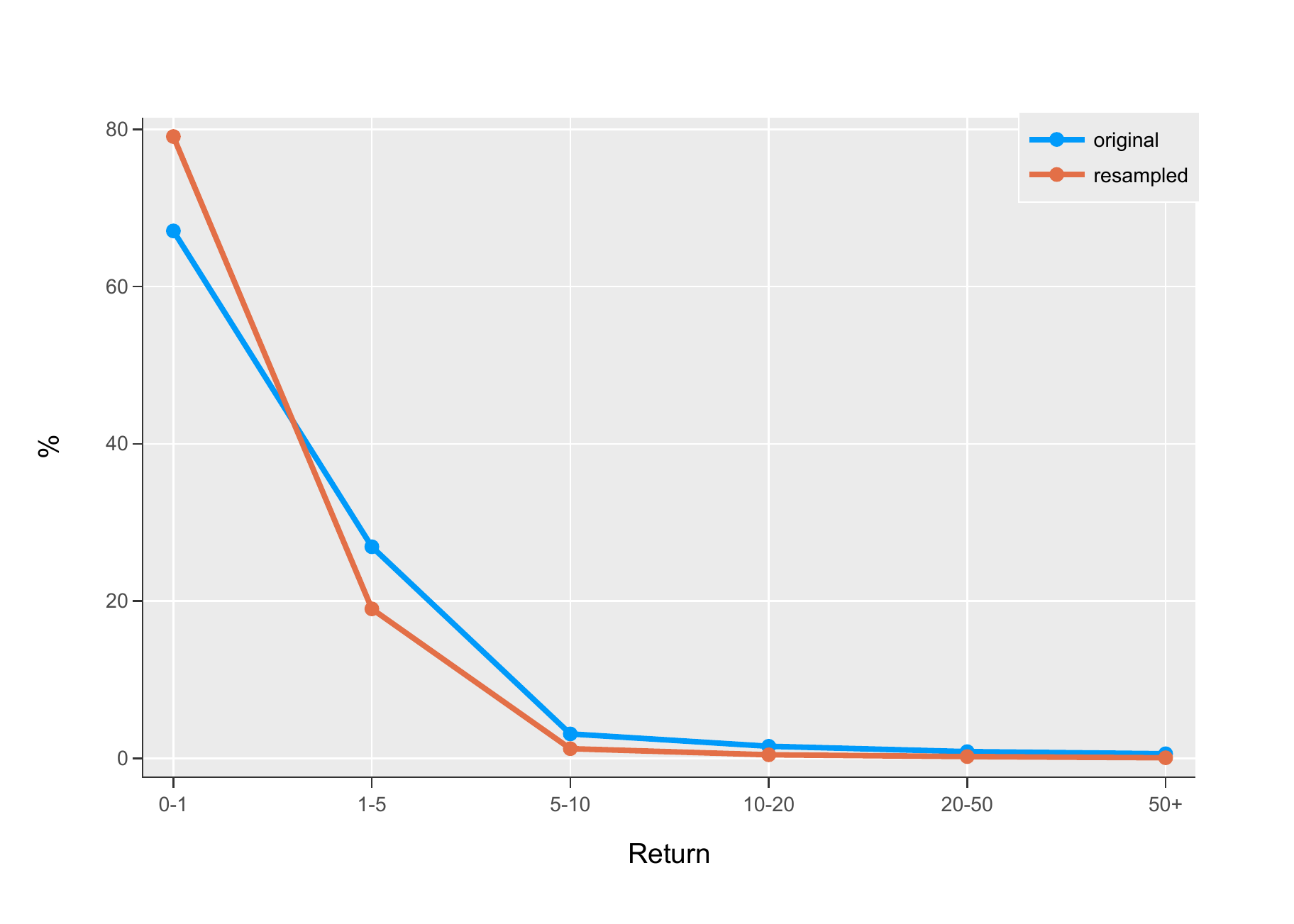}
        \caption{Underperformer vs world distribution.}
        \label{fig:underperformer_distribution}
    \end{figure}

    An under-performing VC will choose bad investments more often than an average one; mathematically it will draw samples from the blue distribution, according to the red one in Figure~\ref{fig:underperformer_distribution}.

\begin{figure}[h!]
        \centering
        \begin{subfigure}[t]{0.32\linewidth}
            \includegraphics[width=\textwidth]{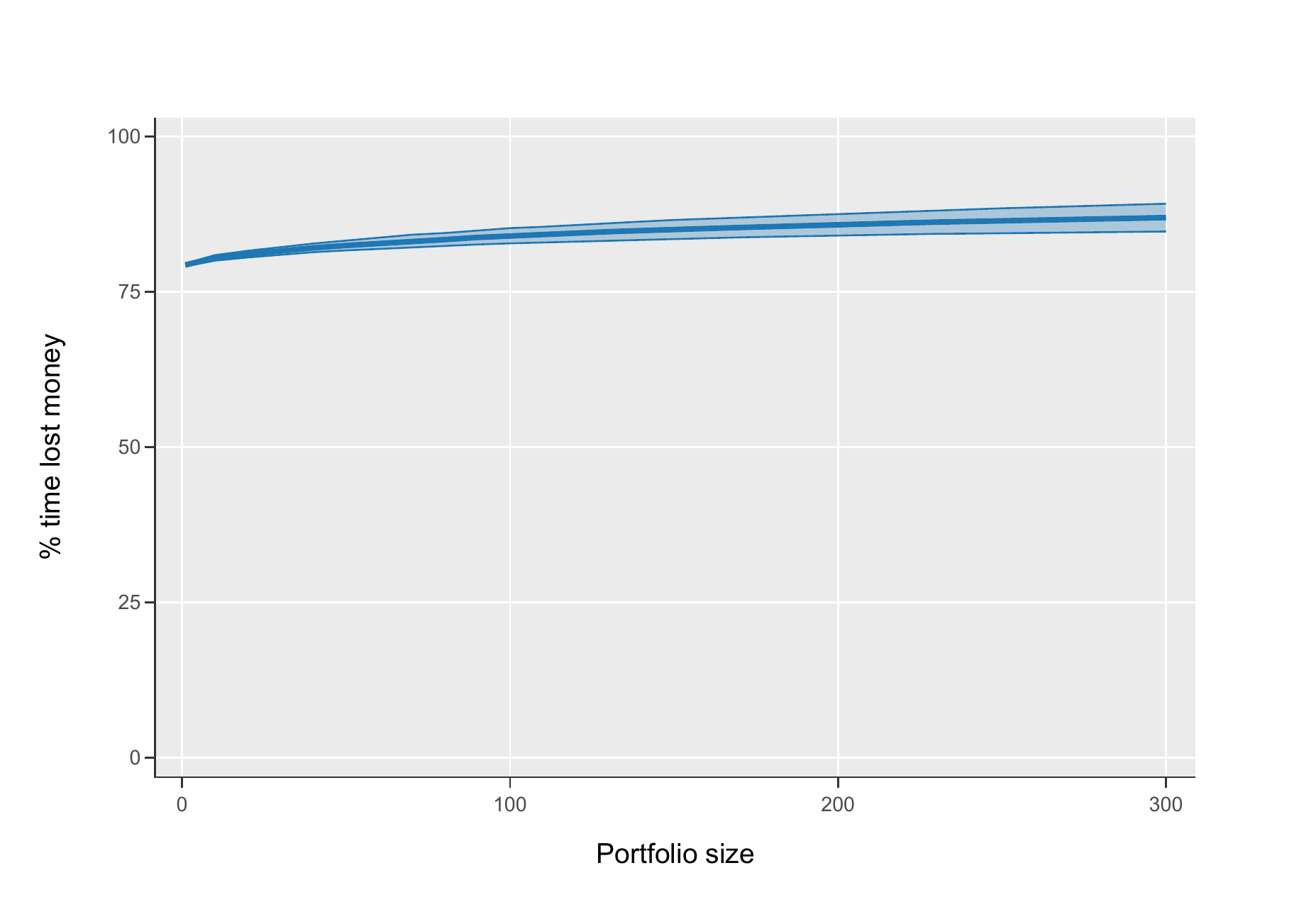}
            \caption{Percentage of portfolios losing money}
            \label{fig:pc_time_lost_money_underperformer}
        \end{subfigure}
        \begin{subfigure}[t]{0.32\linewidth}
            \includegraphics[width=\textwidth]{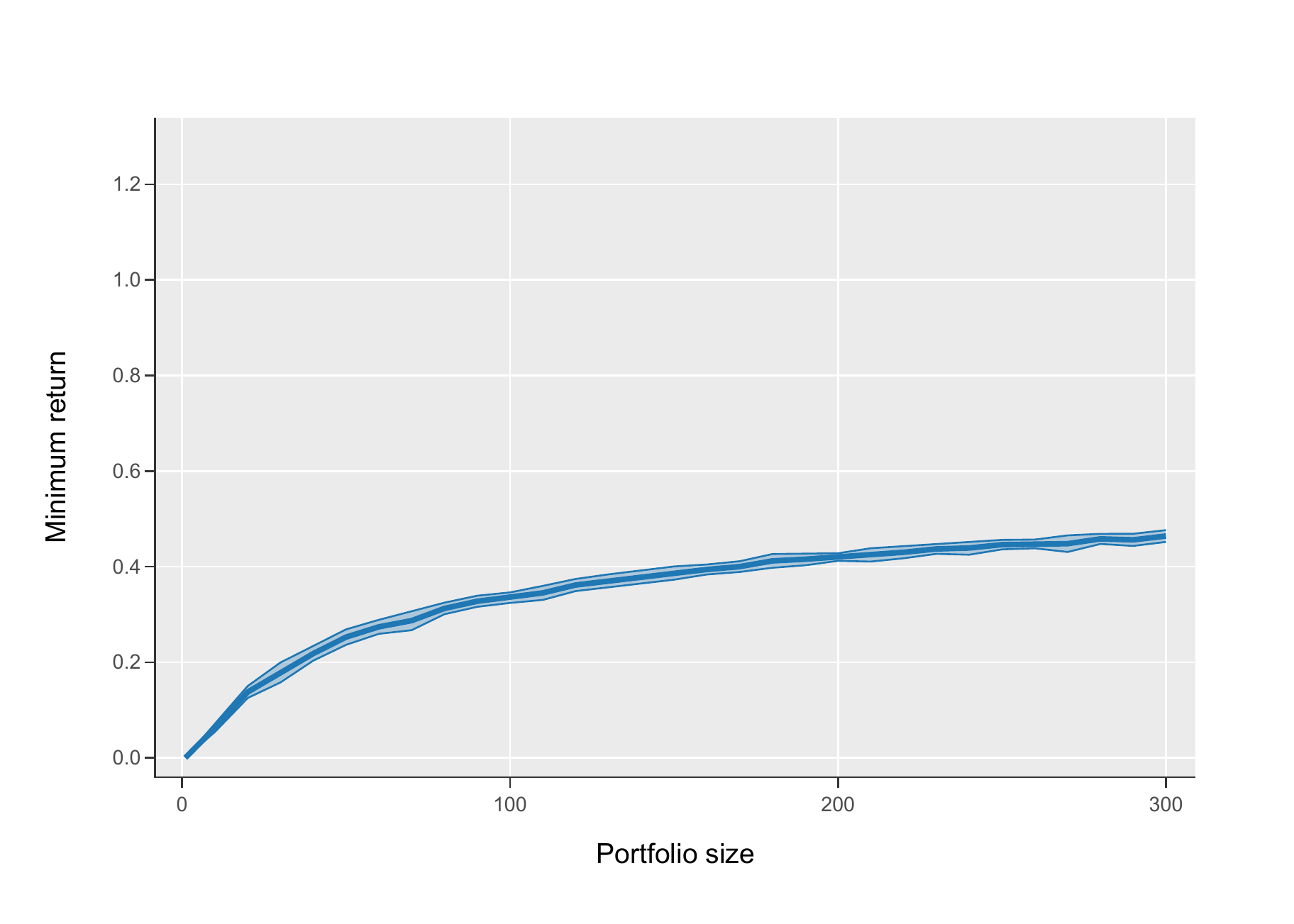}
            \caption{Minimum return}
            \label{fig:min_return_underperformer}
        \end{subfigure}
        \begin{subfigure}[t]{0.32\linewidth}
            \includegraphics[width=\textwidth]{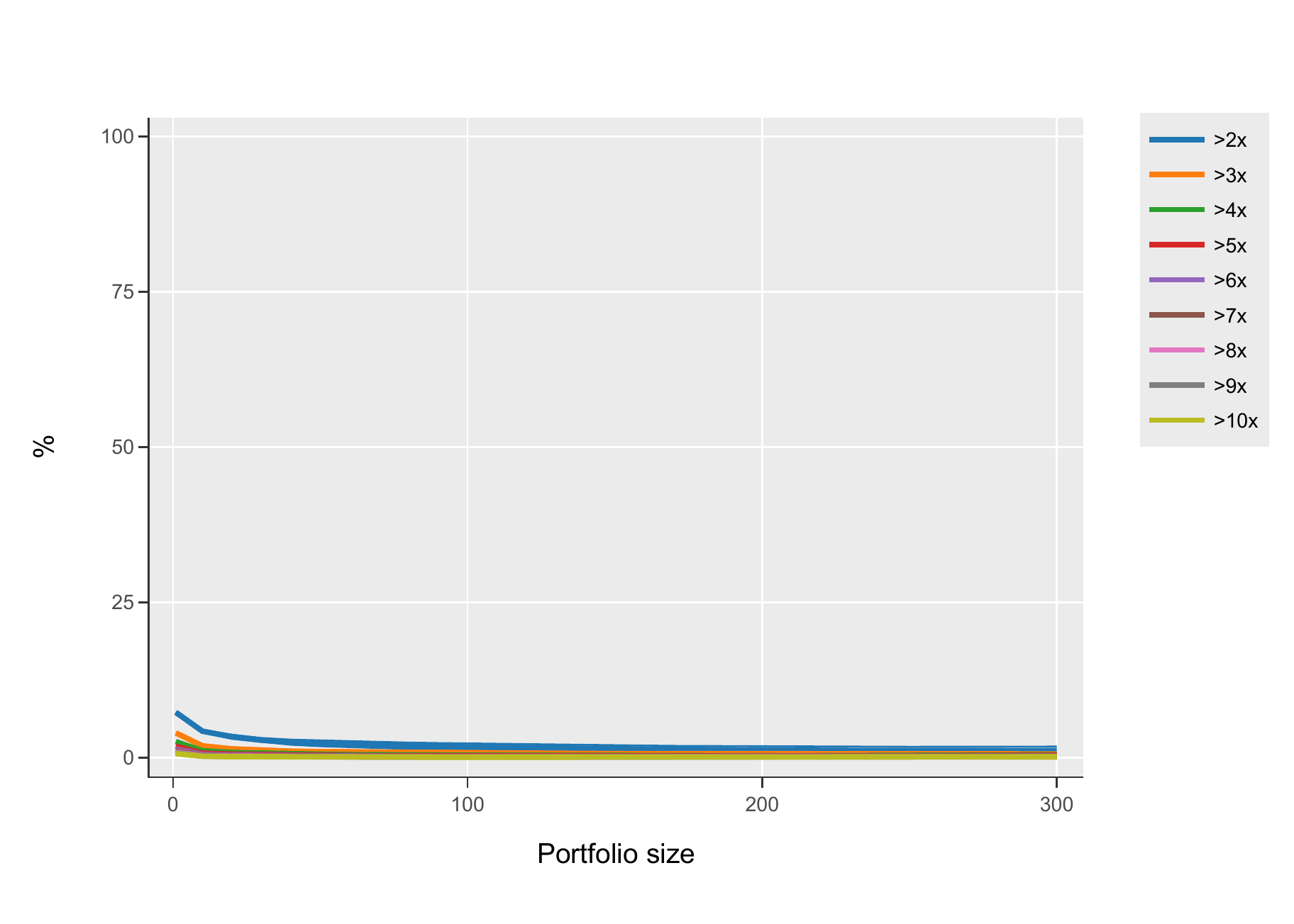}
            \caption{Frequency of 2-10x returns}
            \label{fig:x_returns_underperformer}
        \end{subfigure}
        \caption{Risk profile and returns for an underperformer for different portfolio sizes (mean and standard deviation).}
        \label{fig:underperformer_risk_profile}
    \end{figure}

    In this case, the data suggests that, while minimum return still decreases as the portfolio size increases, you're better off having a portfolio as small as possible if you want to lower your chances of losing money or slightly increase your chances of high returns. But we're talking about small numbers here ($<$6\%) – see Figure~\ref{fig:underperformer_risk_profile}.

    \begin{helpquote}{Why does this happen?}
        Simply because the number of high-return investments is not enough to offset the number of bad investments to obtain any return multiple.
    \end{helpquote}

    \section{Impact of ticket sizing policy}
    So far, we've assumed that the the capital is equally split across all the investments in a portfolio. In reality, that's not often the case since round dynamics often determine investment sizes.

    To simulate variable ticket sizes, we'll run the same set of experiments as before, but, instead of allocating the same fraction of the overall capital to each investment, we'll consider different strategies.

    \subsection{Random ticket sizes}
\begin{figure}[h!]
        \centering
        \begin{subfigure}[t]{0.32\linewidth}
            \includegraphics[width=\textwidth]{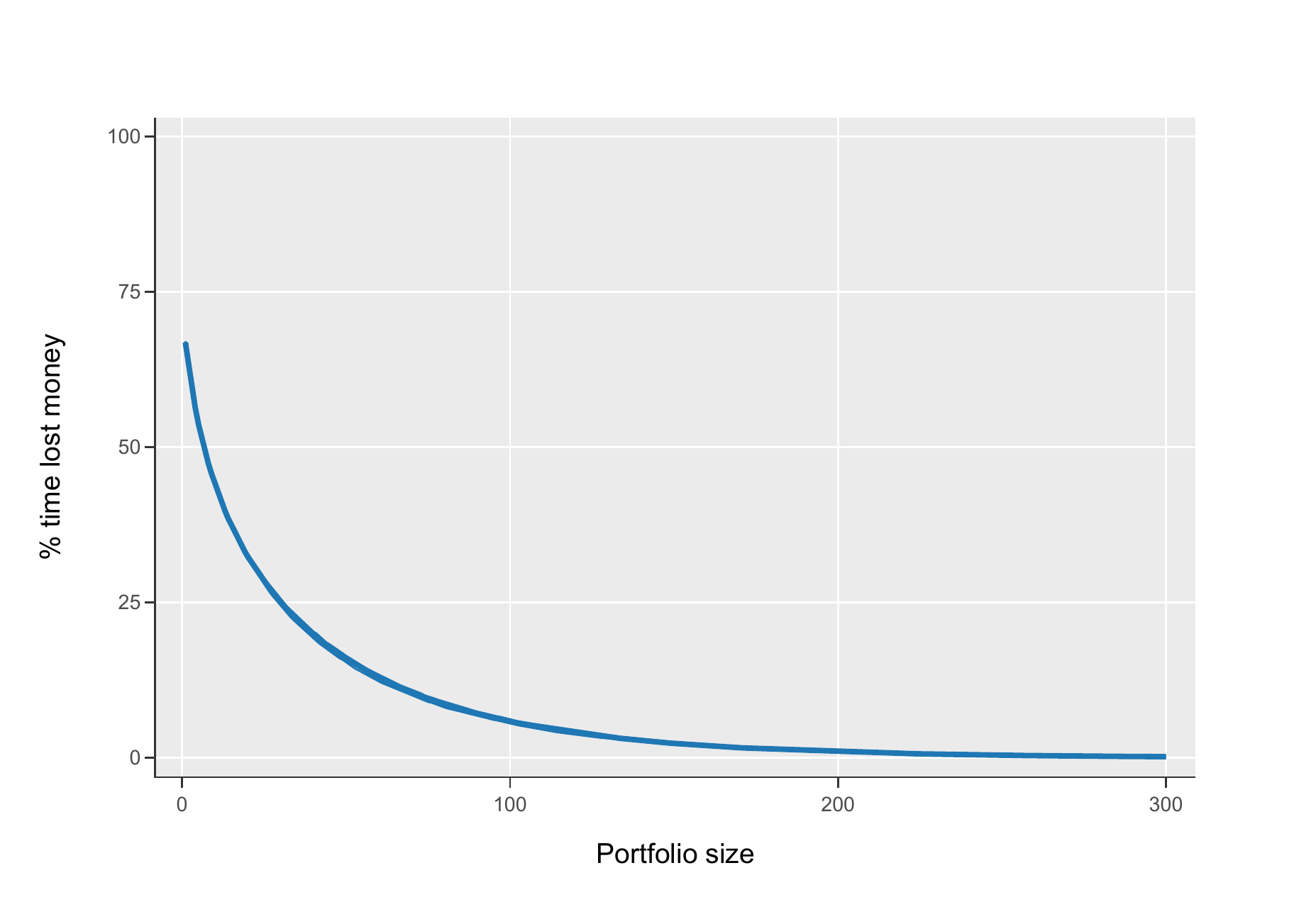}
            \caption{Percentage of portfolios losing money}
            \label{fig:pc_time_lost_money_random_tickets}
        \end{subfigure}
        \begin{subfigure}[t]{0.32\linewidth}
            \includegraphics[width=\textwidth]{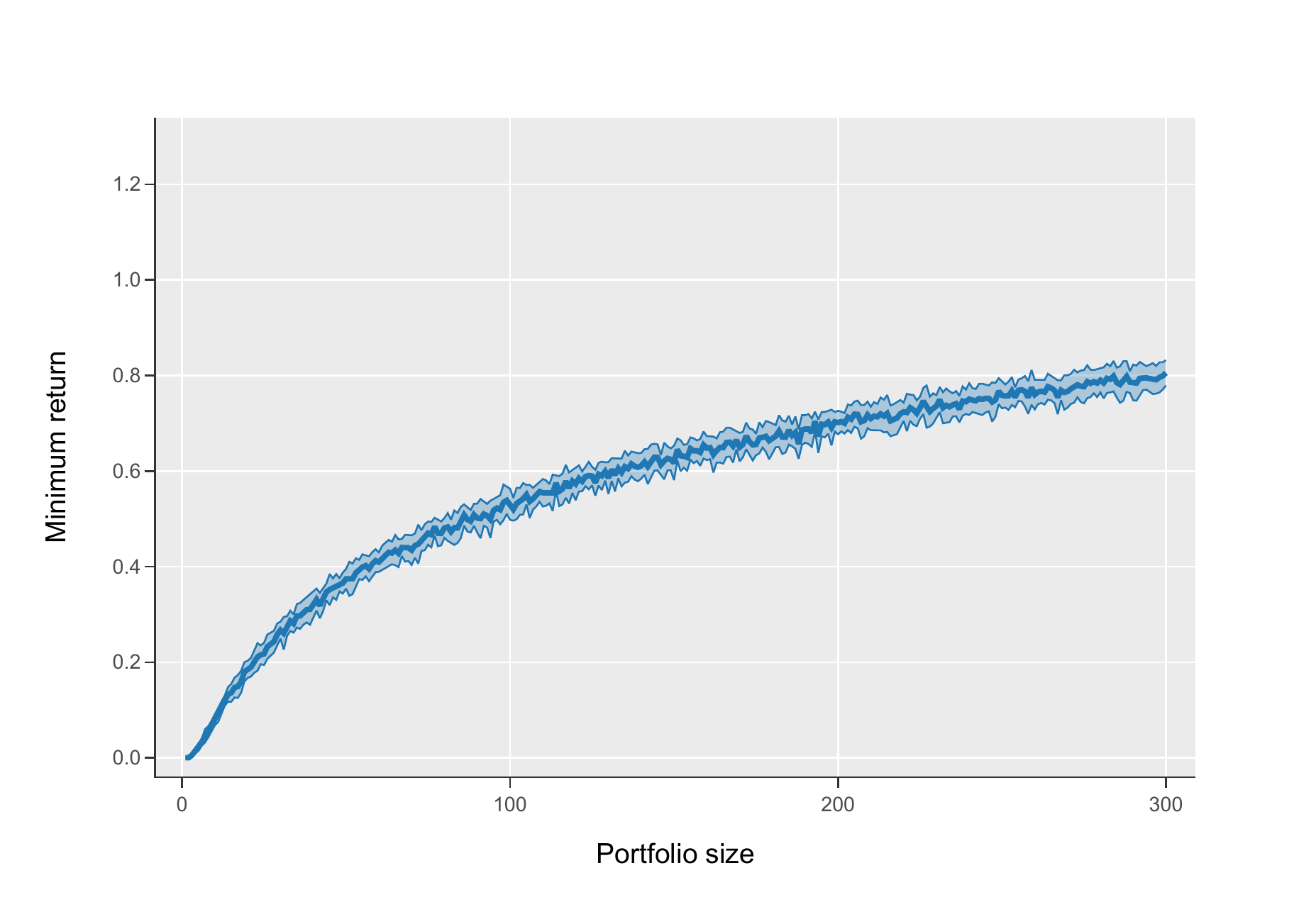}
            \caption{Minimum return}
            \label{fig:min_return_random_tickets}
        \end{subfigure}
        \begin{subfigure}[t]{0.32\linewidth}
            \includegraphics[width=\textwidth]{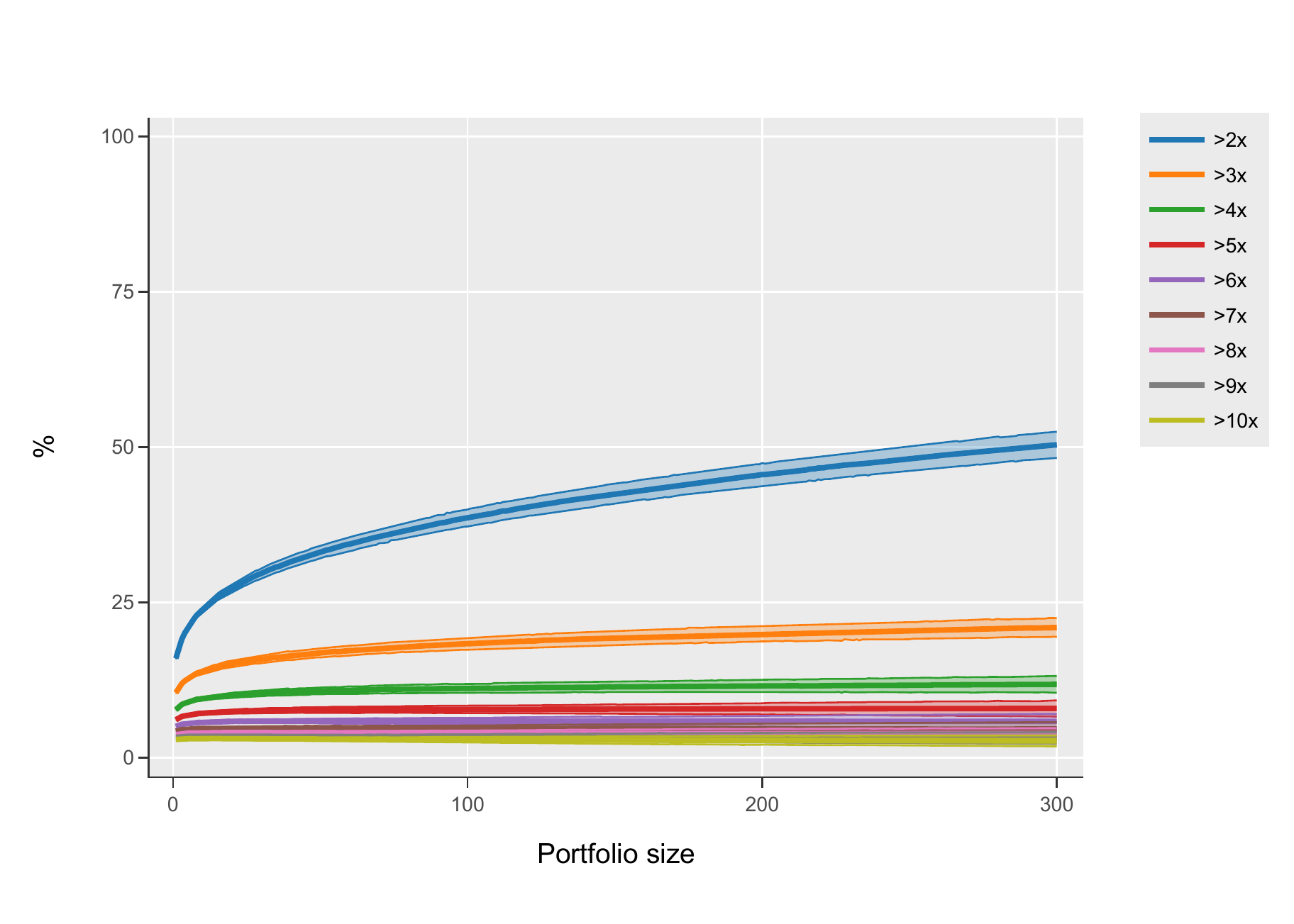}
            \caption{Frequency of 2-10x returns}
            \label{fig:x_returns_random_tickets}
        \end{subfigure}
        \caption{Risk profile and returns for different portfolio sizes (mean and standard deviation) using random ticket sizes.}
        \label{fig:random_tickets_risk_profile}
    \end{figure}

    In this set of simulations, we randomly allocate different amounts to each investment, with a min-max ratio of 10 (meaning that the largest investment out of a fund will be at most 10 times larger than the smallest one). Results are shown in Figure~\ref{fig:random_tickets_risk_profile}.

    As we can see, the overall trend is maintained. However, there's a slight decrease in performance, especially for small portfolio sizes. In Table~\ref{tab:min_return_random_tickets} we report the results for the expected minimum return as an example.

    \begin{table}[h!]
        \scriptsize
        \centering
        \begin{tabular}{r|r|r|r}
            Portfolio size & Uniform tickets & Variable tickets & Difference \% \\ \hline\hline
            10 & 0.0935 & 0.077 & -17.6471 \\ \hline
            50 & 0.4145 & 0.3685 & -11.0977 \\ \hline
            100 & 0.565 & 0.532 & -5.84071 \\ \hline
            200 & 0.7415 & 0.7025 & -5.25961 \\ \hline
            300 & 0.839 & 0.7995 & -4.70799 \\ \hline
        \end{tabular}
        \caption{Minimum return for different portfolio sizes: uniform tickets vs random tickets strategy.}
        \label{tab:min_return_random_tickets}
    \end{table}

    \begin{warningquote}{The impact of bounded ROI}
        Interestingly, by using variable ticket sizes, the impact of bounded returns on the probability of returning large multiples of the fund is smoothed. However, the overall trend is maintained.
        \begin{center}
            \captionsetup{type=figure}
            \begin{subfigure}{0.32\linewidth}
                \includegraphics[width=\textwidth]{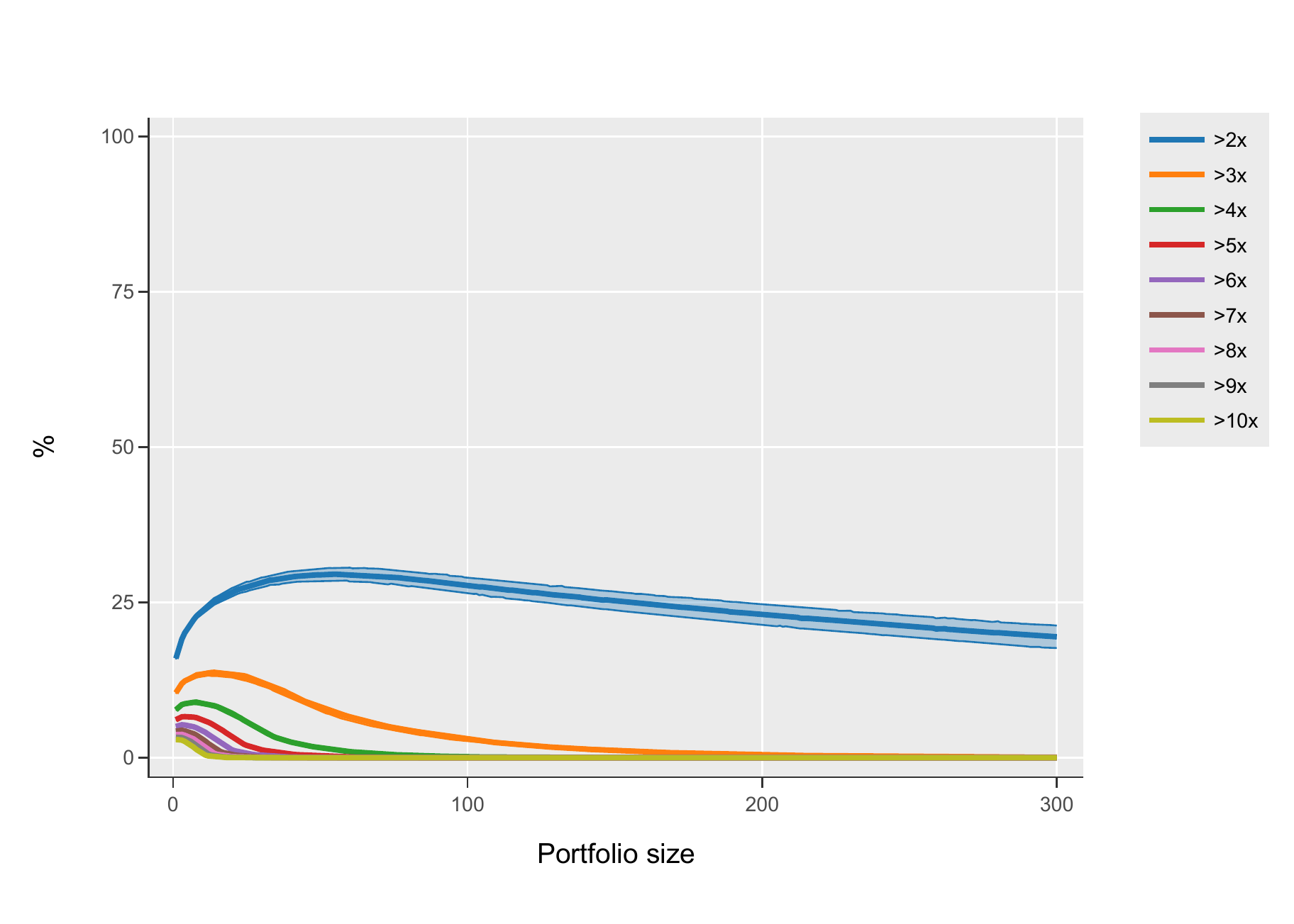}
                \caption{50x bound}
                \label{fig:x_returns_random_ticket_50}
            \end{subfigure}
            \begin{subfigure}{0.32\linewidth}
                \includegraphics[width=\textwidth]{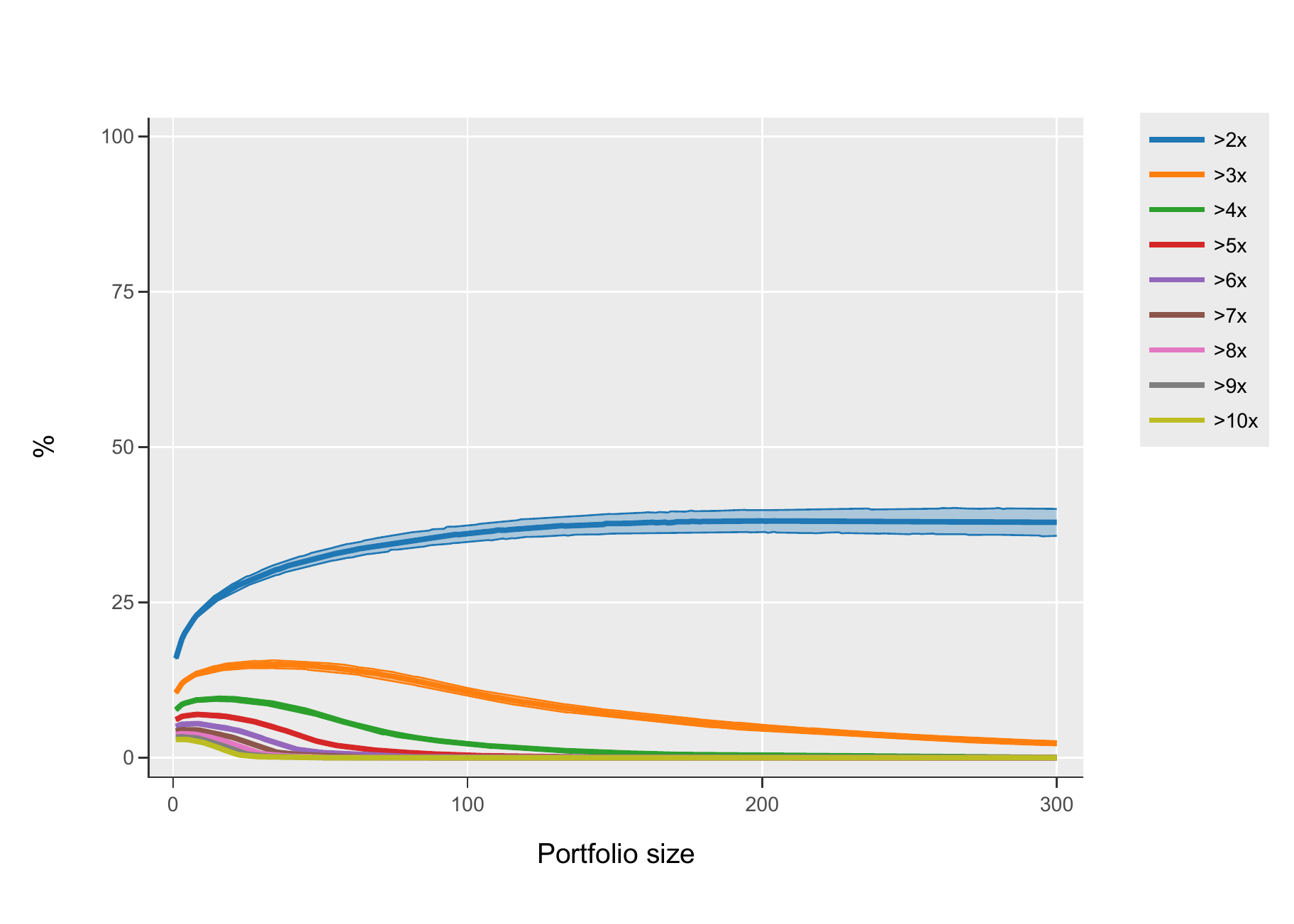}
                \caption{100x bound}
                \label{fig:x_returns_random_ticket_100}
            \end{subfigure}
            \begin{subfigure}{0.32\linewidth}
                \includegraphics[width=\textwidth]{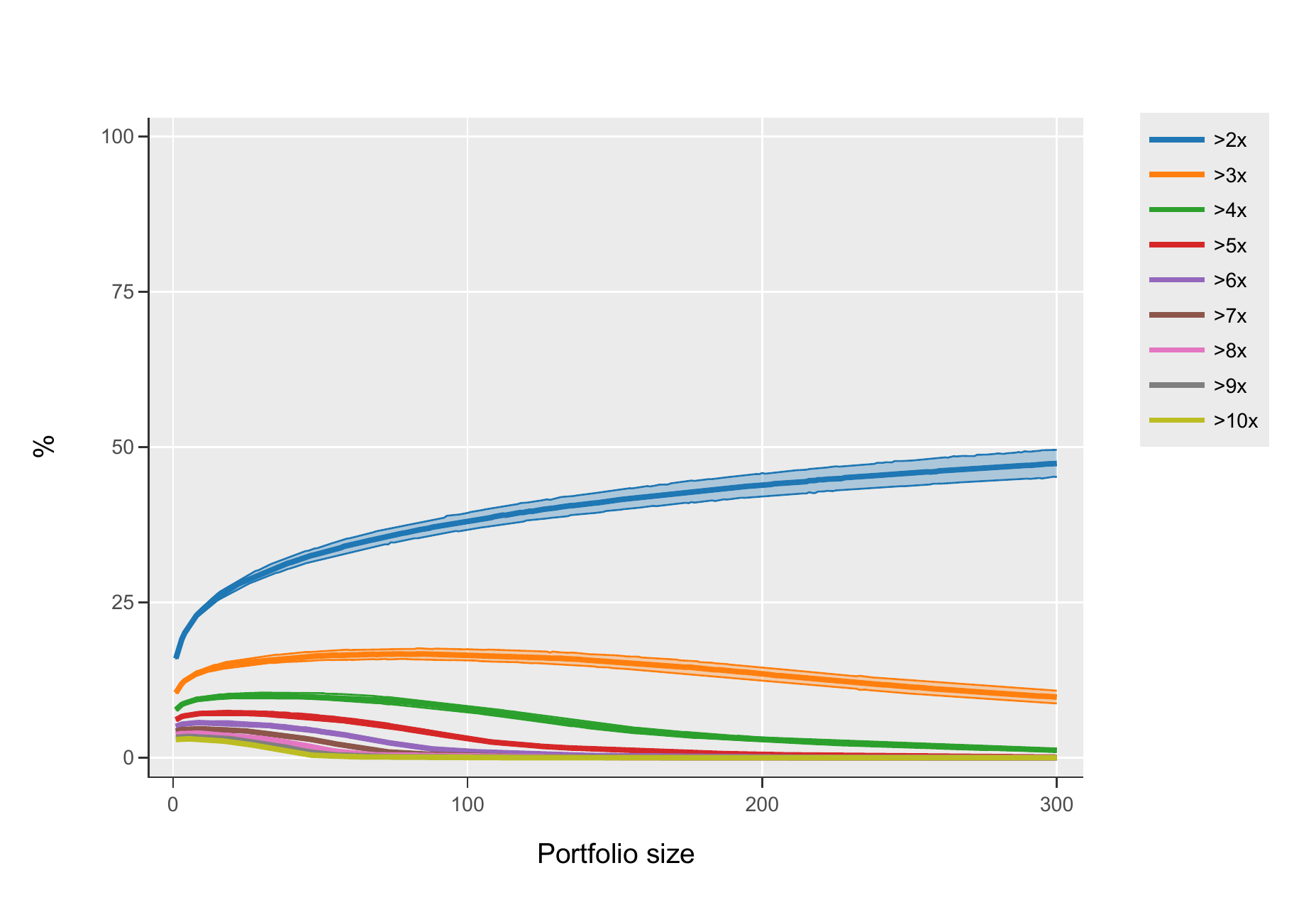}
                \caption{200x bound}
                \label{fig:x_returns_random_ticket_200}
            \end{subfigure}
            \begin{subfigure}{0.32\linewidth}
                \includegraphics[width=\textwidth]{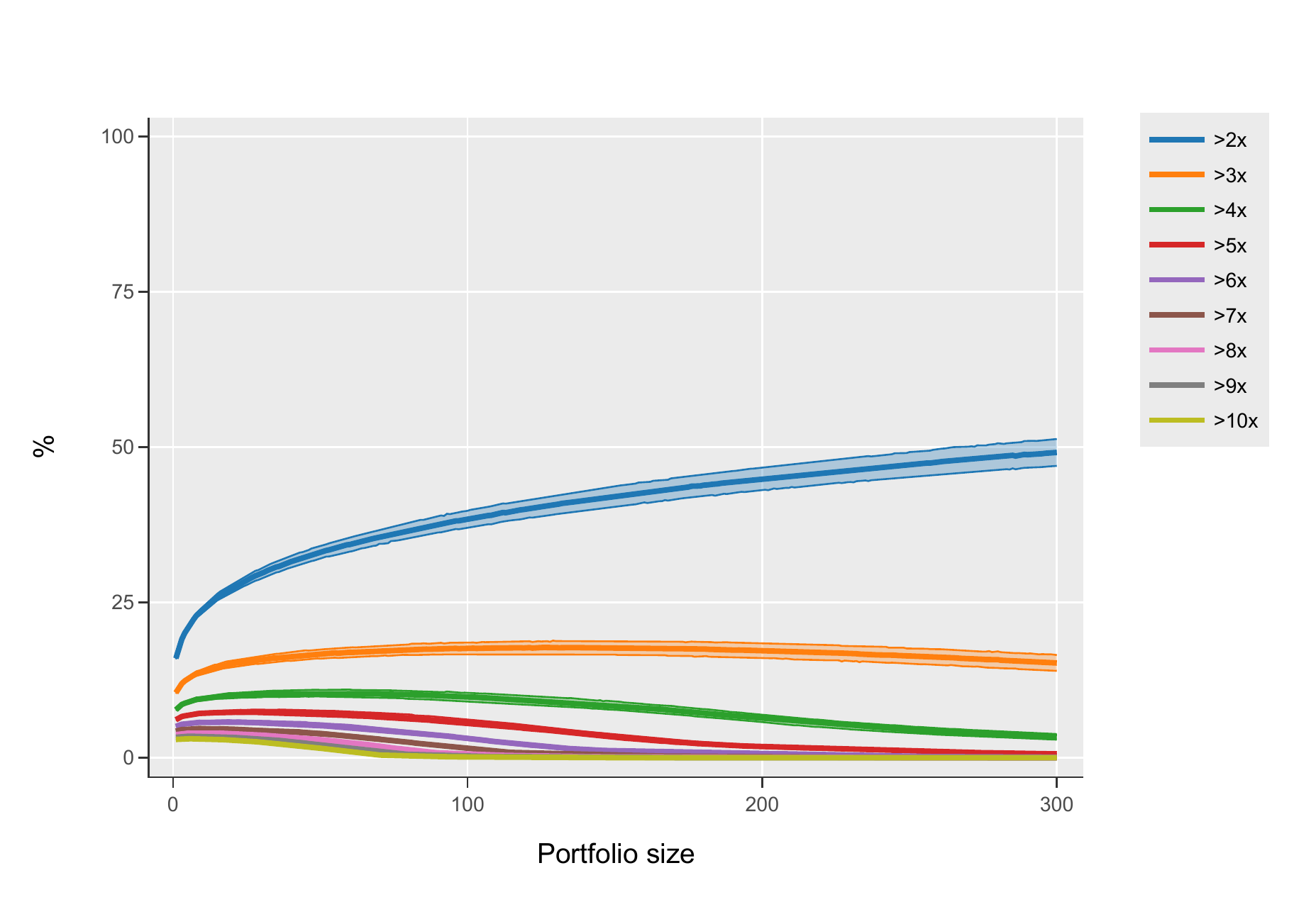}
                \caption{300x bound}
                \label{fig:x_returns_random_ticket_300}
            \end{subfigure}
            \begin{subfigure}{0.32\linewidth}
                \includegraphics[width=\textwidth]{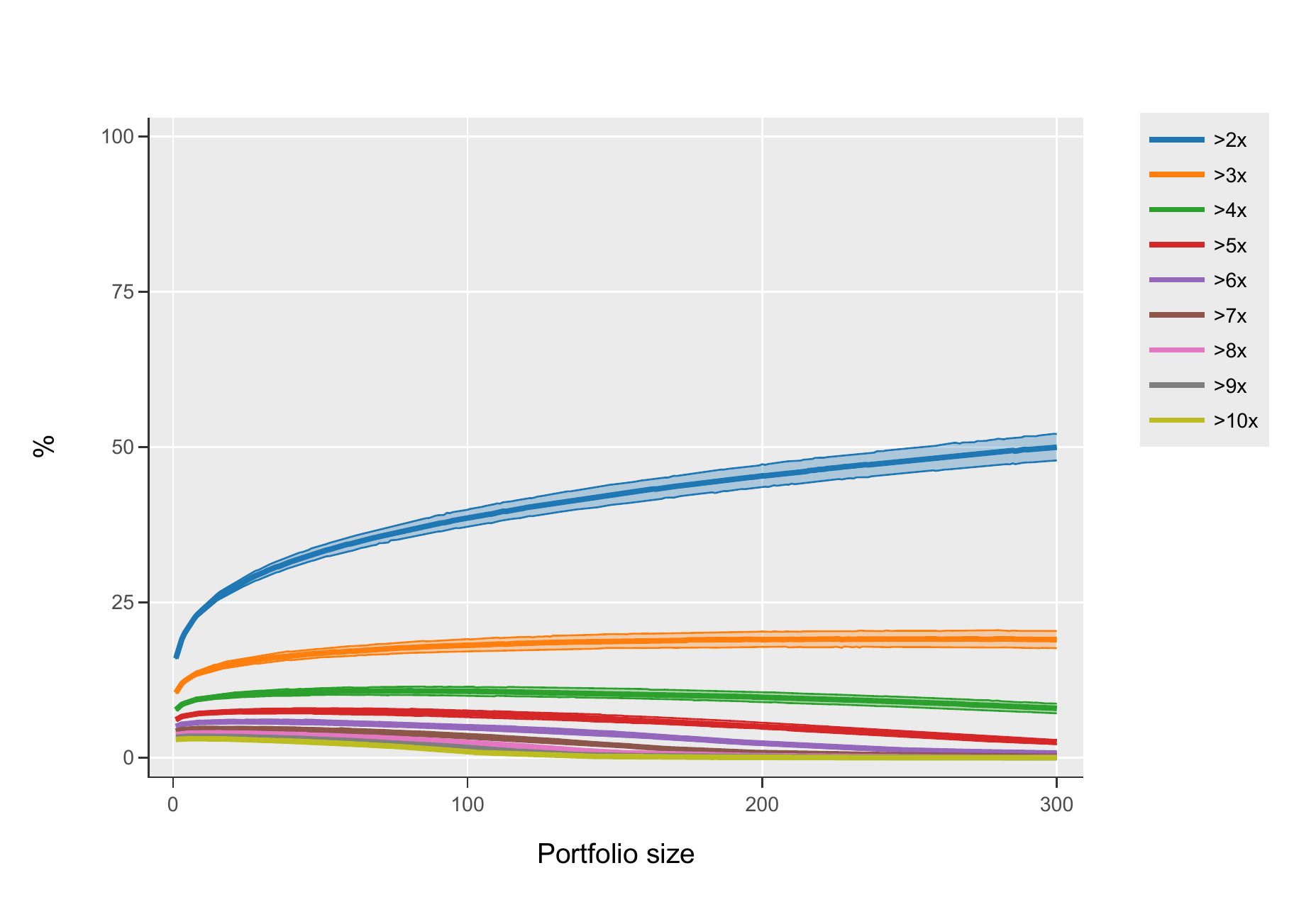}
                \caption{500x bound}
                \label{fig:x_returns_random_ticket_500}
            \end{subfigure}
            \begin{subfigure}{0.32\linewidth}
                \includegraphics[width=\textwidth]{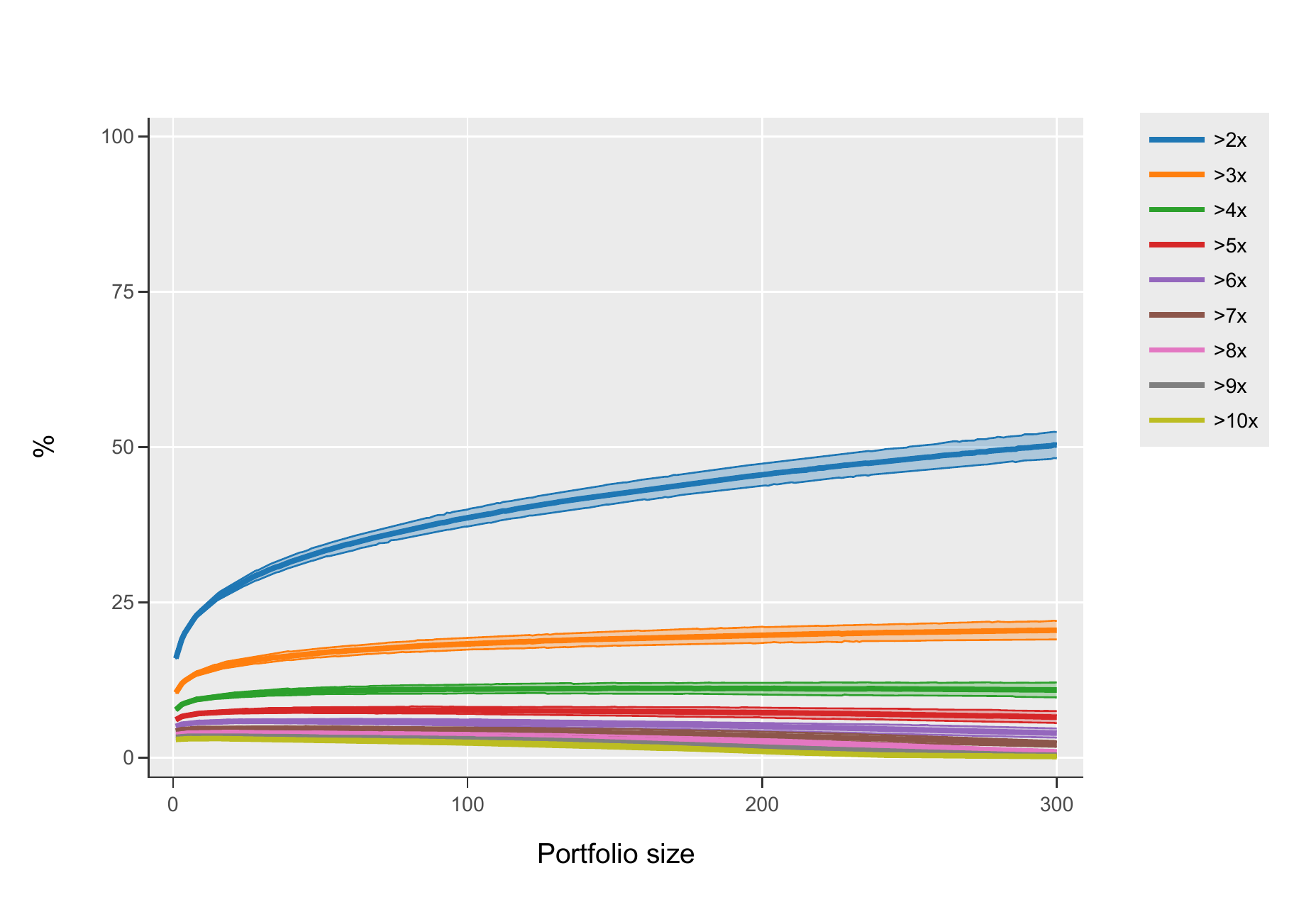}
                \caption{1000x bound}
                \label{fig:x_returns_random_ticket_1000}
            \end{subfigure}
            \caption{Frequency of returns in the range 2-10x for different portfolio sizes and different bounds on the maximum return per investment (mean and standard deviation) using random tickets.}
            \label{fig:x_returns_random_ticket_bounds}
        \end{center}
    \end{warningquote}

    \subsection{Ticket size proportional to deal quality}
\begin{figure}[h!]
        \centering
        \begin{subfigure}[t]{0.32\linewidth}
            \includegraphics[width=\textwidth]{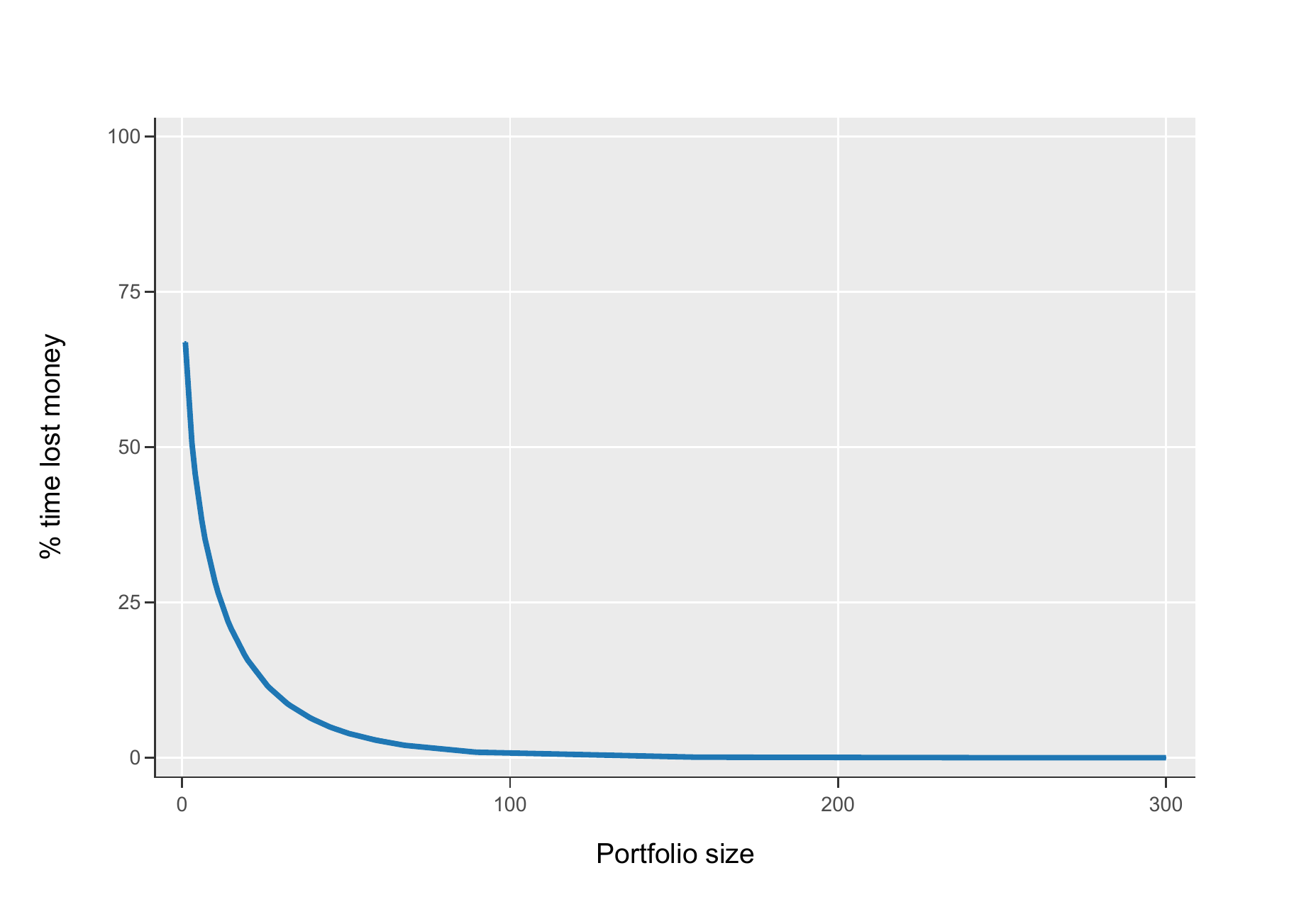}
            \caption{Percentage of portfolios losing money}
            \label{fig:pc_time_lost_money_proportional_tickets}
        \end{subfigure}
        \begin{subfigure}[t]{0.32\linewidth}
            \includegraphics[width=\textwidth]{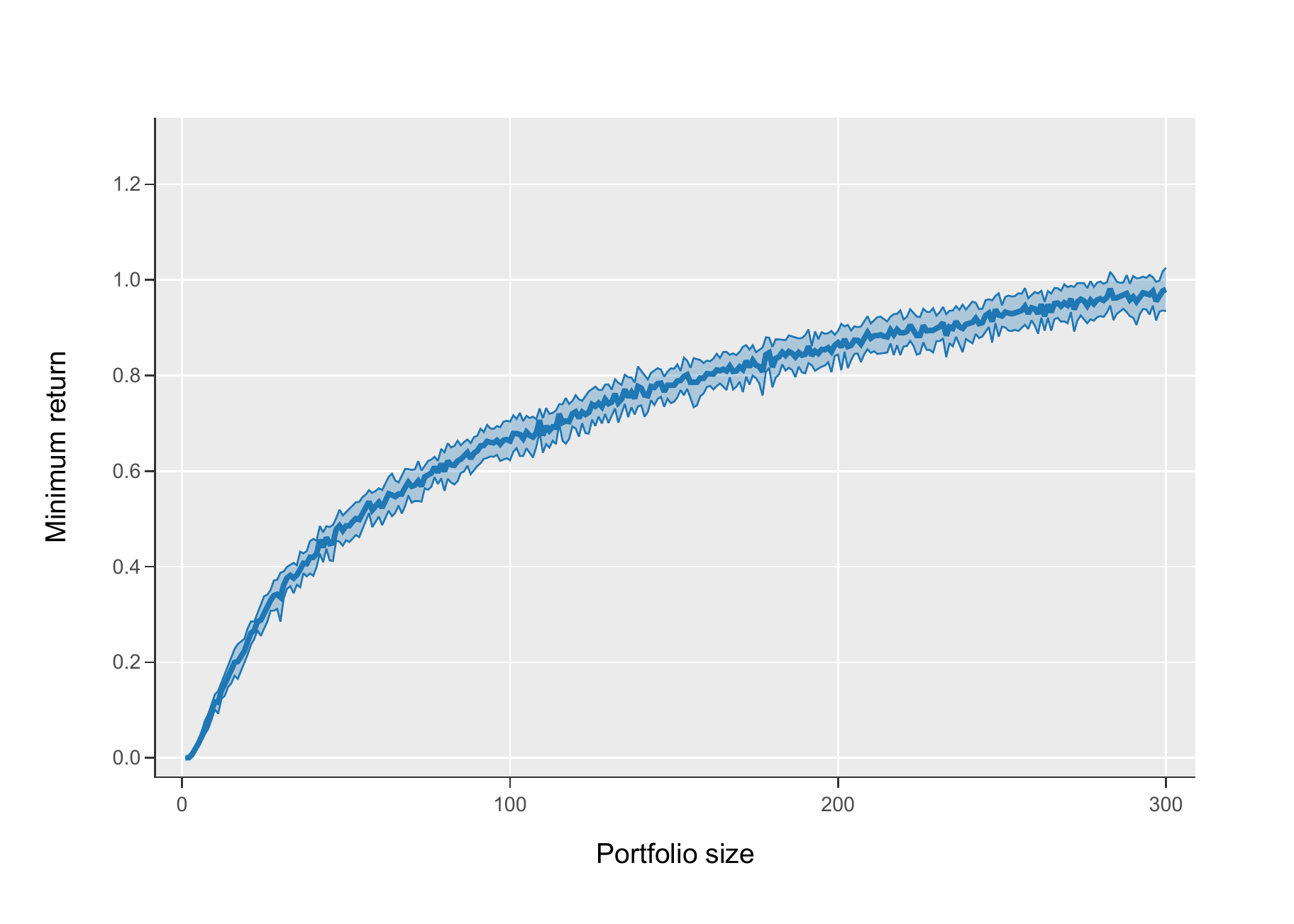}
            \caption{Minimum return}
            \label{fig:min_return_proportional_tickets}
        \end{subfigure}
        \begin{subfigure}[t]{0.32\linewidth}
            \includegraphics[width=\textwidth]{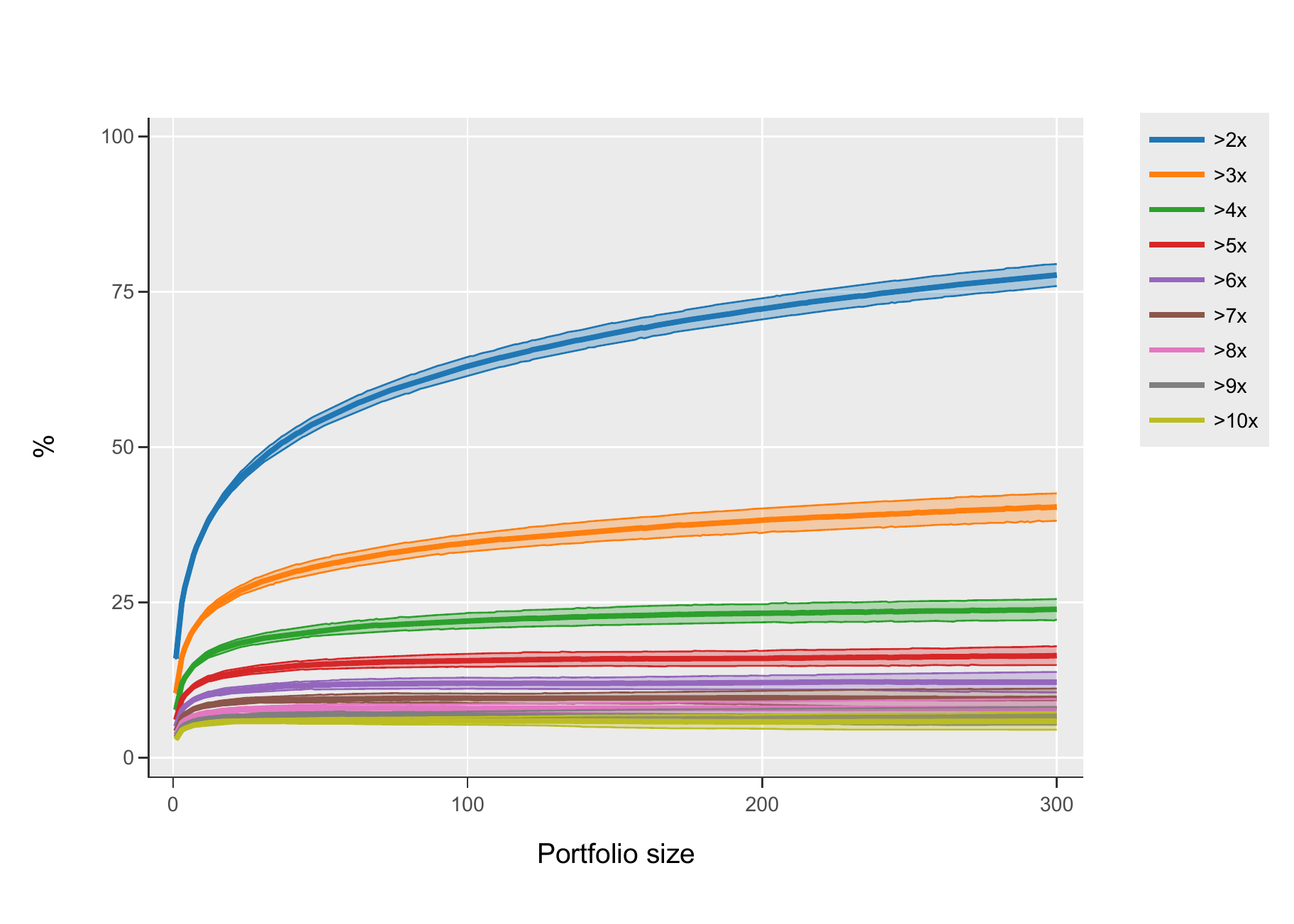}
            \caption{Frequency of 2-10x returns}
            \label{fig:x_returns_proportional_tickets}
        \end{subfigure}
        \caption{Risk profile and returns for different portfolio sizes (mean and standard deviation) using ticket sizes proportional to deal quality.}
        \label{fig:proportional_tickets_risk_profile}
    \end{figure}

    We now assume that we're able to determine with a certain accuracy which investments are more likely to provide high returns and which are most likely to fail. We allocate larger tickets to the former and smaller to the latter, proportionally to their expected return (plus some noise), with a min-max ratio of 2 – see Figure~\ref{fig:proportional_tickets_risk_profile}. As we can see, this strategy clearly allows one to overperform the average VC.

    \begin{warningquote}{The impact of bounded ROI}
        By using a ticket size proportional to quality, the impact of bounded returns on the probability of returning large multiples of the fund is reduced.
        \begin{center}
            \captionsetup{type=figure}
            \begin{subfigure}{0.32\linewidth}
                \includegraphics[width=\textwidth]{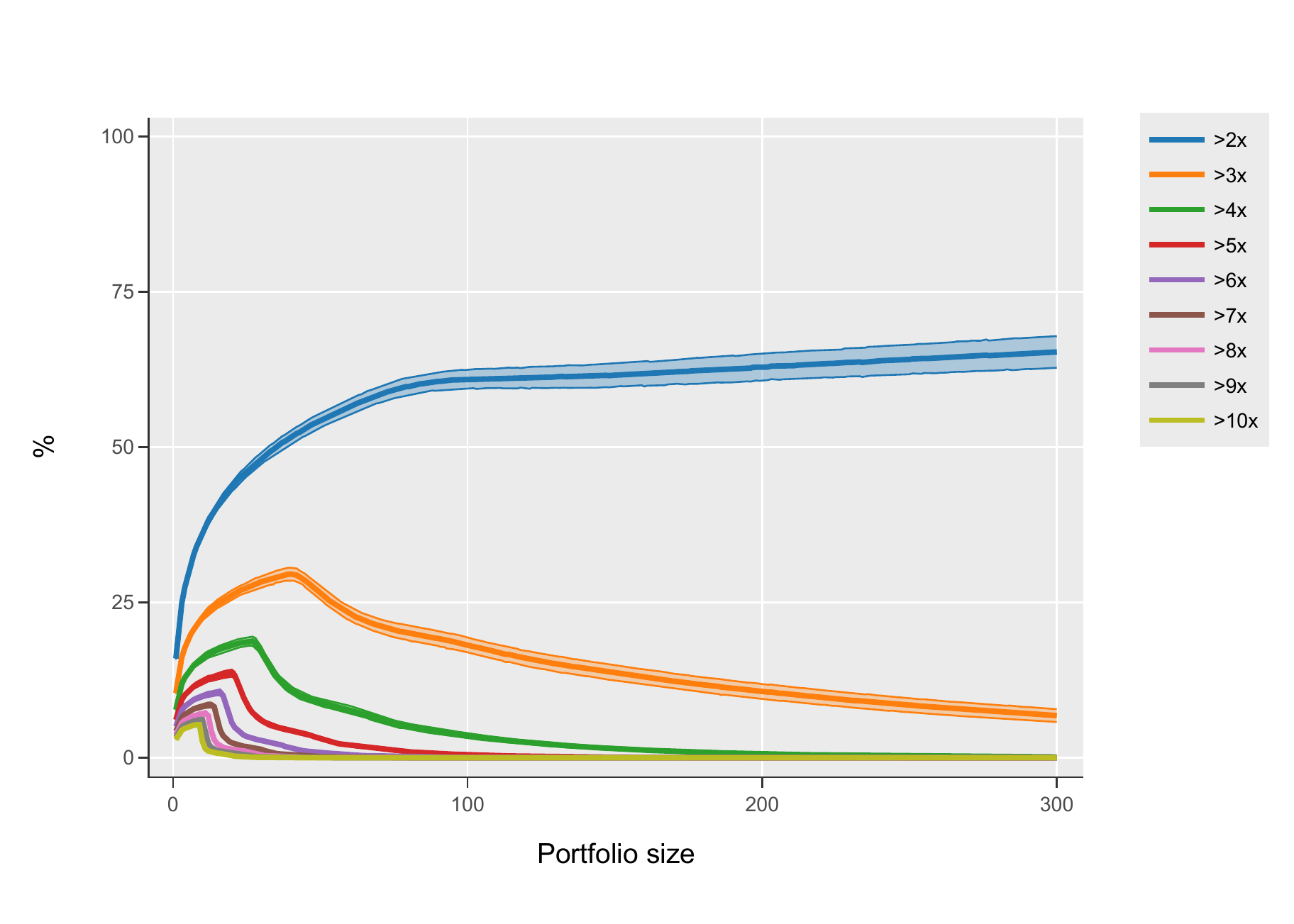}
                \caption{50x bound}
                \label{fig:x_returns_porportional_ticket_50}
            \end{subfigure}
            \begin{subfigure}{0.32\linewidth}
                \includegraphics[width=\textwidth]{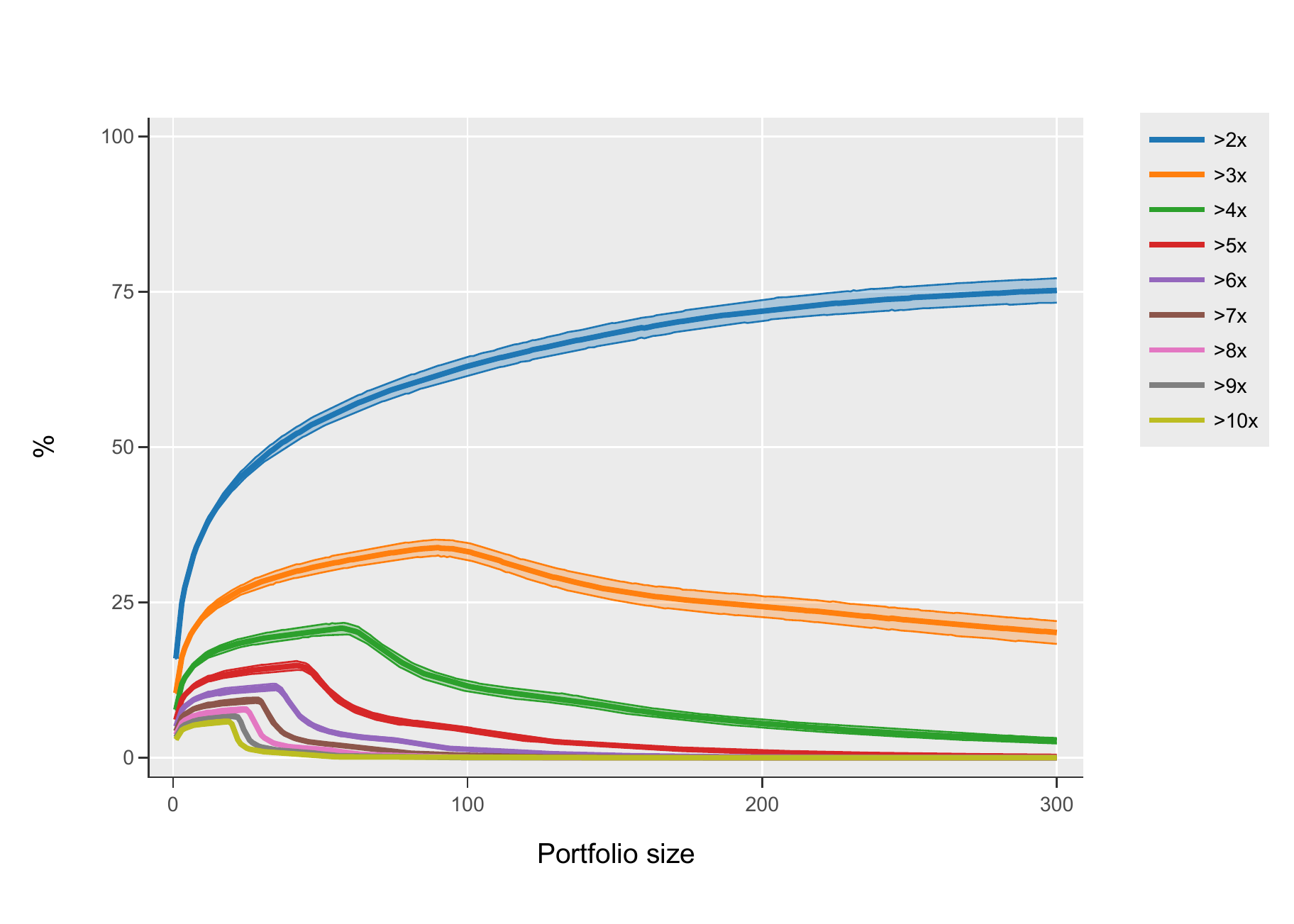}
                \caption{100x bound}
                \label{fig:x_returns_porportional_ticket_100}
            \end{subfigure}
            \begin{subfigure}{0.32\linewidth}
                \includegraphics[width=\textwidth]{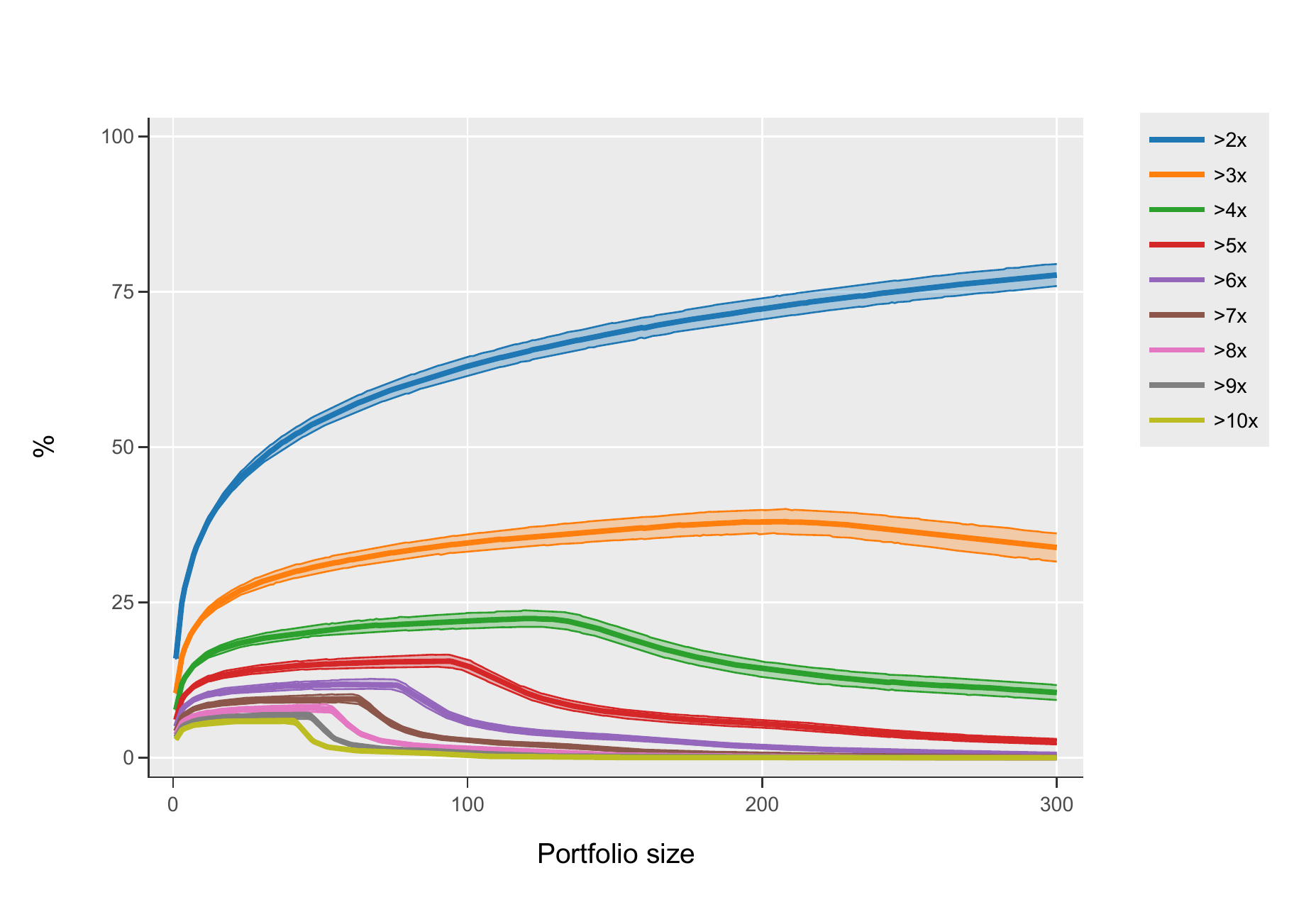}
                \caption{200x bound}
                \label{fig:x_returns_porportional_ticket_200}
            \end{subfigure}
            \begin{subfigure}{0.32\linewidth}
                \includegraphics[width=\textwidth]{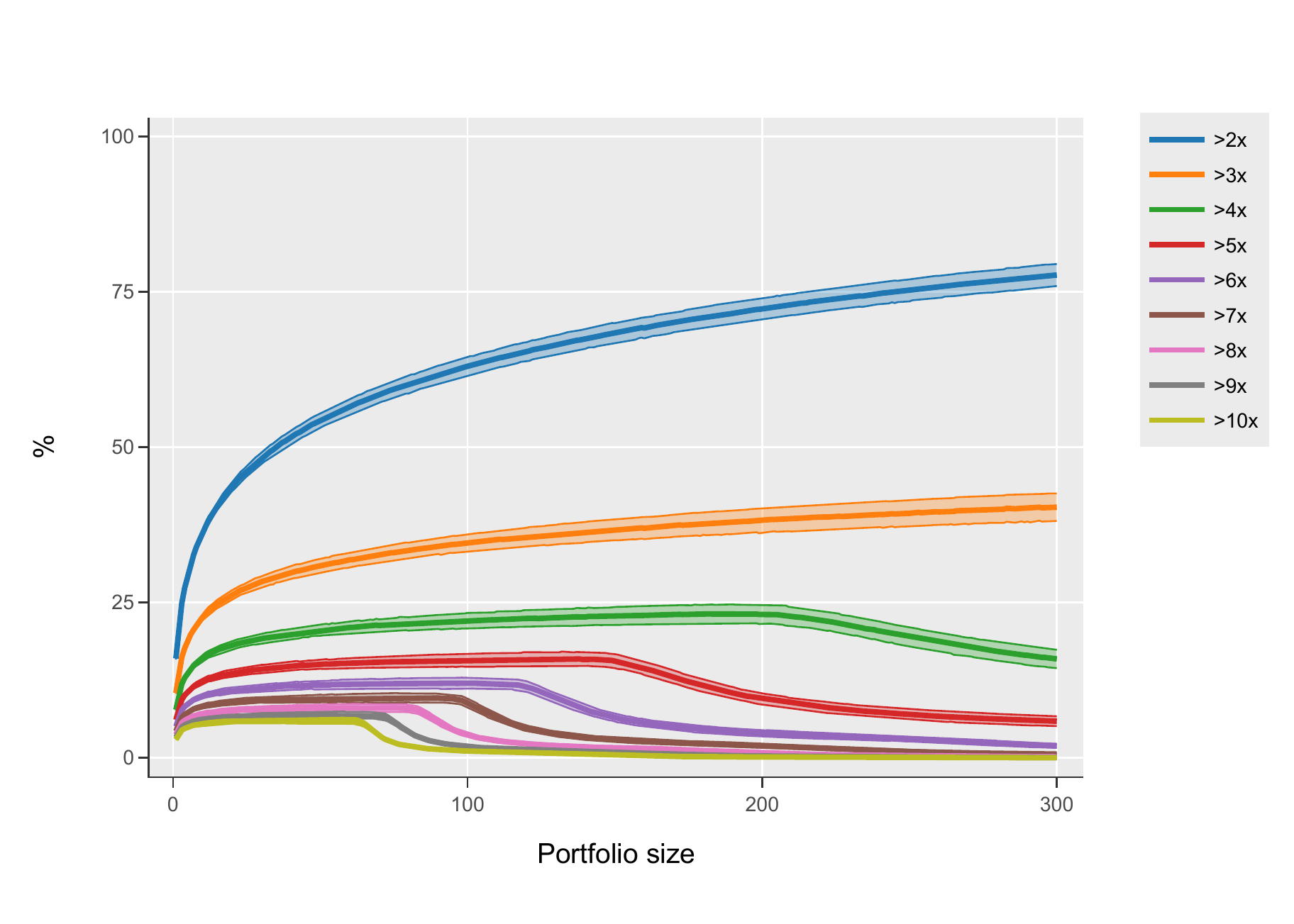}
                \caption{300x bound}
                \label{fig:x_returns_porportional_ticket_300}
            \end{subfigure}
            \begin{subfigure}{0.32\linewidth}
                \includegraphics[width=\textwidth]{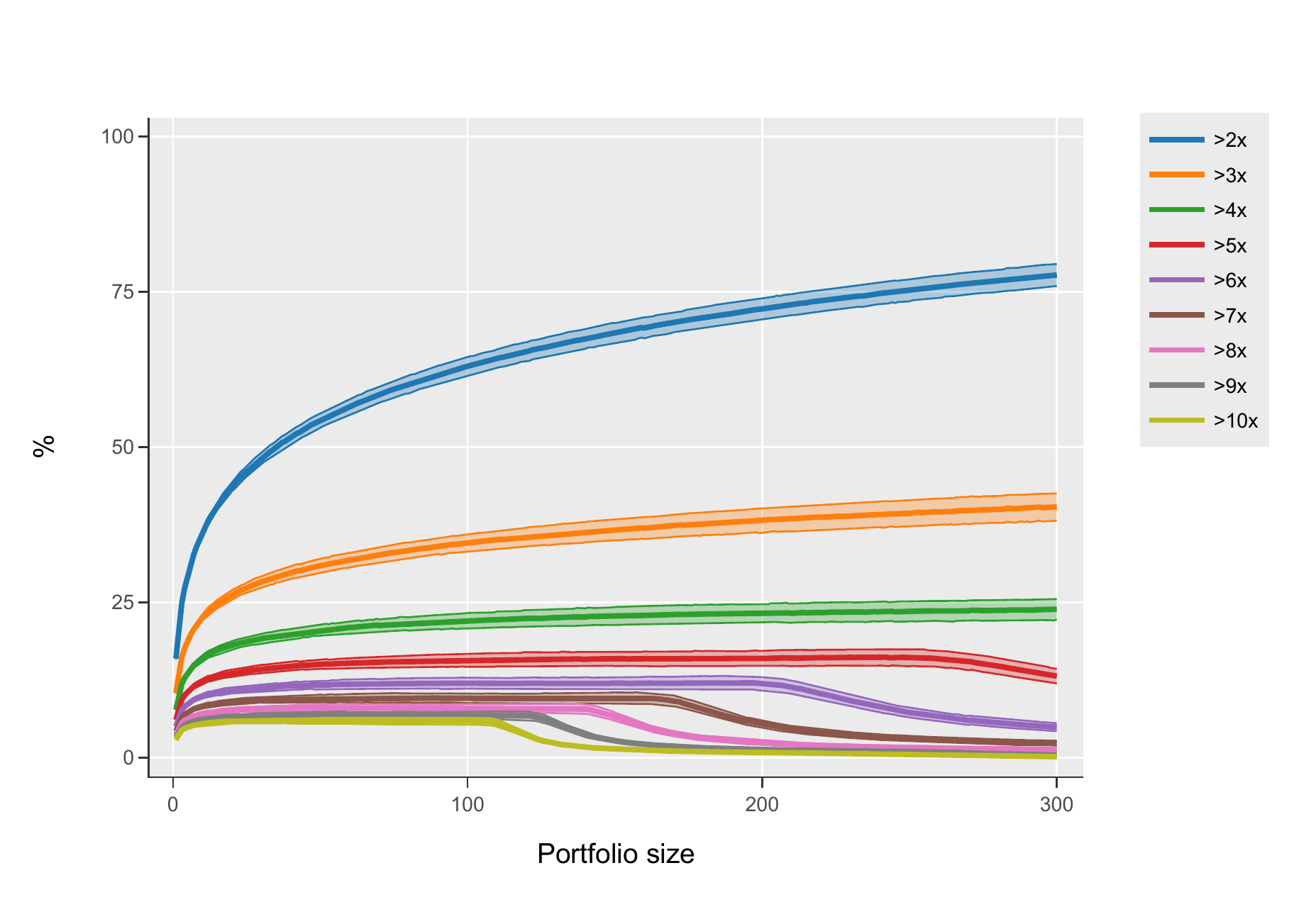}
                \caption{500x bound}
                \label{fig:x_returns_porportional_ticket_500}
            \end{subfigure}
            \begin{subfigure}{0.32\linewidth}
                \includegraphics[width=\textwidth]{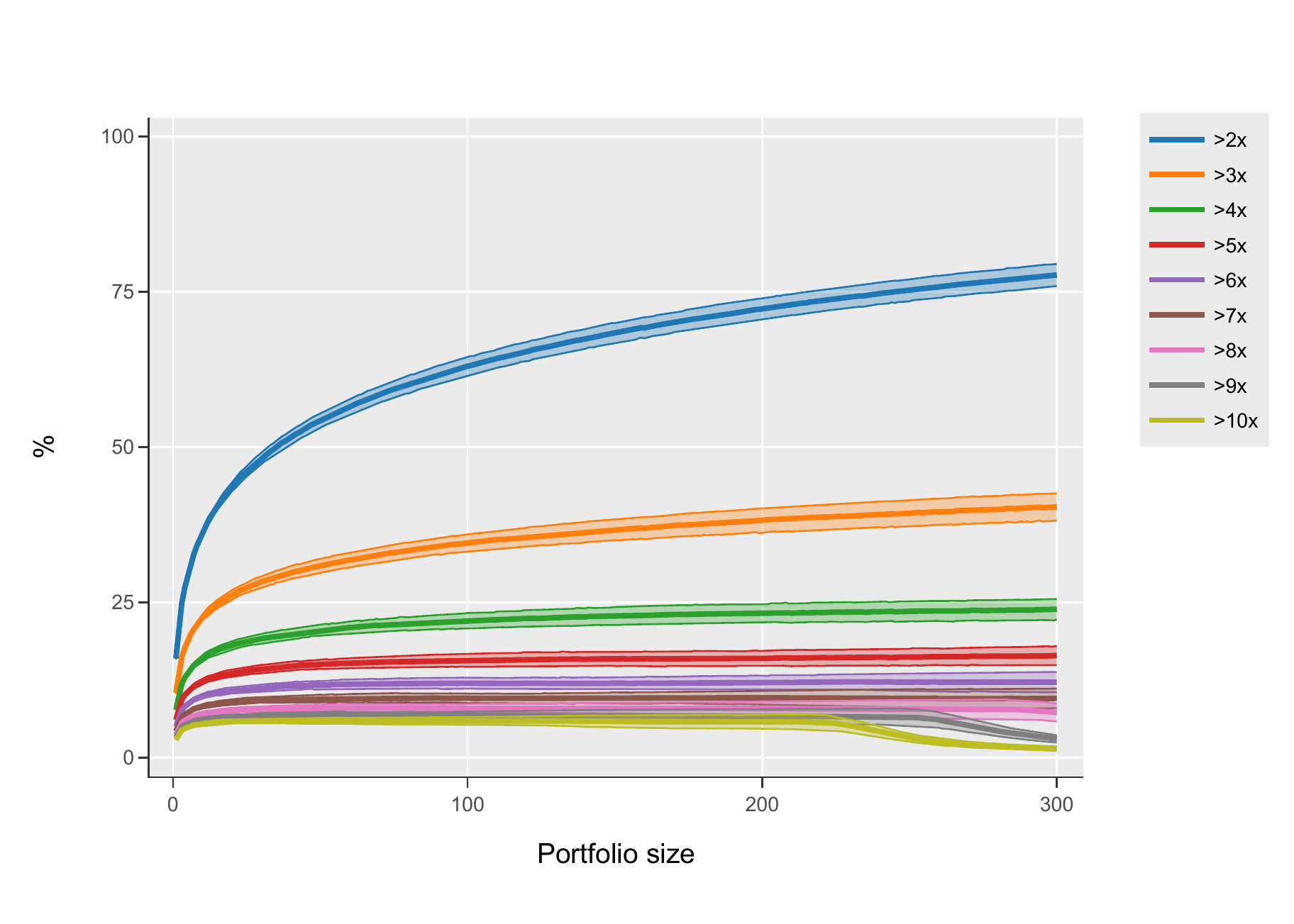}
                \caption{1,000x bound}
                \label{fig:x_returns_porportional_ticket_1000}
            \end{subfigure}
            \caption{Frequency of returns in the range 2-10x for different portfolio sizes and different bounds on the maximum return per investment (mean and standard deviation) using tickets proportional to deal quality.}
            \label{fig:x_returns_porportional_ticket_bounds}
        \end{center}
    \end{warningquote}

    \section{Impact of follow-ons}
    Let's now study the impact of follow-ons on risk and returns. We'll consider an average scenario where a fund manager equally splits the investable capital among initial investments and follow-on reserve. We'll also assume a 3x increase in round size for Series A investments (meaning that with the same amount used for the initial investments, we're able to buy 1/3 of the shares).

    \begin{figure}[h!]
        \centering
        \begin{subfigure}[t]{0.32\linewidth}
            \includegraphics[width=\textwidth]{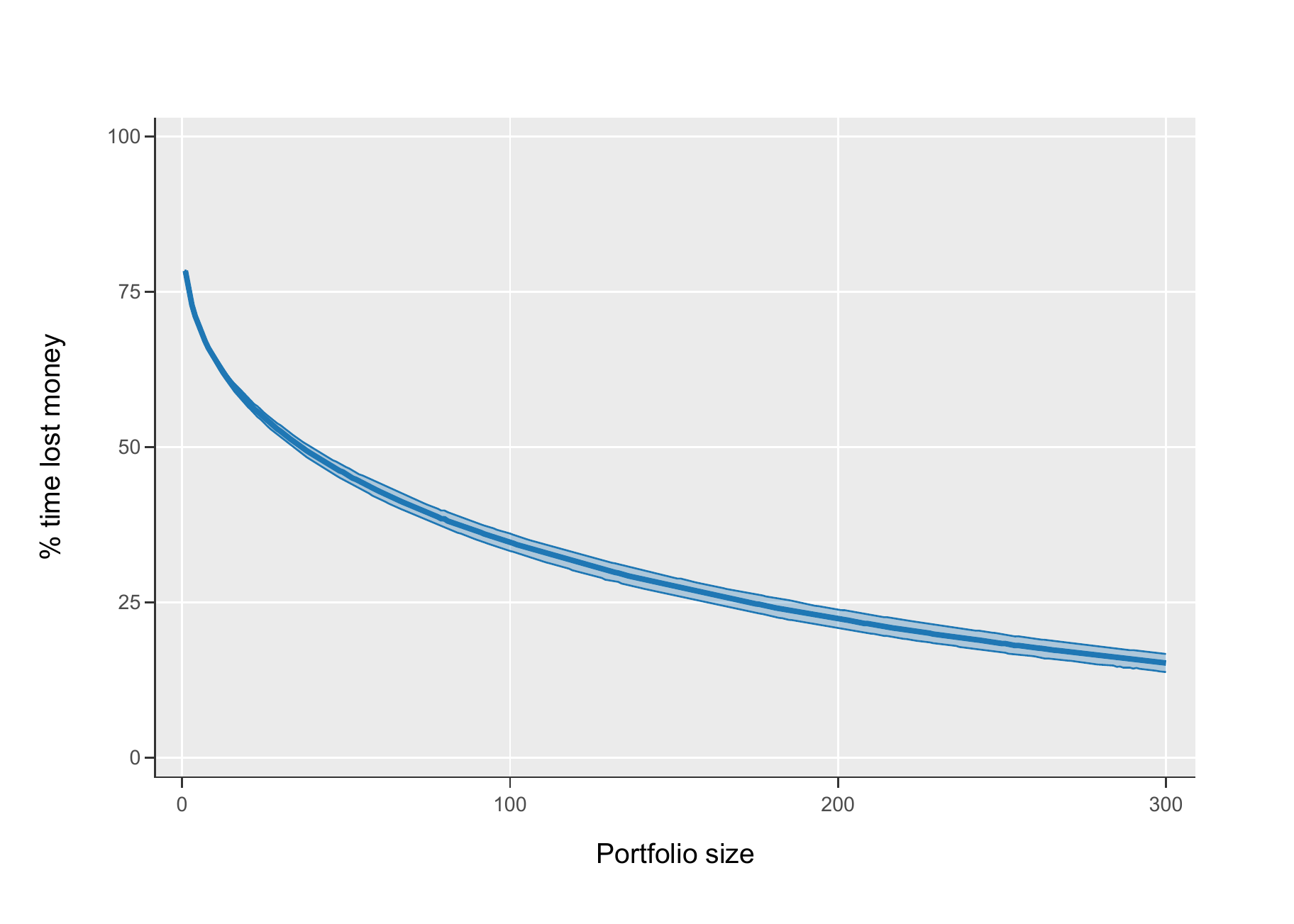}
            \caption{Percentage of portfolios losing money}
            \label{fig:pc_time_lost_money_avg_follow_on_50}
        \end{subfigure}
        \begin{subfigure}[t]{0.32\linewidth}
            \includegraphics[width=\textwidth]{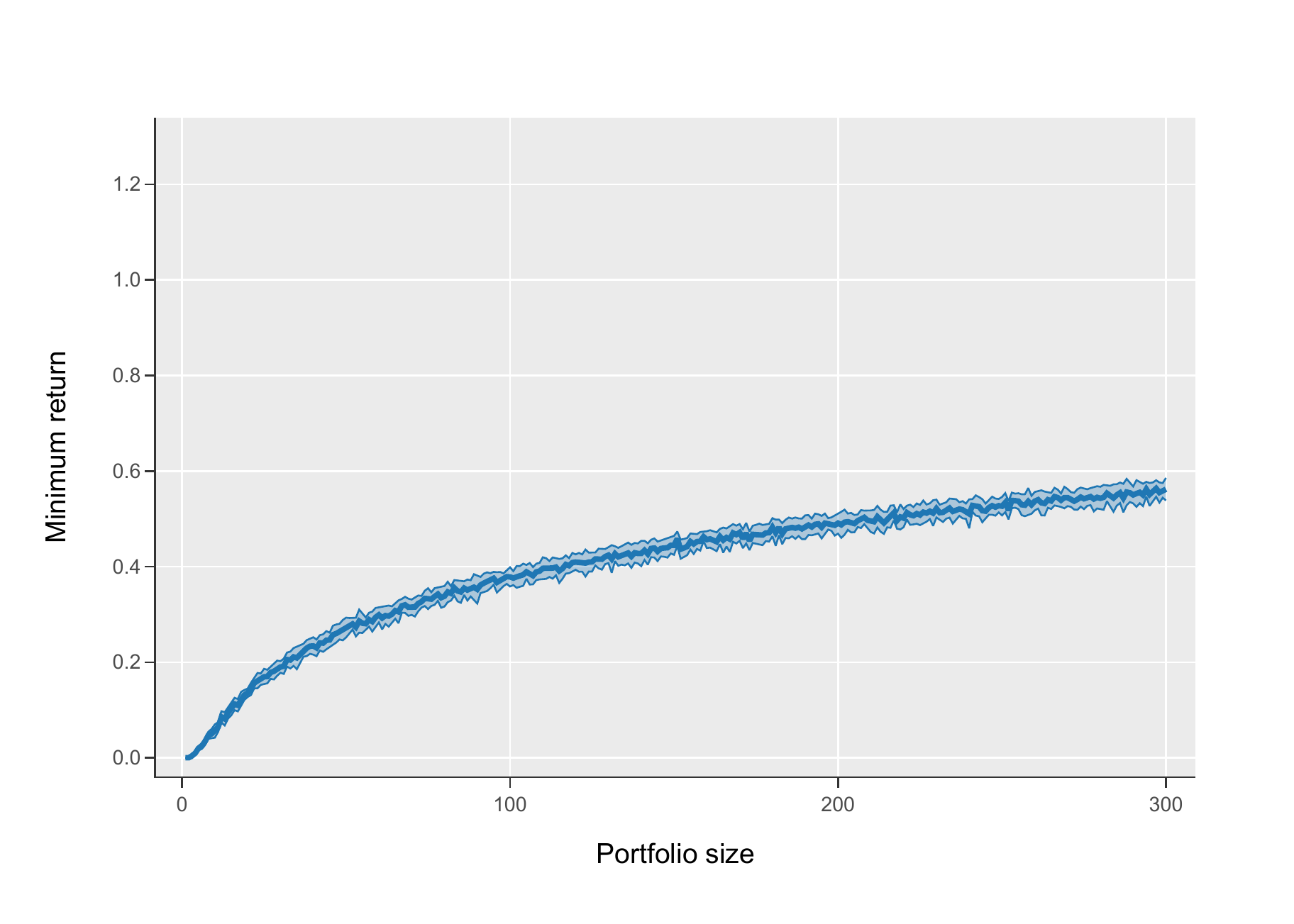}
            \caption{Minimum return}
            \label{fig:min_return_avg_follow_on_50}
        \end{subfigure}
        \begin{subfigure}[t]{0.32\linewidth}
            \includegraphics[width=\textwidth]{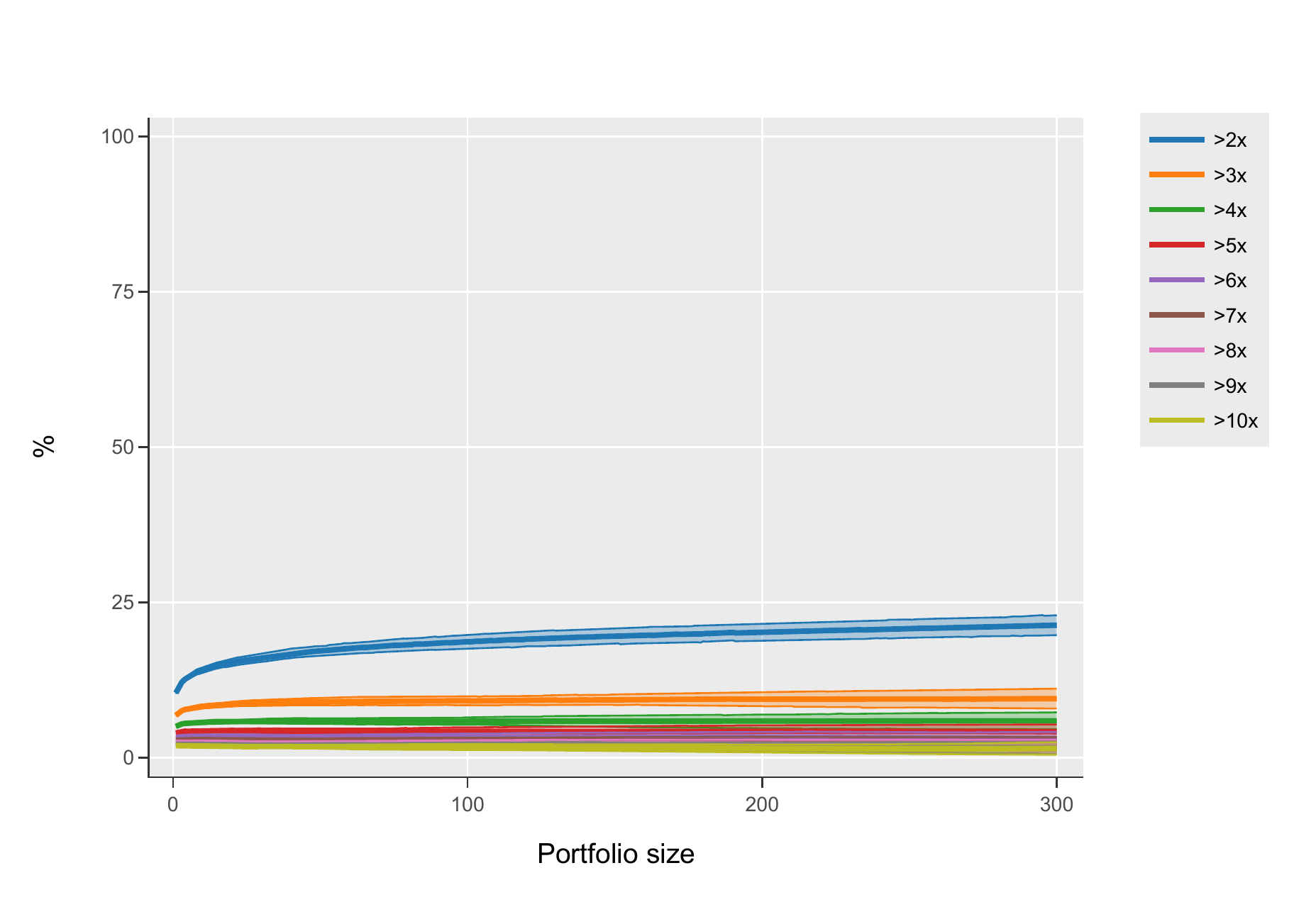}
            \caption{Frequency of 2-10x returns}
            \label{fig:x_returns_avg_follow_on_50}
        \end{subfigure}
        \caption{Risk profile and returns for different portfolio sizes (mean and standard deviation) using a 50\% follow-on reserve.}
        \label{fig:risk_profile_avg_follow_on_50}
    \end{figure}

    As we can see in Figure~\ref{fig:risk_profile_avg_follow_on_50}, the drop in performance is quite dramatic. While the overall trend is the same as before, the probabilities of high returns (and \textbf{not} losing money), are much lower with respect to the scenario of no follow-ons.

    That said, follow-ons are usually quite important to one's strategy. Not doing so can be perceived as a bad signal by other investors, and may negatively affect the chances of future investment rounds. It suggests that one needs to devise strategies to mitigate the influence of follow-ons on fund performance.

    \subsection{Follow-on reserve: how much?}
    The size of the follow-on reserve is probably the main parameter one can tune. 
    Let's consider different reserve sizes $r$, as a fraction of the initial investment. We'll simulate $r=0,0.1, 0.2, \dots, 0.9$.

    \begin{figure}[h!]
        \centering
        \begin{subfigure}[t]{0.49\textwidth}
            \includegraphics[width=\textwidth]{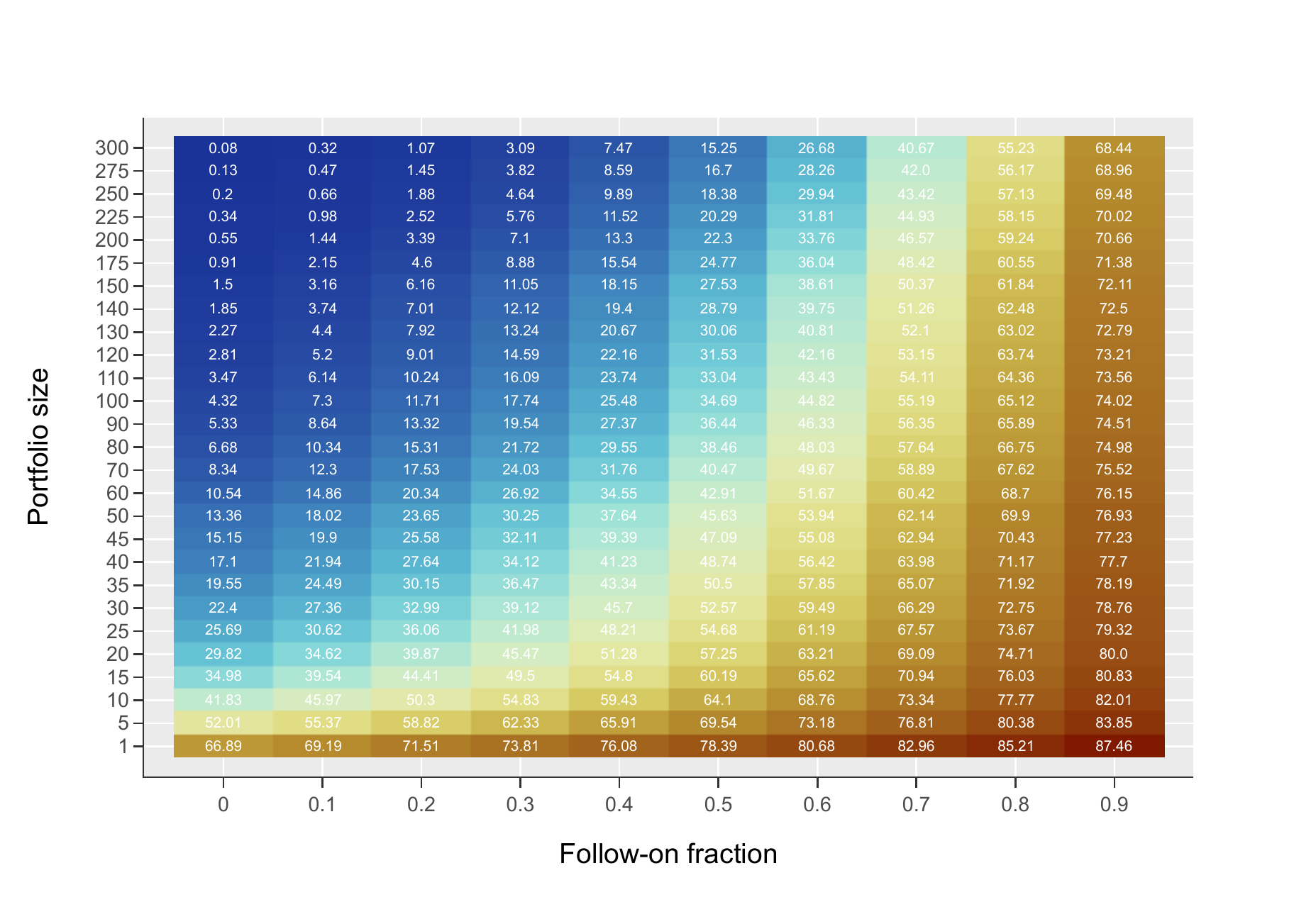}
            \caption{Probability of losing money.}
            \label{fig:follow_on_prob_losing}
        \end{subfigure}
        \begin{subfigure}[t]{0.49\textwidth}
            \includegraphics[width=\textwidth]{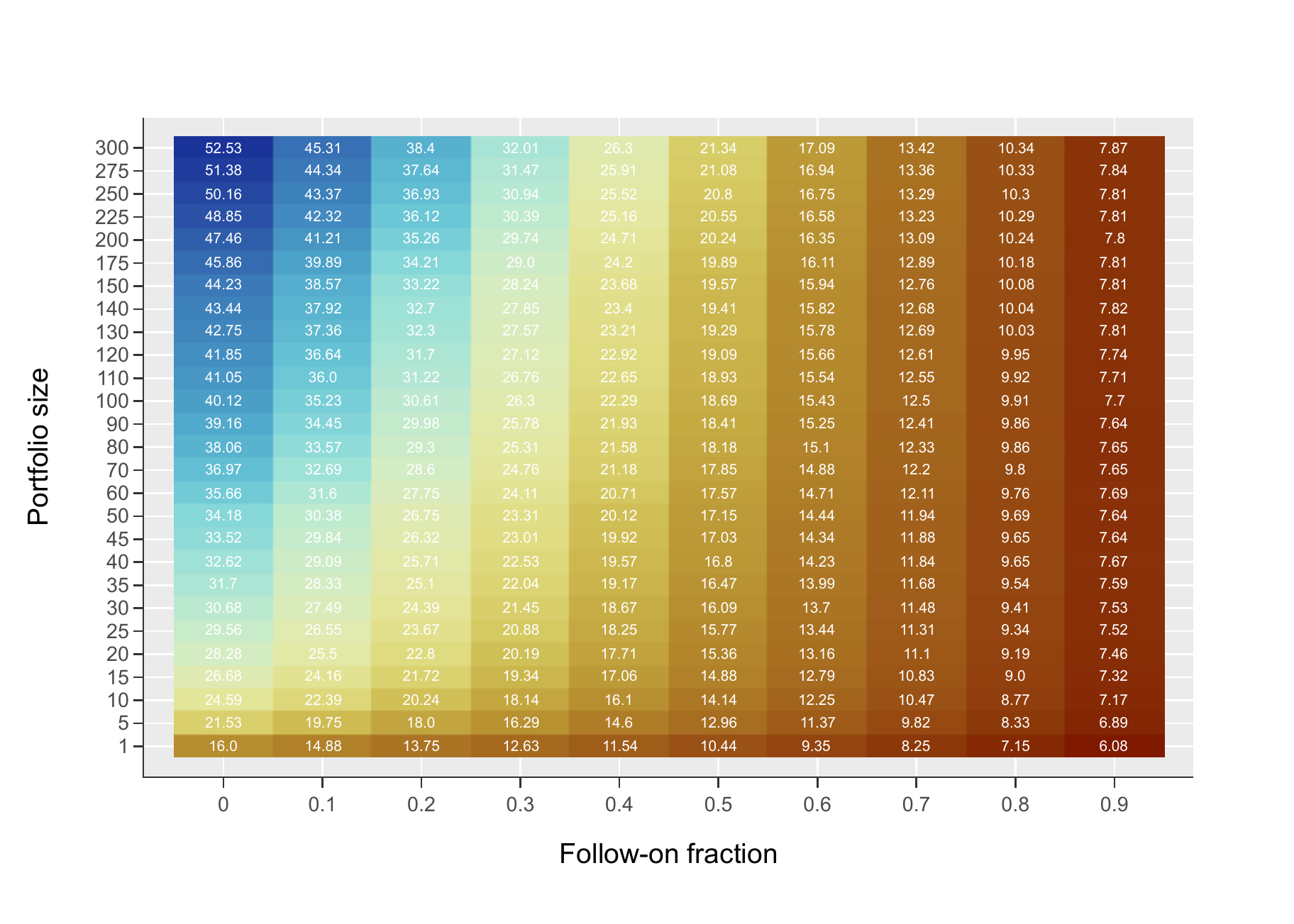}
            \caption{Probability of returning more than 2x the fund.}
            \label{fig:follow_on_prob_2x}
        \end{subfigure}
        \caption{Effect of different follow-on reserve amounts.}
    \end{figure}
    
    Figure~\ref{fig:follow_on_prob_losing} shows the probability of losing money. It's clear that increasing the follow-on allocation increases the probability of losses too. As an example, a portfolio with 50 investments and no follow-on has the same risk of a portfolio of 90 investments with 20\% allocated to follow-ons.

    Now, look at the probabilities of returning 2x the fund (Figure~\ref{fig:follow_on_prob_2x}). Again, allocating a larger portion of the fund to follow-ons reduces the probability of returning 2x the fund. As shown in Figure~\ref{fig:follow_on_grid_heatmaps}, in some cases the effect can be mitigated by increasing the portfolio size. However, when looking at the probability of returning more than 5x the fund, the effect of follow-ons can't be mitigated.

    \subsubsection{Impact of bounded ROI and follow-ons}
\begin{figure}[htbp!]
        \centering
        \includegraphics[width=0.75\textwidth]{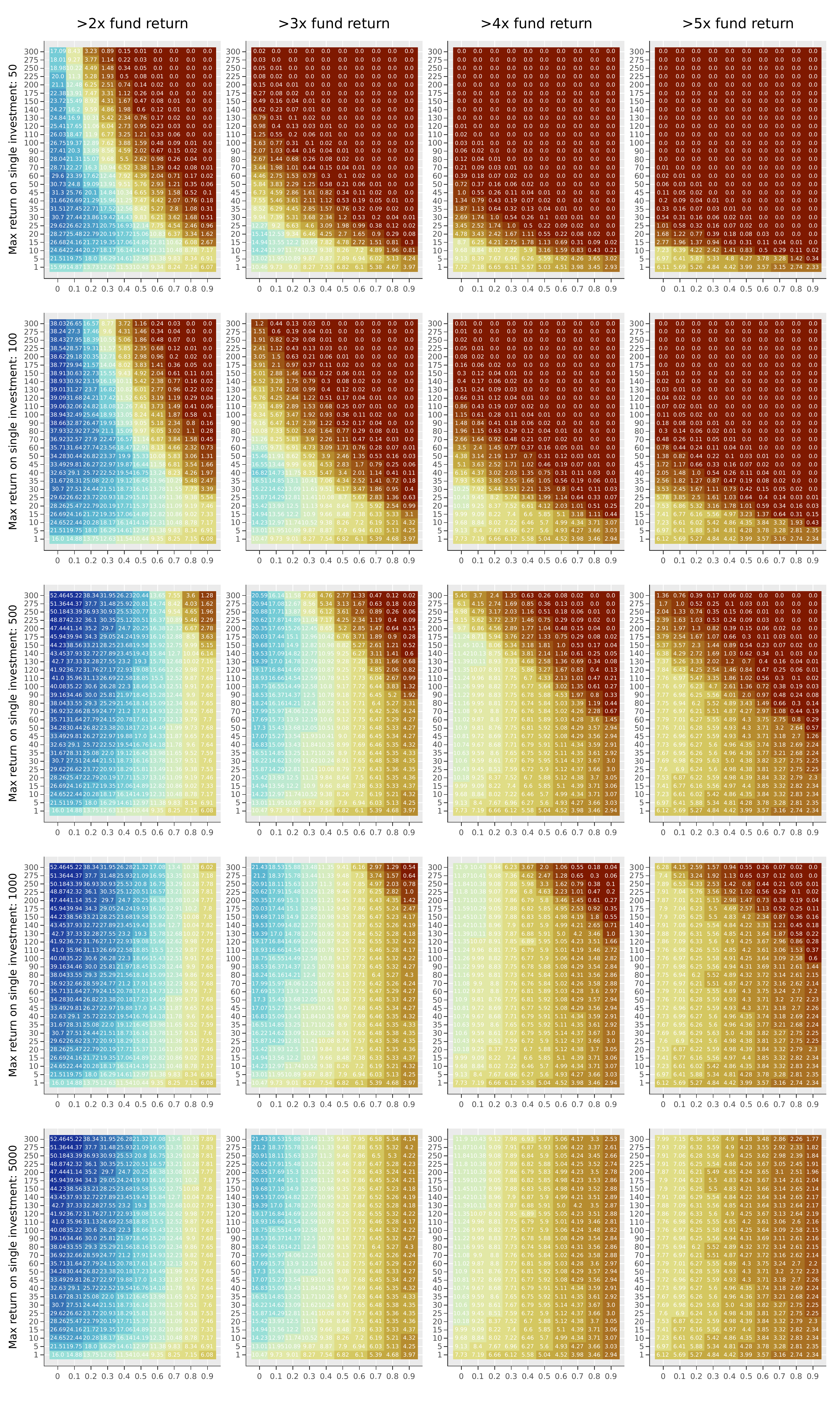}
        \caption{Probability of returning a multiple of the fund for different follow-on fractions.}
        \label{fig:follow_on_grid_heatmaps}
    \end{figure}

    What's the impact of follow-ons when the return per investment is bounded? The different cases we've considered so far are reported in Figure~\ref{fig:follow_on_grid_heatmaps}. As one can see, the behaviour is non linear: depending on the bound, one may need to either increase or decrease the portfolio size in order to mitigate the effect of follow-ons. All in all, the impact of follow-ons is dramatic.
    
    \begin{msgquote}{Take-home}
        In general, one should aim for the lowest amount possible allocated to follow-ons.
    \end{msgquote}

    \subsection{Selective follow-on strategy}
    So far, we've assumed that we're following-on on all the investments. In reality, that's not the case. One usually follows-on just on portfolio companies with high potential. Results show that the distribution of returns for later-stage companies is quite different with respect to early-stage ones~\cite{2017_venture_returns_late}.

\begin{figure}[h!]
        \centering
        \includegraphics[width=0.75\textwidth]{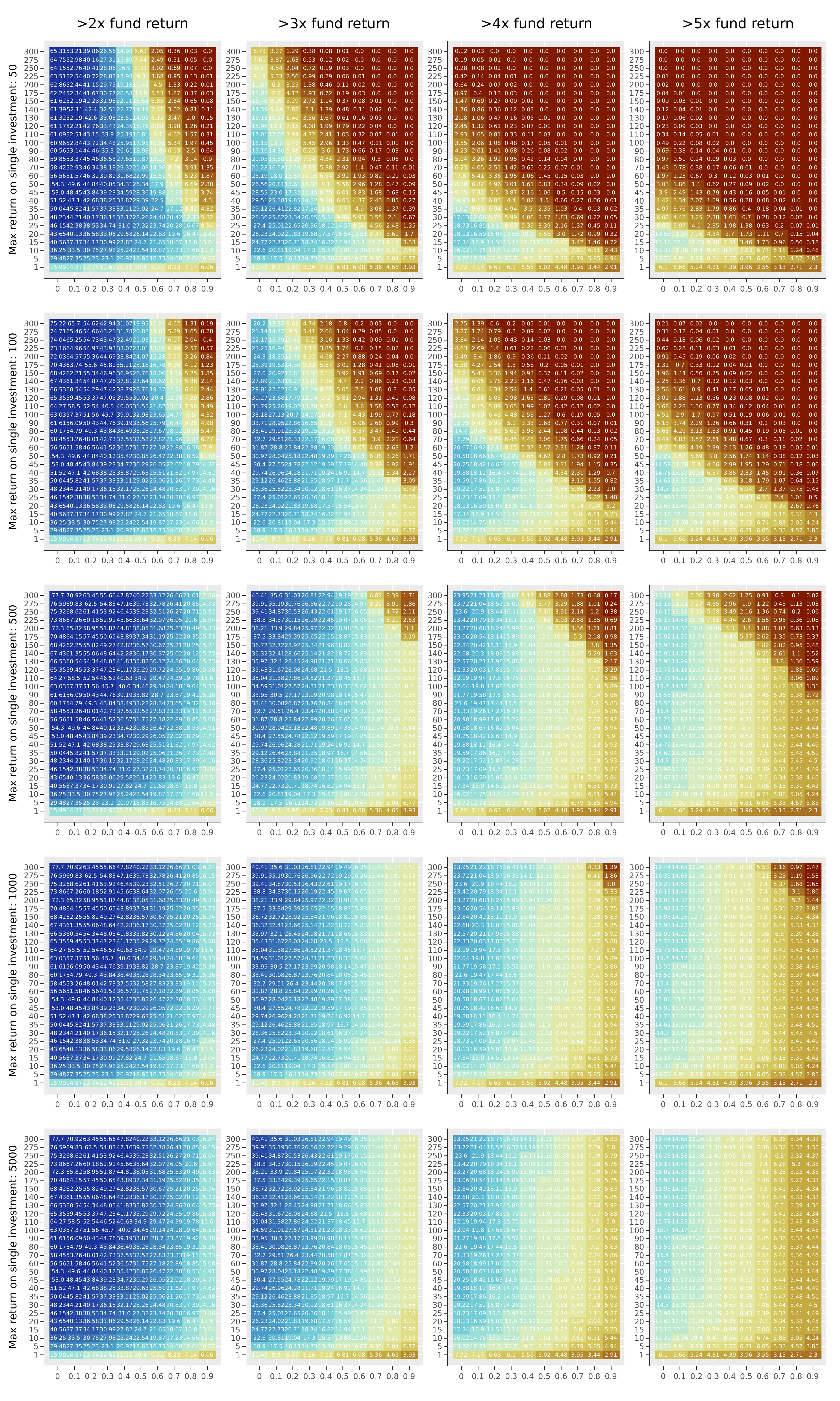}
        \caption{Probability of returning a multiple of the fund for different follow-on fractions using a selective follow-on strategy.}
        \label{fig:follow_on_grid_heatmaps_good}
    \end{figure}

    Let's assume a fund manager has some ability to discern companies that are going to grow from those that are not, resulting in them doing follow-ons on just 70\% of investments that will return 0-1x and on 90\% of those that will return more than 1x. Results are shown in Figure~\ref{fig:follow_on_grid_heatmaps_good}.

    In this scenario, if the bound of the ROI is not too low, larger portfolios usually mitigate the effect of follow-ons on expected returns.

    \begin{msgquote}{Take-home}
        Taking this argument to its extreme, one should follow-on only when doing so is the best use of the available capital – meaning that the return one expects from the follow-on investment is higher than the return one expects from any other initial investment they could make until the end of the fund deployment period.
    \end{msgquote}

    \section{What if we're wrong about the returns distribution?}
    So far we've assumed that returns follow a power law distribution with a specific set of parameters, fitted to returns we saw in the past. What if those parameters need to be changed to fit the future? Let's consider two scenarios: one in which returns are better than what we assumed, and one in which returns are worse than what we assumed.
   
    \subsection{If returns won't be as good as we think}
    Let's consider the scenario in which returns still follow a power law distribution, but 0-1x returns have a higher frequency, meaning that there will be more bad opportunities. To simulate this, we'll set $\alpha=2.3$.

\begin{figure}[h!]
        \centering
        \begin{subfigure}[t]{0.32\linewidth}
            \includegraphics[width=\textwidth]{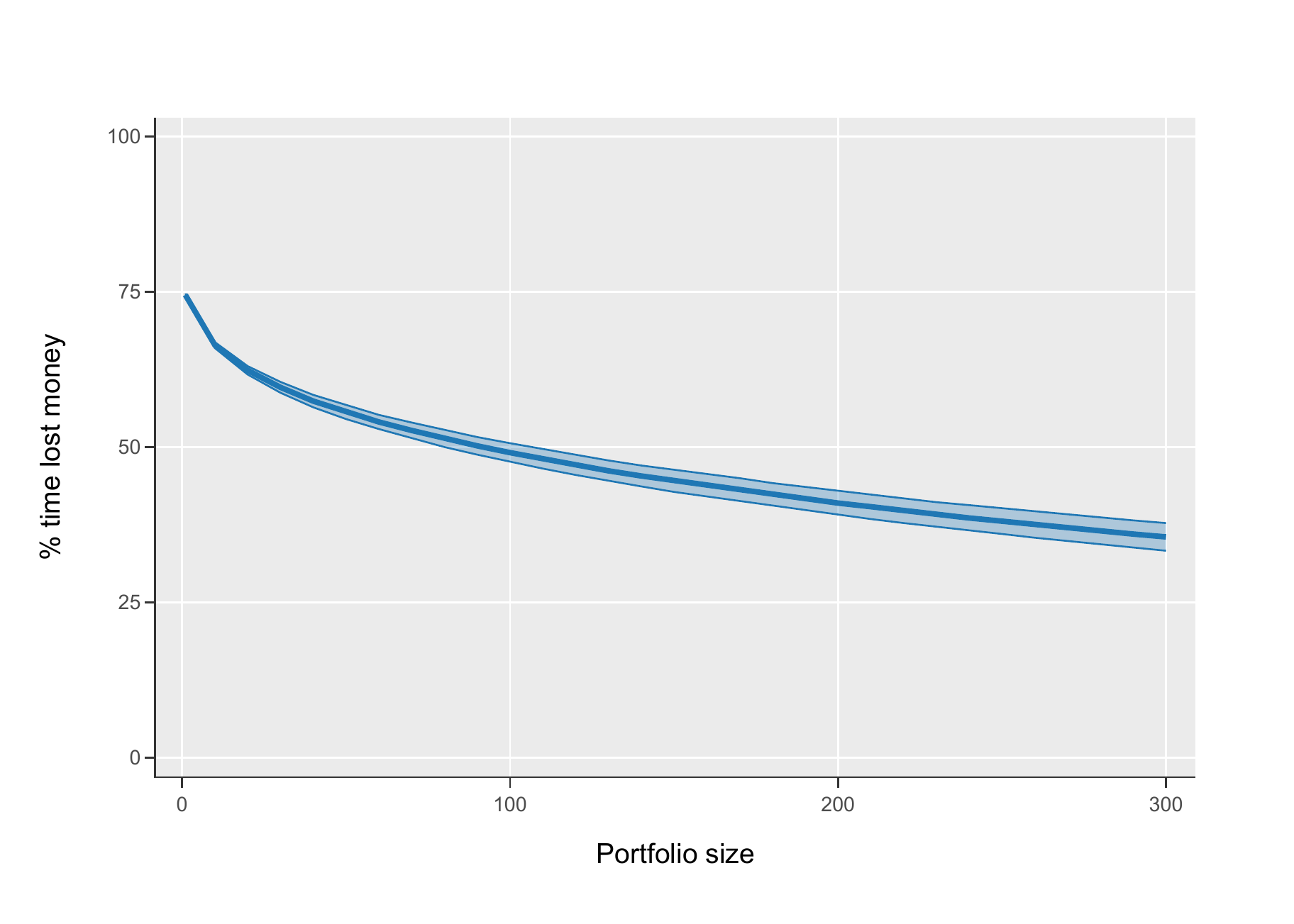}
            \caption{Percentage of portfolios losing money}
            \label{fig:pc_time_lost_money_bad_world}
        \end{subfigure}
        \begin{subfigure}[t]{0.32\linewidth}
            \includegraphics[width=\textwidth]{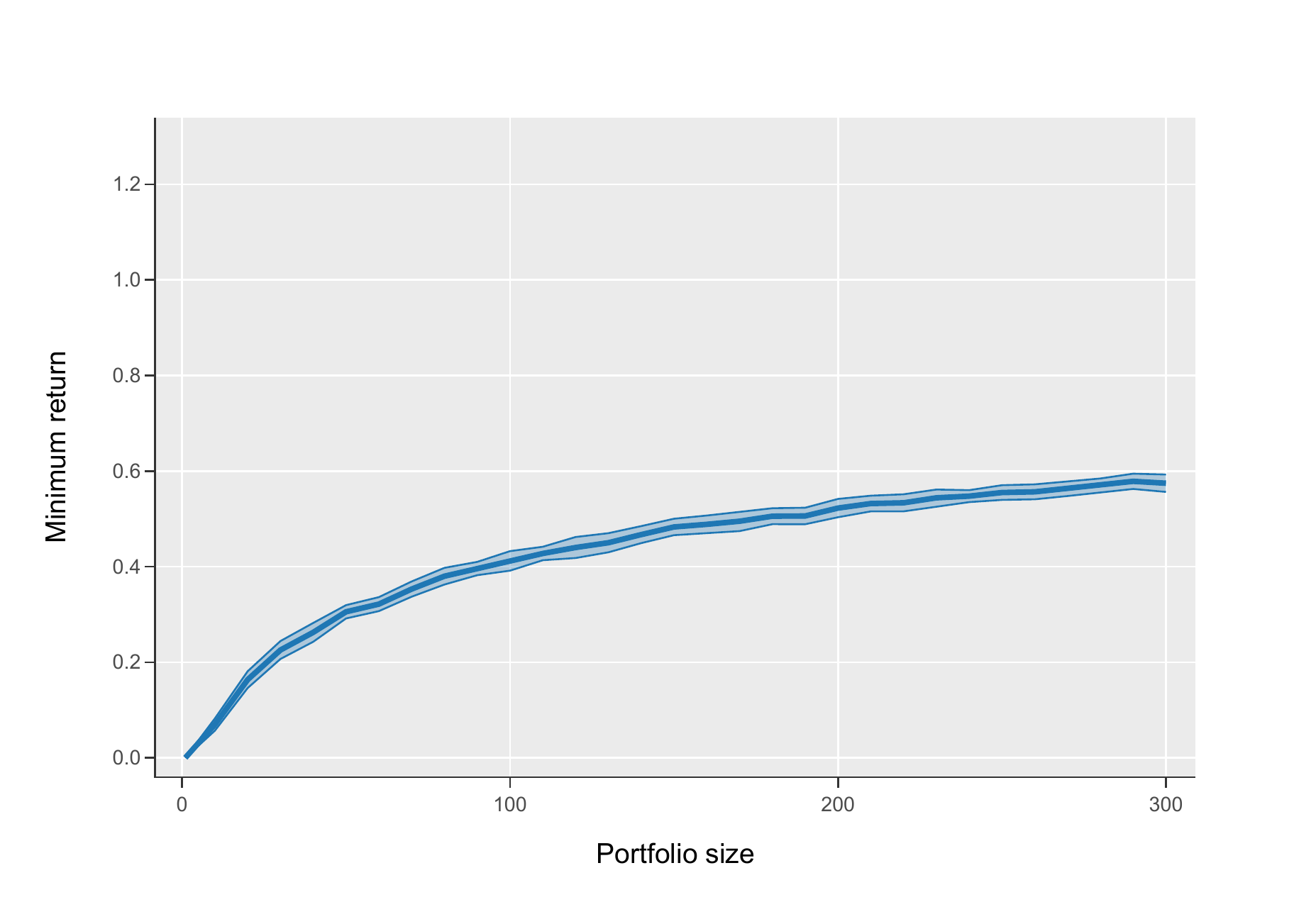}
            \caption{Minimum return}
            \label{fig:min_return_bad_world}
        \end{subfigure}
        \begin{subfigure}[t]{0.32\linewidth}
            \includegraphics[width=\textwidth]{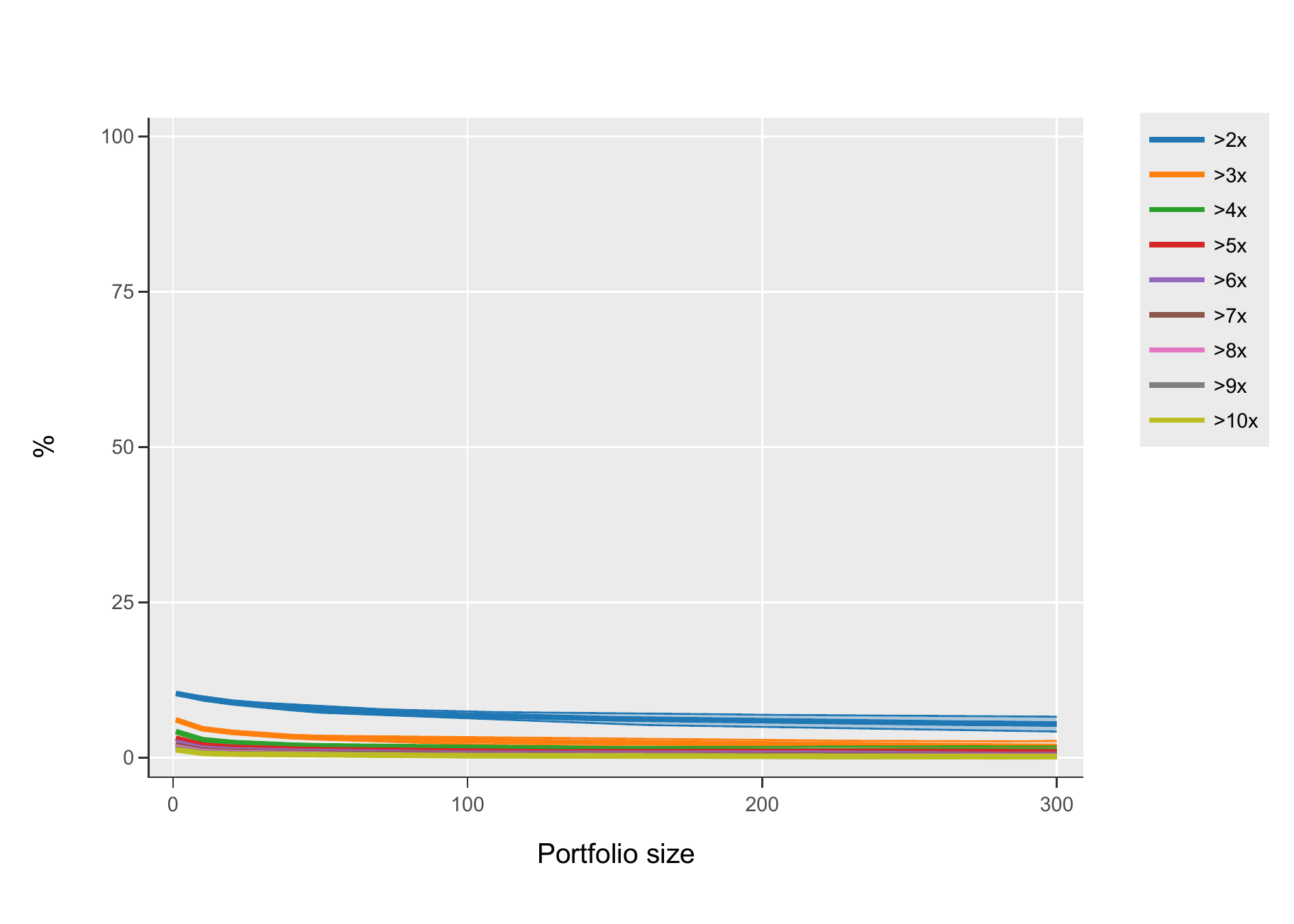}
            \caption{Frequency of 2-10x returns}
            \label{fig:x_returns_bad_world}
        \end{subfigure}
        \caption{Risk profile and returns for different portfolio sizes (mean and standard deviation) in a bad world.}
        \label{fig:risk_profile_bad_world}
    \end{figure}

    Let's look at the average VC performance in this case (Figure~\ref{fig:risk_profile_bad_world}). For the average fund, the risk of losing money increases with portfolio size, but it's always pretty high - returning multiples of the fund is always hard, but harder for large portfolios.

    \begin{helpquote}{Why does this happen?}
        That's because, on average, the number of high-return investments is not enough to offset the number of bad investments.
    \end{helpquote}

    \subsubsection{Overperformers still strive}
    What if, in this bad scenario, one is still able to overperform and pick good deals with a higher frequency? Let's simulate that with $\alpha=2.05$.

\begin{figure}[h!]
        \centering
        \begin{subfigure}[t]{0.32\linewidth}
            \includegraphics[width=\textwidth]{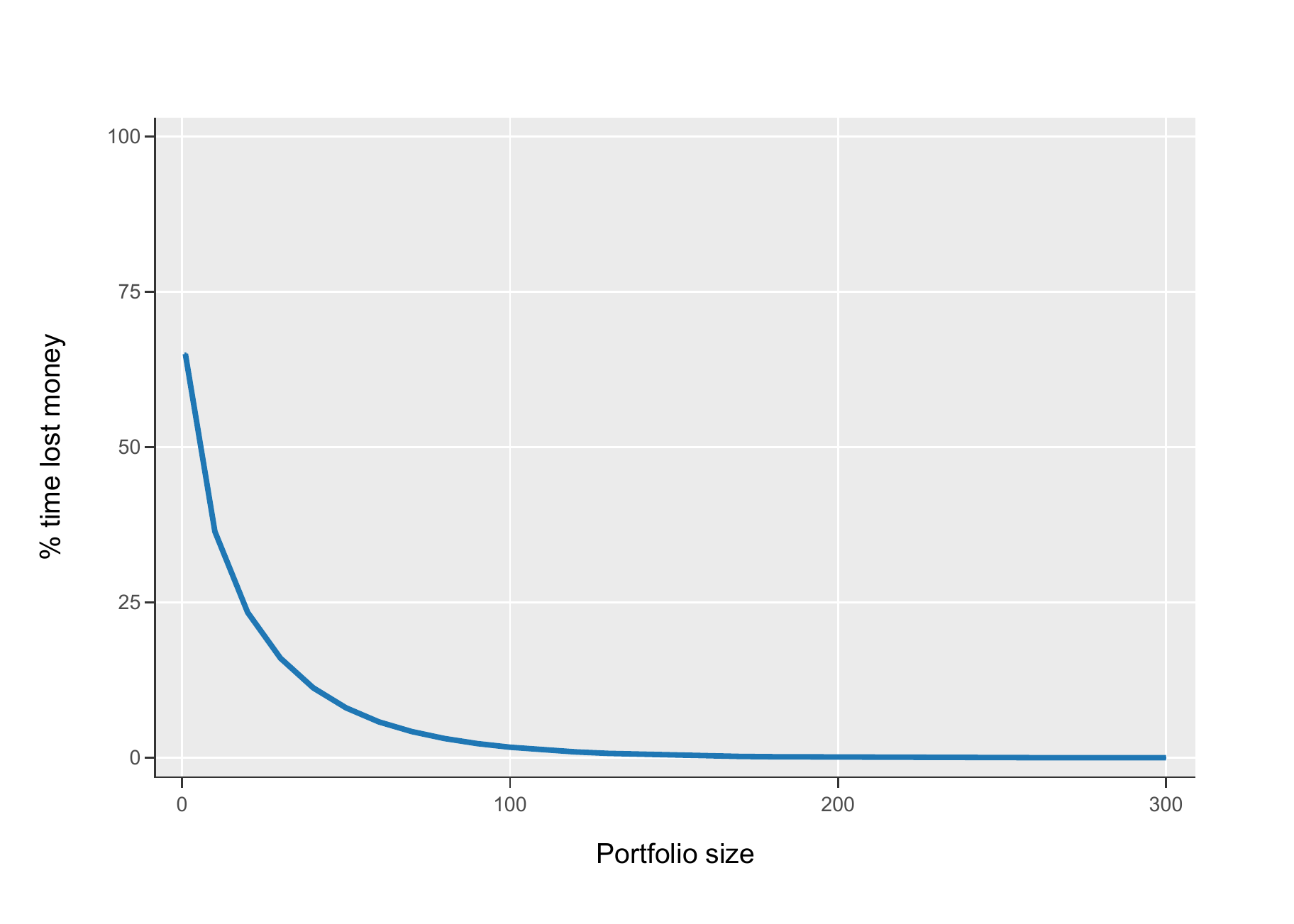}
            \caption{Percentage of portfolios losing money}
            \label{fig:pc_time_lost_money_bad_world_overperformer}
        \end{subfigure}
        \begin{subfigure}[t]{0.32\linewidth}
            \includegraphics[width=\textwidth]{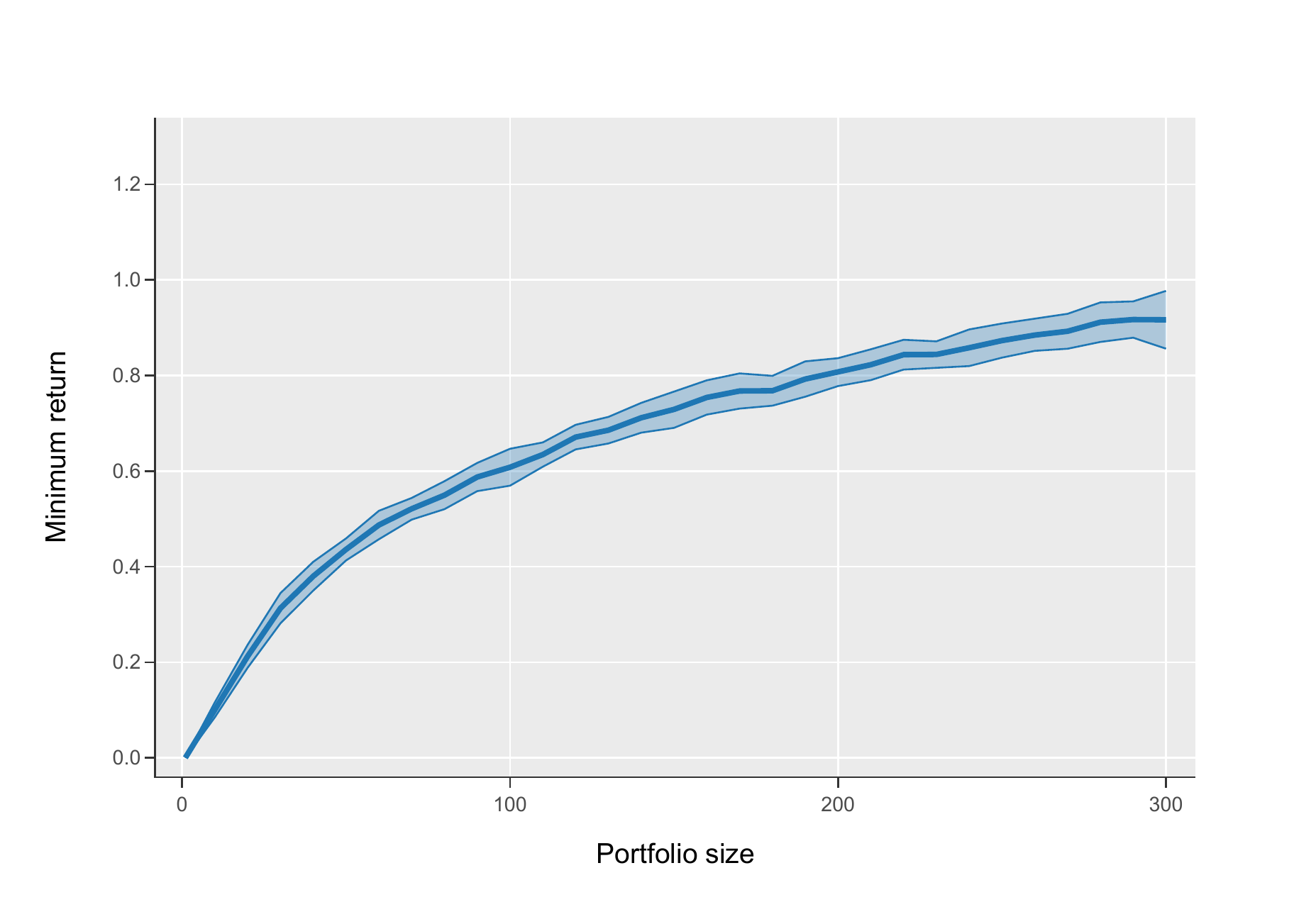}
            \caption{Minimum return}
            \label{fig:min_return_bad_world_overperformer}
        \end{subfigure}
        \begin{subfigure}[t]{0.32\linewidth}
            \includegraphics[width=\textwidth]{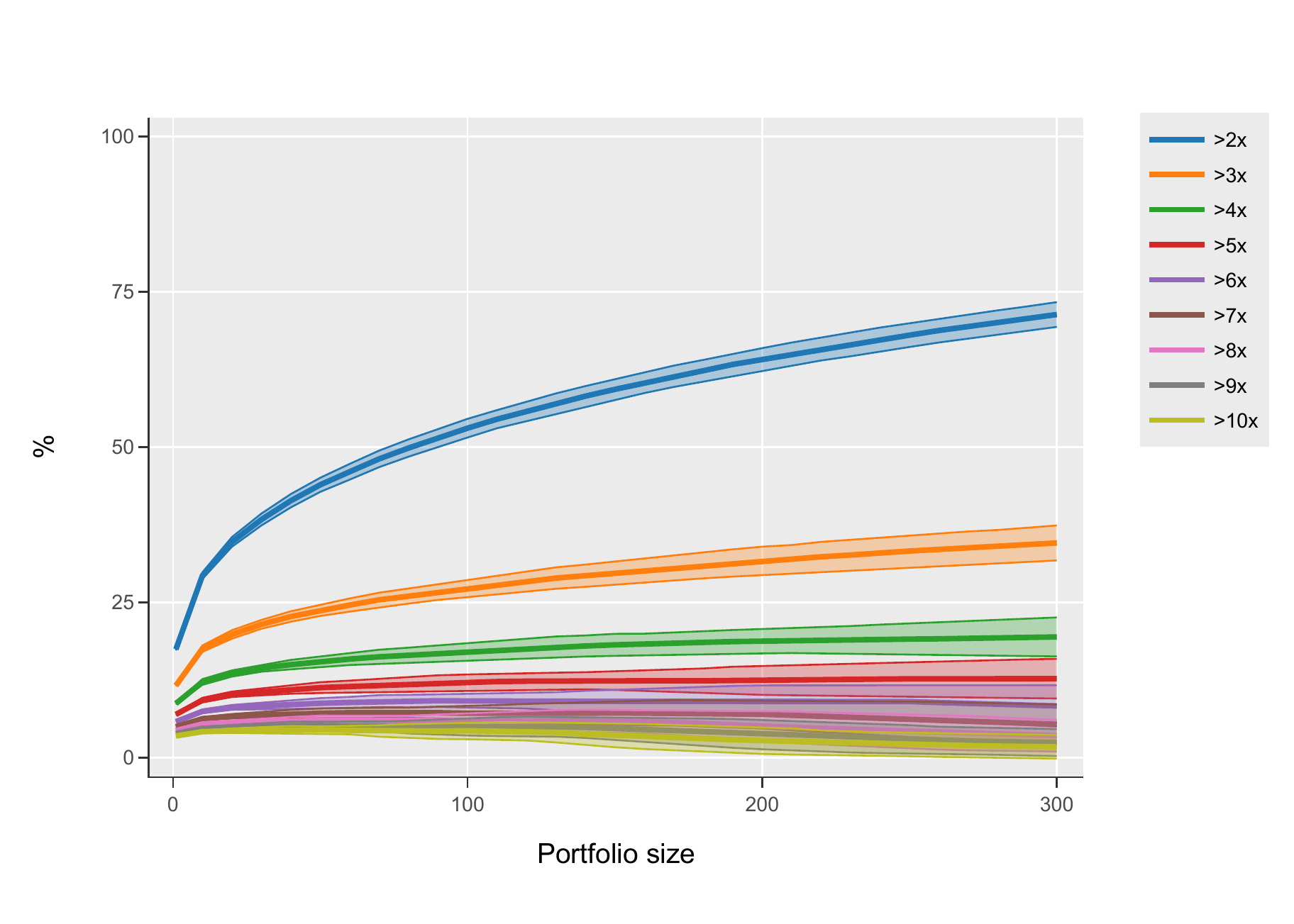}
            \caption{Frequency of 2-10x returns}
            \label{fig:x_returns_bad_world_overperformer}
        \end{subfigure}
        \caption{Risk profile and returns for different portfolio sizes (mean and standard deviation) for an overperfomer in a bad world.}
        \label{fig:risk_profile_bad_world_overperformer}
    \end{figure}

    In Figure~\ref{fig:risk_profile_bad_world_overperformer} we can see that, even in this case, overperformers can dramatically reduce their risk by increasing their portfolio size, while still increasing their probability of returning reasonable multiples of the fund. For example, the probability of a 5x still peaks at around $N=100$, while the probabilities of 2x, 3x and 4x continue to increase with the size. However, the probability of higher returns starts to slightly decrease after $N=150$.

    \subsection{If returns will be better than we think}
    Now, let's assume that 0-1x returns happen with a lower frequency, meaning that there will be fewer bad opportunities. To simulate this, we'll set $\alpha=1.85$.

\begin{figure}[h!]
        \centering
        \begin{subfigure}[t]{0.32\linewidth}
            \includegraphics[width=\textwidth]{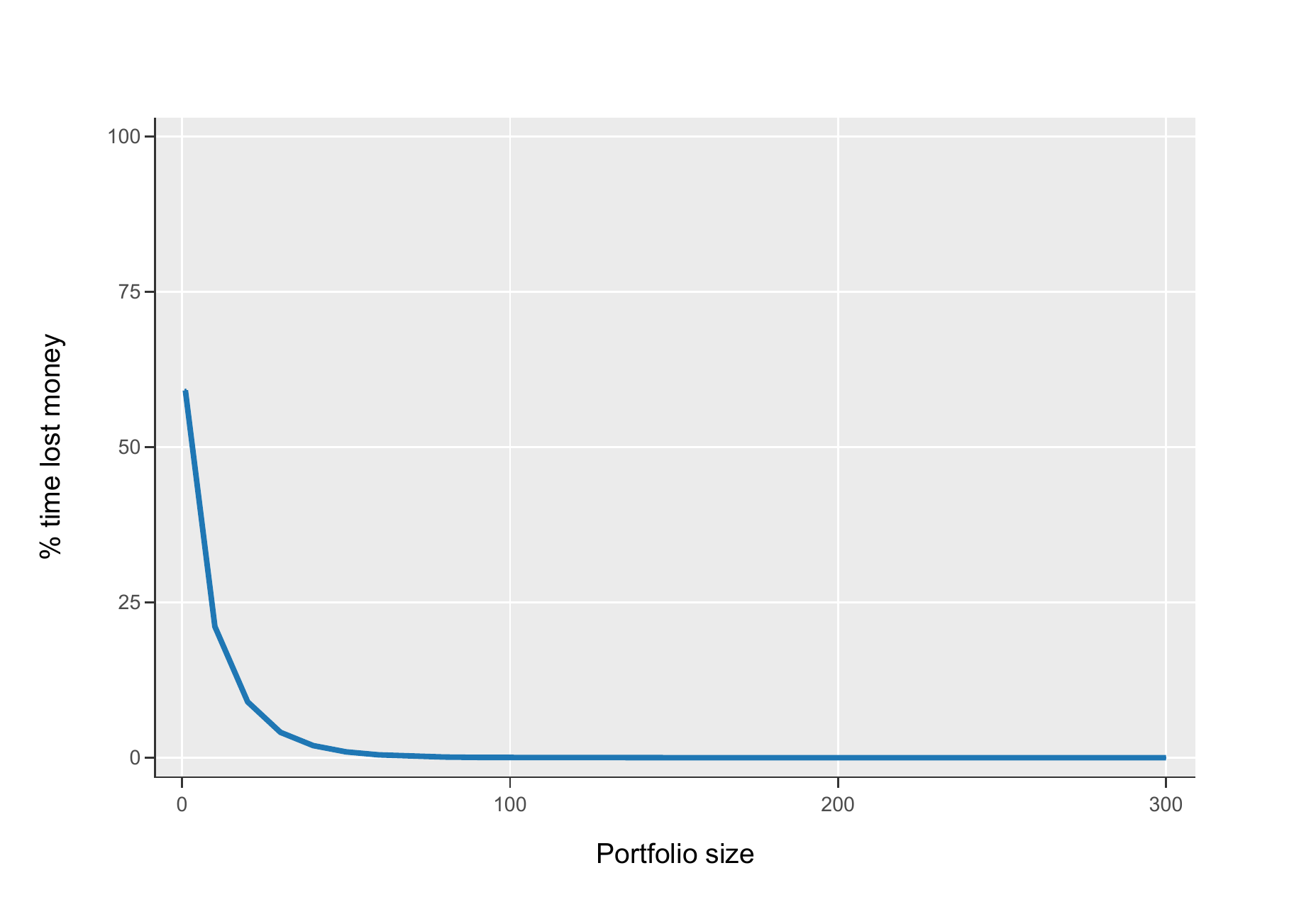}
            \caption{Percentage of portfolios losing money}
            \label{fig:pc_time_lost_money_good_world}
        \end{subfigure}
        \begin{subfigure}[t]{0.32\linewidth}
            \includegraphics[width=\textwidth]{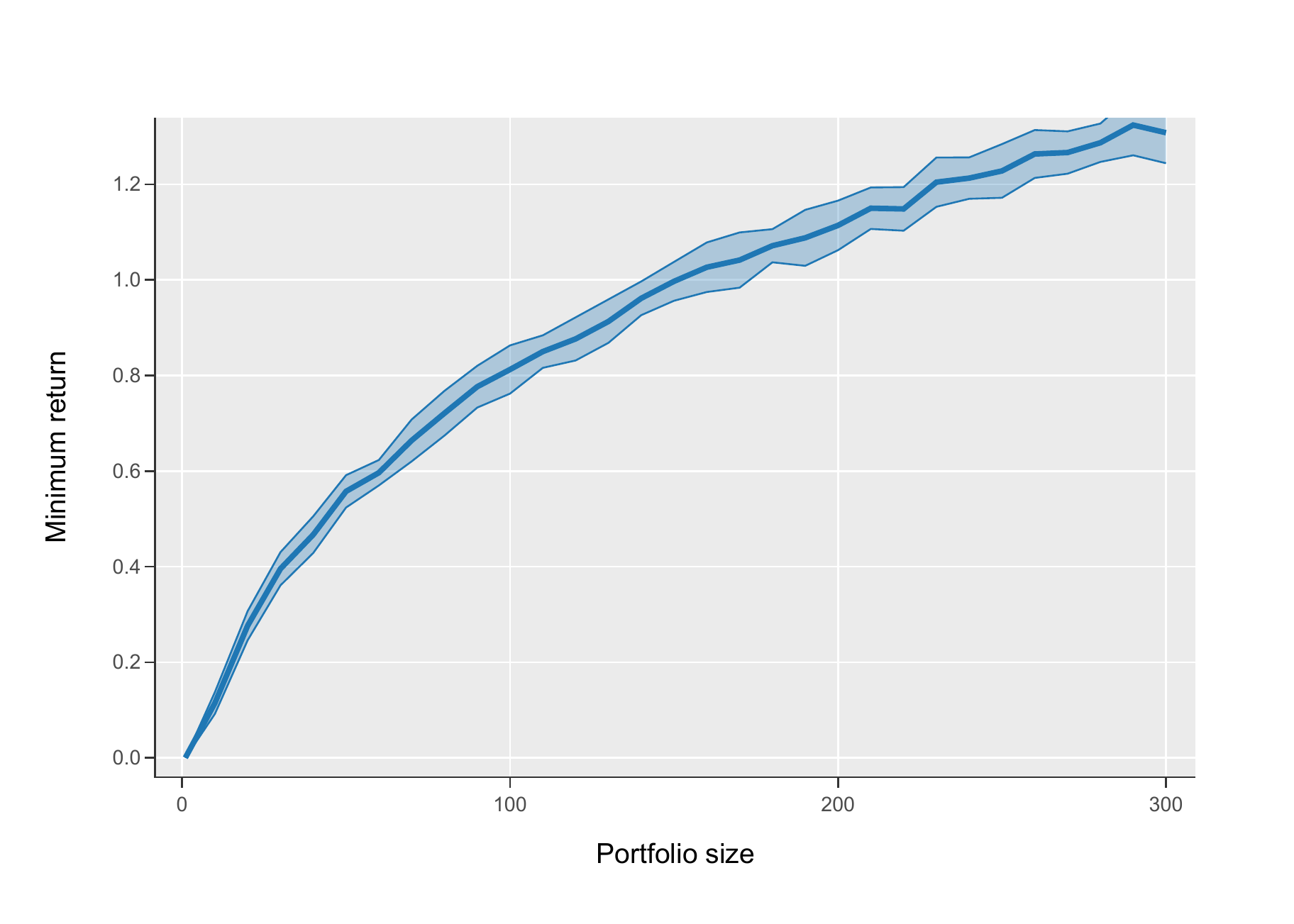}
            \caption{Minimum return}
            \label{fig:min_return_good_world}
        \end{subfigure}
        \begin{subfigure}[t]{0.32\linewidth}
            \includegraphics[width=\textwidth]{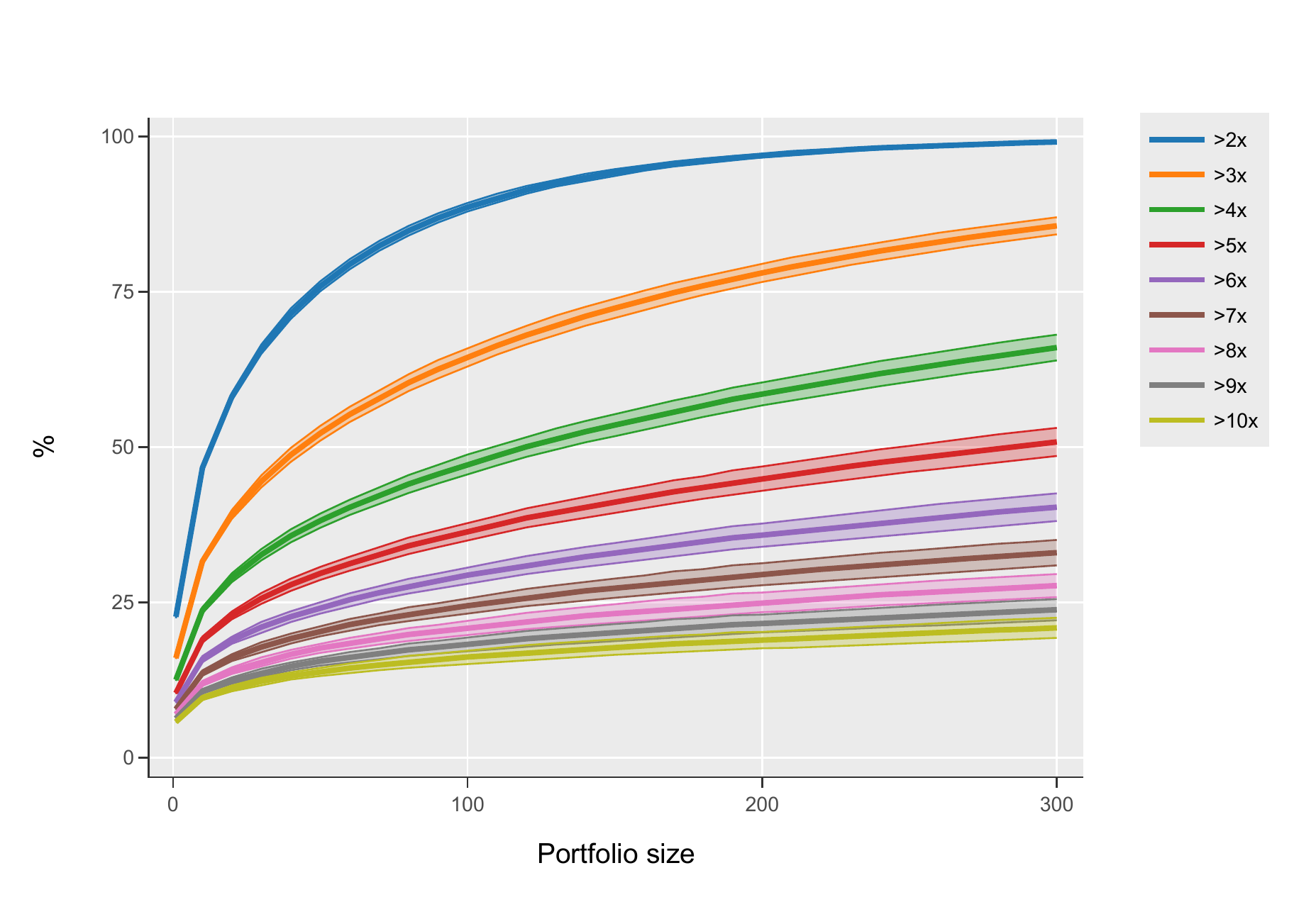}
            \caption{Frequency of 2-10x returns}
            \label{fig:x_returns_good_world}
        \end{subfigure}
        \caption{Risk profile and returns for different portfolio sizes (mean and standard deviation) in a good world.}
        \label{fig:risk_profile_good_world}
    \end{figure}

    Let's look at the average VC performance in this case (Figure~\ref{fig:risk_profile_good_world}). In this scenario, everyone behaves as they were an over-performer in an average world.
    
    \pagebreak

    \section{Conclusion}
    As we've discussed in this article, five main factors affect portfolio performance: decision quality, portfolio size, ticket size, follow-on policy, and the upper bound on ROI on a single investment. Of these factors, the upper bound on returns and decision quality appear to have the greatest impact.
    
    There are, of course, a lot more variables that one may wish to take into account. For example, one could add a time horizon to the model, which would add nuance to questions about strategy. The optimal strategy for achieving 10x in two years will be very different to that for achieving 10x in five years.
    
    One could also model deal flow. The upper bound on ROI acts like a very rough proxy for deal flow, if we assume that VCs have a good ability to predict which companies will be the most significant returners. But this is something that could be modelled more accurately to account for investors’ different levels of access.
    
    Being primarily a pre-seed and seed-stage investor, early-stage venture portfolios were our focus with this research. Later-stage venture and private equity follow completely different probability distributions, marked by lower risk and lower upside, and would require different core parameters and a different model. This would be worthy of exploration.
    
    Finally, it is worth reiterating that, regardless of the number of variables modelled, it will always be hard to settle on a general formula for success. It is a function of many factors, so one should start with the objective of the fund: what do you want it to achieve? With that in place, you can use the known effects of these variables to navigate a more mathematically sound course.
    
    If you want to explore our models in a more fine-grained, interactive way, try our portfolio simulator tool~\cite{2023_portfolio_simulator}.
  
    \newpage

    \bibliographystyle{unsrt}
    \bibliography{biblio.bib}

    \newpage
    \appendix
    \section{Power law statistics}
    \paragraph{Expected Value}
    The expected value of a random variable distributed according to~\eqref{eq:powerlaw} is
    \begin{align}
        \mathbb{E}[X] &= \int_{\xmin}^{\infty} x f(x) \partial x \nonumber \\
        &= (\alpha-1)\xmin^{\alpha-1} \int_{\xmin}^{\infty} x^{1-\alpha}\partial x \nonumber \\
        &= \frac{\alpha-1}{2-\alpha}\xmin^{\alpha-1} \left[ x^{2-\alpha}\right]_{\xmin}^\infty \label{eq:expected_value_pre}
    \end{align}
    Notice that~\eqref{eq:expected_value_pre} becomes infinite for $\alpha \leq2$. In fact, power law distributions with $\alpha\leq 2$ has no finite expected value. If we assume $\alpha> 2$, then the expected value is
    \begin{equation}
        \label{eq:powerlaw_expected}
        \mathbb{E}[X] = \frac{\alpha-1}{\alpha-2} \xmin
    \end{equation}
    \paragraph{Median}
    The median of a random variable distributed according to~\eqref{eq:powerlaw} is defined as the value $x_{1/2}$ such that $P[X<x_{1/2}]=0.5$. This can be easily derived by solving the equation
    \begin{equation}
        \label{eq:powerlaw_median}
        F(x_m) = \frac{1}{2}
    \end{equation}
    which gives
    \begin{equation}
        \label{eq:powerlaw_median_sol}
        x_{1/2} = 2^{1/(\alpha-1)}\xmin
    \end{equation}
    \paragraph{Higher moments}
    The higher moments of a power law distribution can be computed similarly to the mean and are defined as 
    \begin{equation}
        \label{eq:powerlaw_moments}
        \mathbb{E}[X^k] = \frac{\alpha-1}{\alpha-k-1} \xmin^k
    \end{equation}
    In general, moments for which $k\geq \alpha-1$ are infinite.
    \paragraph{Expected value of the maximum in a sample}
    The expected value of the maximum in a sample of $n\gg 1$ independent random variables distributed according to~\eqref{eq:powerlaw} is
    \begin{equation}
        \label{eq:powerlaw_max}
        \mathbb{E}[\text{max}\lbrace X_1,\dots, X_n\rbrace] \sim n^{\tfrac{1}{\alpha-1}}
    \end{equation}
    
\end{document}